\documentclass[12pt]{iopart}
\pdfoutput=1
\usepackage{iopams}
\usepackage{ulem}

\expandafter\let\csname equation*\endcsname\relax
\expandafter\let\csname endequation*\endcsname\relax
\usepackage{amsmath,amssymb,mathtools}

\usepackage[colorlinks,linkcolor=blue,urlcolor=blue,citecolor=blue]{hyperref}

\usepackage{graphicx,color}
\usepackage{epsfig}
\usepackage{cite}

\usepackage{amsfonts}
\usepackage{euscript}
\usepackage{amsthm}
\usepackage{newlfont}
\usepackage{mathrsfs}
\usepackage{subfigure}

\usepackage{titlesec}
\titleclass{\subsubsubsection}{straight}[\subsection]

\newcounter{subsubsubsection}[subsubsection]

\renewcommand\thesubsubsubsection{\thesubsubsection.\arabic{subsubsubsection}}

\titleformat{\subsubsubsection}
  {\normalfont\normalsize\bfseries}{\thesubsubsubsection}{1em}{}
\titlespacing*{\subsubsubsection}
{0pt}{3.25ex plus 1ex minus .2ex}{1.5ex plus .2ex}

\def\be{\begin{equation}}
\def\ee{\end{equation}}
\def\bse{\begin{subequations}}
\def\ese{\end{subequations}}
\def\bcs{\begin{cases}}
\def\ecs{\end{cases}}
\def\bea{\begin{eqnarray}}
\def\eea{\end{eqnarray}}

\def\ba{\begin{array}}
\def\ea{\end{array}}

\definecolor{dgreen}{rgb}{0,0.7,0}

\newcommand{\Aref}[1]{Appendix}%
\newcommand{\aref}[1]{Appendix~\ref{#1}}%

\begin{document}

\title[Dynamics of a randomly kicked particle]{Dynamics of a randomly kicked particle}
\author{Santanu Das, Anupam Kundu}

\address{International Center for Theoretical Sciences, TIFR, Bangalore 560089, India}

\begin{abstract}
L\'evy walk (LW) process has been used as a simple model for describing anomalous diffusion in which the mean squared displacement of the walker grows non-linearly with time in contrast to the diffusive motion described by simple random walks or Brownian motion. In this paper we study a simple extension of the LW model in one dimension by introducing correlation among the velocities of the walker in different (flight) steps. Such correlation is absent in the LW model. The correlations are introduced by making the velocity at a step dependent on the velocity at the previous step in addition to the usual random noise (`kick') that the particle gets at random time intervals from the surrounding medium as in the LW model.  Consequently the dynamics of the position becomes non-Markovian. We study the  statistical properties  of  velocity and position of the walker at time $t$, both analytically and numerically. We show how different choices of the distribution of the random time intervals and the degree of correlation, controlled by a parameter $r$, affect the late time behaviour of these quantities.



\end{abstract}

\section{Introduction}
\label{intro}

Dynamics of a particle kicked at random by the surrounding medium is an exciting and widely investigated problem for studying wide varieties of stochastic processes in the field of physics, chemistry and biology \cite{van1992stochastic,ricciardi2013diffusion,chandrasekhar1943stochastic,taylor1922diffusion,
chong2020vortices,jaeger1996granular,barrat2005granular,mcwilliams1987transport,bechinger2016active}.
There are numerous examples of such dynamics, which include the movement of a pollen grain in water or a dust particle in air  \cite{chandrasekhar1943stochastic,ricciardi2013diffusion},
tracer particle dynamics  in turbulent flows \cite{taylor1922diffusion,chong2020vortices}, grains in inelastic gases \cite{jaeger1996granular,barrat2005granular}, charged particle in plasmas \cite{mcwilliams1987transport} and active particles in crowded environment \cite{bechinger2016active} among many others.
In the simplest setting, the dynamics of a randomly kicked particle can be modelled by random walk and Brownian motion and their different variants \cite{haus1987diffusion,bouchaud1990anomalous} which provide a suitable  conceptual approach. Random walk is a simple process that can be used to study almost any stochastic dynamics at the basic level. The common feature of a particle performing random walk or a Brownian motion is that its mean squared displacement $\sigma^2_x(t)$ grows linearly with time and the late time distribution of the position of the particle is Gaussian. This behaviour is known as diffusion. On the contrary, in case of anomalous diffusion the mean square displacement is characterised by a non-linear growth with time $\sigma^2_x(t) \sim t^\nu$ with $\nu \neq 1$. The motion is called super-diffusive for $\nu>1$ and sub-diffusive for $\nu<1$. Super-diffusive regime is interesting and can be observed in different contexts such as energy transport phenomena in one-dimensional systems \cite{dhar2013exact,spohn2014nonlinear,zaburdaev2011perturbation, dhar2019anomalous}, Josephson junction \cite{shlesinger1985comment, geisel1985accelerated}, turbulent diffusion \cite{okubo1984chaos} and in  fluctuation of end-to-end distance of a polymer \cite{holm2004polyelectrolyte} to name a few. On the other hand sub-diffusive phenomena appears in the motion of atoms in optical lattice \cite{marksteiner1996anomalous}, tracer particle motion in turbulent flow \cite{solomon1993observation}, in food searching process by long-range hoping by animals \cite{viswanathan1996levy,shlesinger2009random} among many others.

One of the simplest model that leads to anomalous diffusion is L\'evy walk dynamics \cite{shlesinger1985comment,shlesinger1987levy, klafter1994levy,zaburdaev2015levy,klafter2011first,dhar2019anomalous}. In one dimension the L\'evy walk process can be described as follows \cite{zaburdaev2015levy}: a walker moves with a velocity $v$ for some time duration $\tau$ chosen from some distribution $\rho(\tau)$. As a result, in this duration the walker makes a displacement $v \tau$ (which we call a `jump') and at the end of this duration the walker choses a new velocity $v'$ from some distribution $p(v)$  and moves with that velocity for another random duration of time $\tau'$ again chosen (independently)  from the distribution $\rho(\tau')$. This continues till the observation time $t$ at which one is usually interested in the position $x(t)$ of the walker. Depending on the choices of the distributions $p(v)$ and $\rho(\tau)$ the mean square displacement $\sigma_x^2(t)$ of the walker can exhibit diffusive and super-diffusive growth with time. 
Although in most of the studies of L\'evy walk in one dimension one considers velocity distribution with fixed speed $v_0$ and random direction \cite{zaburdaev2015levy}, some of early studies consider random speed as well  \cite{denisov2012levy,zaburdaev2011perturbation,froemberg2015asymptotic,
zaburdaev2008random}. In all these studies, the common feature is that the velocity at different `jumps' are completely uncorrelated. In this paper we consider an extension of this model of L\'evy walk in which the velocities at a jump is correlated to the velocities at previous `jumps'.

More elaborately, in this paper we study a simple model of random walk by considering correlation of velocity at different `jump' steps which can describe a wide range of dynamics of a particle that is kicked at random by the surrounding medium. In our model, the walker/particle makes a `jump' {\it i.e.} moves with a velocity for some random interval of time $\tau$ chosen from distribution $\rho(\tau)$. At the end of this time duration its velocity gets changed to a new one due to a `kick' from the surrounding medium and another random time interval $\tau'$ is chosen independently (as in the L\'evy walk model) for which it moves ballistically with the new velocity. Only difference is now that the velocity in the new `jump' step also depends explicitly on the velocity in the previous `jump' step. Let $v_{i}$ and $v_{i-1}$  are the velocities of the  particle at the $i^{\textrm{th}}$ and $(i-1)^{\textrm{th}}$ `jump' steps (of different random durations) respectively. They are related via
\bea
v_i = -r v_{i-1} + \eta_{i},~~\textrm{for}~~i=1,2,...~~\textrm{with}~~v_0=\eta_0, 
\label{modl_1}
\eea
where $\eta_i$s are independent and identically distributed (i.i.d) random variables that represent the 'kicks' from the medium. These i.i.d. random variables are each chosen from a mean zero Gaussian distribution with variance $\sigma^2$ 
\bea
~~~~~~~~~~~~~~~~~~~p(\eta)=\frac{e^{-\frac{\eta^2}{2\sigma^2}}}{\sqrt{2 \pi \sigma^2}}.
\label{p(eta)}
\eea
Here $r$ in Eq.~\eref{modl_1} is a dimensionless parameter which takes values within $[-1:1]$. Note that for $r \neq 0$ the velocities of the particle at different `jump' steps get correlated.
The parameter $r$ controls the degree of correlation in the problem similar to the Hurst exponent $H \in [0:1]$ in fractional Brownian motion \cite{norros1995use,decreusefond1999stochastic}.
It is clear from the dynamics that the correlations among velocities become extreme in the limits $r \rightarrow \pm 1$ while it is  zero at $r = 0$. 
In particular, in the domain $-1 < r < 0$, the process becomes positively correlated like $H$ in the range $1/2 < H \le 1$. 
Similarly, the correlation is negative within $0 < r < 1$ as one observes in case of fractional Brownian motion for $0 < H < 1/2$. This correlated dynamics of the velocity, particularly in the extreme limits of $r = \pm 1$, is expected to yield many intriguing and non-trivial outcomes of any observable associated with the velocity. 
This motivates us to mainly focus on these two extreme limits and the $r = 0$ case.

The parameter $r$ can also be interpreted as a restitution coefficient in a collision problem \cite{prasad2014high,prasad2014driven,majumdar2007inelastic}. Imagine a granular particle is being driven and  dissipating energy via inelastic collisions with a massive vibrating wall.
The post collision velocities of this particle and the wall, $v'$ and $V'$ respectively, are related to the pre-collision velocities $v$ and $V$ through the relation $(v' - V')= -r (v - V)$ where $r$ serves as  the coefficient of restitution. 
Notably, this coefficient characterises the degree of inelasticity in the collision process. 
For example, for \textit{$r = 1$}, collision is elastic while for \textit{$r = 0$} it is perfectly inelastic. 
Between these two limits $(0 < r < 1)$ collisions are inelastic.
Conservation of momentum of the particle and wall with masses $m$ and $M$, respectively, implies that 
$mv'  + M V' = m v + MV$.
Solving the above two relations in the limit $M \gg m$ yields $V' = V$ and $v' = -r v + (1+r) V$. 
In addition, if one assumes that the velocity of the vibrating wall is random, uncorrelated (in time) and independent of the motion of the particle then it allows to consider $V (1+r) = \eta$, a random noise and to write $v' = -r v + \eta$ \cite{majumdar2007inelastic} which is same as Eq.~\eref{modl_1}. Only thing  we added to this dynamics is that the collisions are occurring after random interval of times chosen independently from some distribution and this way we have introduced time in our model which allows us to study the displacement of the particle in addition to its velocity.

In this paper, we consider the following three choices for the distribution $\rho(\tau)$ of the `jump' durations:
\be
\rho(\tau) =
\bcs
\delta(\tau - a); \;\;\;\;\;\; \; a>0,  & \text{case I}\\
\beta \;e^{-\beta \tau};  \;\;\;\;\; \; \; \; \beta>0; \; \tau \in [0, \infty),   & \text{case II}\\
\frac{\alpha}{\bar{\tau}}\left(\frac{\bar{\tau}}{\tau}\right)^{1+\alpha}; \;\;\alpha>0; \; \tau \in [\bar{\tau}, \infty), & \text{case III}.
\ecs
\label{time_dist_1}
\ee
The first choice is a simple case in which all the `jump' durations are equal to $a$.
In the $2$nd case, we consider an exponential distribution characterised by the parameter $\beta > 0$.
The third one is a power-law distribution characterized by the exponent $\alpha > 0$.
For convenience, throughout this paper, we consider $\bar{\tau} = 1$.
Note that the exponential distribution has all moments finite, while the power-law distribution has diverging moments which depends on exponent $\alpha$.

In this paper, we study  the velocity $v(t)$ and the position $x(t)$ of the particle at time $t$.
Precisely, we study the distribution of these two quantities in the limit of large $t$. 
For a trajectory of duration $t$ having $m$ `jump' steps of durations $\tau_0,\tau_1,....,\tau_{m-1}, \tau_m^*$ 
the velocity $v(t)$ at time $t$ can be written from Eq.~\eref{modl_1} as 
\begin{equation}
v(t) = \sum_{i=0}^{m}  \eta_{m-i}\; (-r)^{i} = \sum_{i=0}^{m}  \eta_{i}\; (-r)^{m-i},
\label{modl_2}
\end{equation}
such that 
\be
t = \sum_{j=0}^{m-1}\tau_i + \tau^{*}_{m} ,
\label{modl_4}
\ee
where $\tau_m^*$ is the duration of the last incomplete step. Note that  $v(0)=v_0=\eta_0$. The time $\tau_m^*$ is known as the backward recurrence time  in the context of the renewal process \cite{godreche2015statistics}. In this context various extreme statistics of i.i.d time intervals $\tau_0,\tau_1,...,\tau_m^*$, also known as waiting times, has been studied in detail \cite{mainardi2007beyond,godreche2015statistics,niemann2016renewal}. 
The position $x_i$ of the particle after $i^{\textrm{th}}$ `jump' step satisfies the equation
\be
x_i = x_{i-1}+v_i\tau_i,~~\textrm{for}~~i=0,1,2,... \textrm{with}~x_{-1}=\bar{x},
\label{modl_5}
\ee
where $\bar{x}$ is the initial position. 
From this equation and Eqs.~\eref{modl_1} and \eref{modl_4}, the position at time $t$ can be written as 
\be
x(t) =  \sum_{j=0}^{m-1} v_j \tau_j + v_{m} \tau^{*}_{m} +\bar{x},
\label{modl_3}
\ee
Without any loss of generality, we choose $x(0)=\bar{x}=0$ in this paper. 
Eqs.~\eref{modl_1} and \eref{modl_5} along with Eq.~\eref{modl_4} constitute the equations of motion for the particle. For specific choices of $r$ and $\rho(\tau)$, these equations of motion lead to well known dynamics. For example, in case I and II it is easy to realise that the dynamics for $r=-1$ leads to the Random acceleration process (RAP) at large times. Similarly, in these two cases the evolution of the velocity for $r \to -1$ can be described by an Ornstein-Uhlenbeck process in  limits of $a \to 0$ or $\beta \to \infty$. Also note that for general $r \neq 0$, the evolution for the position of the particle becomes non-markovian. In the following we study  the variance and the distribution of the velocity and the position of the particle both theoretically and numerically for the three choices of the `jump' time distribution $\rho(\tau)$ and different values of $r$.



The paper is organised as follows: 
In sec.~\ref{p_vt}, we study distribution of the velocity of the particle at time $t$ where we obtain various scaling properties of this distribution for three choices of $\rho(\tau)$ in Eq.~\eref{time_dist_1}. In the next sec.~\ref{var_xt} we study the variance of the position in detail again for these three choices of `jump' time distributions. In this section we also study velocity-velocity correlation and discuss its relation with the variance of the position. This section is followed by the study of the distribution of the position $x(t)$ in sec.~\ref{pdf_dis} where we again discuss the three choices of $\rho(\tau)$  and for each choices, we discuss the cases for $r=0$ and $r=\pm1$ separately. 
Finally, in sec.~\ref{discussion} we summarise our results with a brief discussion and conclusion.
For clarity of the presentation, we provide some details of the analysis in the \Aref{appendix}.

\section{Distribution of the velocity}
\label{p_vt}

The distribution of velocity $v$ at time $t$, denoted by $P(v,t)$ can be written as
\bea
P(v,t)  
= \sum_{m=0}^{\infty} \left\langle \delta \left(v - v_m\right) \right\rangle \left\langle \delta \left(t-\sum_{j=0}^{m-1}\tau_{j} - \tau^{*}_m \right) \right\rangle
\label{dist_vt}
\eea
where we have used the fact that $v_m$ and $\tau_m$ are independent random variables.
Performing Fourier-Laplace transform on both sides of the above equation  and simplifying we get {(see \ref{furlap_v_app})}
\bea
\nonumber
\widetilde{P_v}(k,s) = \int_{v=-\infty}^{\infty} dv \; e^{ikv} \int_{t=0}^{\infty} dt \; e^{-st} P(v,t) \\
= \sum_{m=0}^{\infty} \exp\left[- \frac{1}{2} k^2 \sigma^2 \frac{1 - r^{2+2m}}{1 - r^2}\right] \; \tilde{\psi}^m(s) \; \tilde{\Psi}(s),
\label{furlap_v}
\eea
where $\tilde{\psi}(s)$ and $\tilde{\Psi}(s)$ are the Laplace transforms of $\rho(\tau)$ and $\Psi(\tau)=\int_{\tau}^{\infty} du \; \rho(u)$ respectively:
\bea
\tilde{\psi}(s) &= \mathcal{L}(\rho(\tau)) =\int_0^\infty d\tau e^{-s \tau}\rho(\tau), \label{psi(s)} \\ 
\tilde{\Psi}(s) &= \mathcal{L}\left(\int_{\tau}^{\infty} du \; \rho(u)\right) = \int_0^\infty d\tau e^{-s \tau}\int_{\tau}^{\infty} du \; \rho(u) = \frac{1-\tilde{\psi}(s)}{s}. \label{Psi(s)}
\eea
Using the last equation, one can simplify the expression of $\widetilde{P_v}(k,s)$ in Eq.~\eref{furlap_v} to get 
\bea
\widetilde{P_v}(k,s) &= \frac{1}{s} \;e^{-k^2 \sigma^2/2},~~~~~~~~~~~~~~~~~~~~\textrm{for},~~~r=0, \label{fl_r0} \\ 
&= \frac{1 - \tilde{\psi}(s)}{s \left( e^{k^2 \sigma^2/2} - \tilde{\psi}(s)\right)},~~~~~~~~~\textrm{for},~~~r=\pm1,\label{fl_r1} 
\eea
from which one can get $P(v,t)$ by performing inverse Fourier-Laplace transform.  One can see that the distribution $P(v,t)$ for $r=0$, is same as that of $\eta$ {\it i.e.,} mean zero Gaussian with variance $\sigma^2$ for any $\rho(\tau)$. This is expected and can be easily seen from Eq.~\eref{modl_1}. However for $r \neq 0$, the distribution $P(v,t)$ should depend on $\rho(\tau)$. In the following, we perform the inverse Fourier-Laplace transform of $\widetilde{P_v}(k,s)$ [given in Eq.~\eref{fl_r1}] for the three choices of $\rho(\tau)$ separately for $r=\pm 1$. In particular we are interested in the large $t$ limit where $P(v,t)$ often satisfies a scaling form after rescaling the velocity $v$ with respect to (some power of) time $t$. 
It is easy to see that for $r \neq 0$ the magnitude of typical velocity is large for large $t$ and this suggests us to look at the behaviour of $P(v,t)$ for large $v$ and $t$ which can be obtained by looking at the $k \to 0$ and $s \to 0$ limit. For small $k$ we approximate $e^{k^2 \sigma^2/2} \approx 1 +k^2 \sigma^2/2$ and for small $s$ we approximate $\tilde{\psi}(s)$~as 
\begin{align}
\tilde{\psi}(s) &
\simeq 1 - a s+ \mathcal{O}(s^2),~~~~~~~~~~~~~~~~~~
~~~~~~~~~~~~~~~~~~~~~~~~~~~~~~~~~~~~~~~~~~
\text{Case~I}\\
\tilde{\psi}(s) &
\simeq 1 - \frac{s}{\beta} + \mathcal{O}(s^2),~~~~~~~~~~~~~~~~
~~~~~~~~~~~~~~~~~~~~~~~~~~~~~~~~~~~~~~~~~~~~
\text{Case~II}\\
\tilde{\psi}(s)  & \simeq \begin{cases}
1 - \Gamma(1-\alpha)\; s^{\alpha}  + \mathcal{O}(s)           &\;      \text{for}\      0 < \alpha < 1 \\
1 - (1 - \gamma_e - \ln(s)) s  +\mathcal{O}(s^2)          &\;      \text{for}\          \alpha = 1 \\
1 - \frac{\alpha s}{\alpha - 1} - \Gamma(1-\alpha) s^{\alpha} + \mathcal{O}(s^2) &\;      \text{for}\   1 < \alpha < 2,  \\
1 - 2s + (\frac{3}{2} - \gamma_e - \ln(s)) s^2  + \mathcal{O}(s^3)  &\;      \text{for}\   \alpha = 2 \\
1 - \frac{\alpha s}{\alpha - 1} + \frac{\alpha s^2}{2(\alpha - 2)}- \Gamma(1-\alpha) s^{\alpha} + \mathcal{O}(s^3) &\;      \text{for}\  \alpha > 2, \\
\end{cases}
~~~~\text{Case~III}
\label{psi_s->0}
\end{align}
where $\Gamma(x)$ is the Gamma function and $\gamma_e$ denotes the Euler constant.
In the following we use these approximations of $e^{k^2 \sigma^2/2} $ and $\tilde{\psi}(s)$ to analyse the inverse Fourier-Laplace transform of $\widetilde{P}(k,s)$ for the three choices of $\rho(\tau)$.

\subsection{Case I: $\rho(\tau)=\delta(\tau-a).$}
For this case it easy to see from Eq.~\eref{modl_2} that $v(t)= \sum_{i=0}^{\lfloor \frac{t}{a}\rfloor}  \eta_{i}\; (-r)^{\lfloor \frac{t}{a}\rfloor-i}$ where $\lfloor x \rfloor$ represents the
largest integer but not larger than $x$. This expression of $v(t)$ represents a weighted  sum of i.i.d. Gaussian random variables. Hence by central limit theorem, the distribution of $P(v,t)$ is also Gaussian with zero mean and variance 
\bea
\hspace{-1cm} \sigma_v^2(t)=\langle v(t)^2 \rangle_c=\langle v(t)^2 \rangle - \langle v(t) \rangle^2 = \sigma^2 \sum_{i=0}^{\lfloor \frac{t}{a}\rfloor} (r^2)^{\lfloor \frac{t}{a}\rfloor-i} 
= \sigma^2 \frac{1-(r^2)^{\lfloor \frac{t}{a}\rfloor+1}}{1-r^2},
\eea
where subscript `c' represents cumulant.

\subsection{Case II : $\rho(\tau)=\beta \;e^{-\beta \tau}.$}
\label{C-2-vel-dist}
Using $\tilde{\psi}(s) = \beta/(\beta + s) \simeq 1 - s/\beta$ for $s \rightarrow 0$ in Eq.~\eref{fl_r1} we write
$\widetilde{P_v}(k,s) \simeq \left( s + {k^2 \beta \sigma^2}/{2} \right)^{-1},$
which after performing inverse Fourier-Laplace transform provides 
$P(v,t) \simeq \frac{1}{\sqrt{\beta \sigma^2 t}}~\mathcal{F}\left( \frac{v}{\sqrt{\beta \sigma^2 t}}\right)$ for large $t$ where 
$\mathcal{F}(u)=G(u)$ with  
\bea
~~~~~~~~~~~~G(u)=\frac{1}{\sqrt{2 \pi}}\exp\left(-\frac{u^2}{2}\right). \label{Gaussian-SF}
\eea
The fact that $P(v,t)$ for large $t$ in this case is Gaussian can be understood very simply from the following: the average time duration between two successive velocity `jump's is $\beta^{-1}$ and the number of `jump's till time $t$ is typically of the order of $m \sim \beta t$ for large $t$. The velocity after these many `jump's, each of duration $\beta^{-1}$ can again be written as a 
weighted sum of Gaussian i.i.d. random variable $\eta$s as in the previous case. Hence the distribution of the velocity at time $t$ would be given by a mean zero Gaussian distribution with variance $\sigma^2 \beta t$.

\subsection{Case III: $\rho(\tau) \sim \tau^{-(1+\alpha)}$}
Performing the inverse Fourier-Laplace transform of $\widetilde{P_v}(k,s)$ for this case is relatively harder. In this case one expects different results in different regimes of $\alpha$. Below we discuss the following regimes 
$0 < \alpha <1$, $\alpha  = 1$, and $\alpha > 1$ separately.

\subsubsection{ For $0 < \alpha < 1$\\\\}

\begin{figure*}[t!]
\includegraphics[width = 1.\hsize]{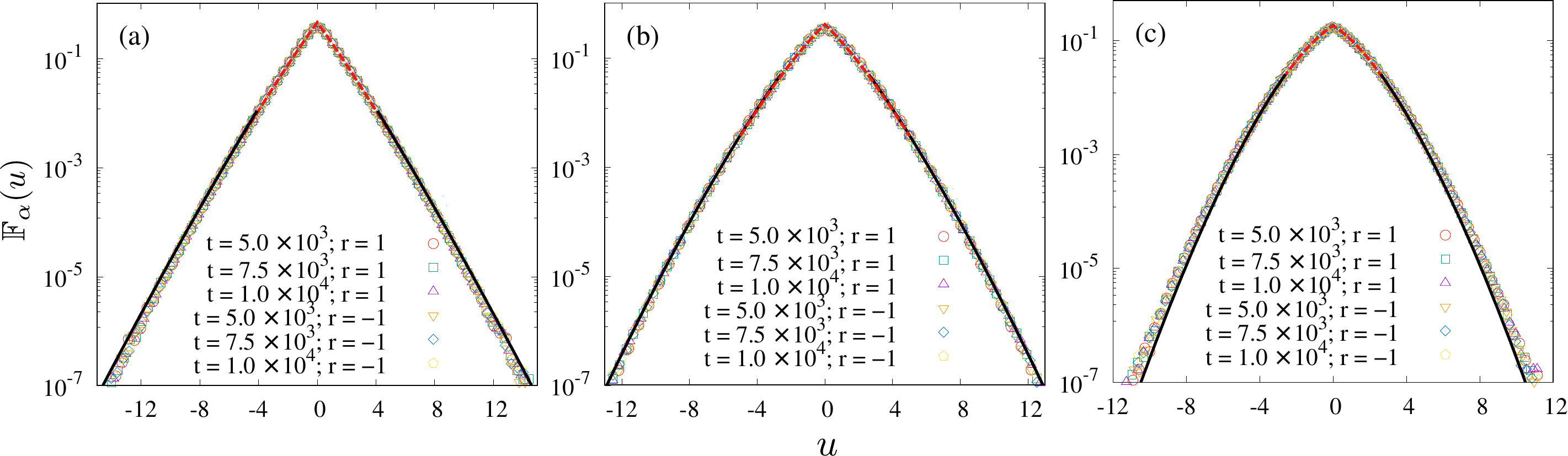}
\caption{\label{pdf_vel_1} (Color online). Plot of the scaling distribution of the velocity as given in Eq.~\eref{sc-P(v,t)-0<a<1} at different times for $r = \pm 1$ with (a) $\alpha = 0.3$, (b) $\alpha = 0.5$ and (c) $\alpha = 0.7$. Discrete points in each plot are obtained from the numerical simulation whereas the (red) dashed lines correspond to the theoretical result obtained by performing the integral in Eq.~\eref{scl_dis_vel} numerically.
The (black) bold lines correspond the approximate analytical expression of the scaling  $\mathbb{F}_\alpha(u)$ valid for large $|u|$ as given in Eq.~\eref{F_alpha}. We have taken $\sigma=1$ for all the figures in this paper.} 
\end{figure*}

For small $s$ and $k$, using $\tilde{\psi}(s) \simeq 1 - \Gamma(1-\alpha)\;s^{\alpha}$ and $e^{k^2 \sigma^2/2} \simeq 1- \frac{k^2 \sigma^2}{2}$  in Eq.~\eref{fl_r1} we get
\bea
\widetilde{P_v}(k,s) \simeq \frac{1}{s}\frac{1 }{1+\frac{k^2 \sigma^2}{2\Gamma(1 - \alpha)s^\alpha}} 
\simeq \sum_{l=0}^{\infty} \frac{1}{s^{\alpha l+1}} \left( -\frac{k^2 \sigma^2}{2\Gamma(1-\alpha)}\right)^l .
\eea
This expression suggests that the distribution $P(v,t)$ satisfies a scaling form 
\bea
P(v,t)=\sqrt{\frac{2\Gamma(1-\alpha) }{\sigma^{2} t^{\alpha}}}F_\alpha\left(\frac{\sqrt{2\Gamma(1-\alpha)}~v}{ \sqrt{ \sigma^{2} t^{\alpha}}}\right),
\label{sc-P(v,t)-0<a<1}
\eea
 for large $t$. To see this we first 
perform inverse Laplace transform (with respect to $s$) of each terms in the above series individually and then perform the infinite series sum to get, 
\bea
\widetilde{P_v}(k,t) \simeq \sum_{l=0}^{\infty} \frac{1}{\Gamma(\alpha l + 1)} \left( -\frac{t^{\alpha}k^2 \sigma^2}{2\Gamma(1-\alpha)} \right)^l  \simeq E_{\alpha} \left( -\frac{t^{\alpha}k^2 \sigma^2}{2\Gamma(1-\alpha)} \right),
\label{ml_fn1}
\eea
with $E_{\alpha}(u)$ being the Mittag-Leffler function \cite{haubold2011mittag}.
Next we perform the inverse Fourier transform with respect to $k$ and get 
\bea
\mathbb{F}_\alpha(u) = \frac{1}{2\pi} \int_{-\infty}^{\infty} d\widetilde{k}\;e^{-i \widetilde{k} u} \;E_{\alpha}(-\widetilde{k}^2),~~\textrm{where},~~~u =\frac{\sqrt{2\Gamma(1-\alpha)}~v}{ \sqrt{ \sigma^{2} t^{\alpha}}}.
\label{scl_dis_vel}
\eea
For $\alpha = 0$, using $E_{0}(- \widetilde{k}^2) = 1/(1  + \widetilde{k}^2)$ \cite{haubold2011mittag} in Eq.~\eref{scl_dis_vel} it is possible to perform the inverse Fourier transform exactly to get $\mathbb{F}_0(u) = e^{-|u|}/2$.
For any arbitrary $0 < \alpha < 1$ performing the integral in Eq.~\eref{scl_dis_vel} analytically seems difficult though it can be performed numerically [red solid lines in figs. \ref{pdf_vel_1} (a), (b) and (c)]. However, the behaviour of $\mathbb{F}_\alpha(u)$ for large $u$ can be obtained using saddle point approximation and we find that the tails of $\mathbb{F}_\alpha(u)$ are given by the following stretched exponential form  (see \ref{saddle_point_alpha} for details)
\bea
\hspace{-1cm}\mathbb{F}_\alpha(u) \simeq \frac{\left( \alpha|u|/2\right)^{-\frac{1-\alpha}{2-\alpha}}}{\sqrt{4 \pi (2 - \alpha)}} \;\exp\left[-\left( 2/\alpha - 1 \right) \left( \alpha|u|/2\right)^{2/(2-\alpha)}\right], ~~\textrm{for~large}~u.
\label{F_alpha}
\eea
This result is verified with our simulation data  in figs. \ref{pdf_vel_1} (a), (b) and (c) for $\alpha=0.3,~0.5$ and $\alpha=0.7$, respectively, where we observe excellent agreement. The red solid lines in these plots are obtained by performing the integral in Eq.~\eref{scl_dis_vel} numerically in Mathematica. Throughout this paper, we have considered $\sigma=1$ for all our numerical computations.\\

\subsubsection{ For $ \alpha = 1$\\\\}

\begin{figure*}[t!]
\centering{\includegraphics[width = 1.0\hsize]{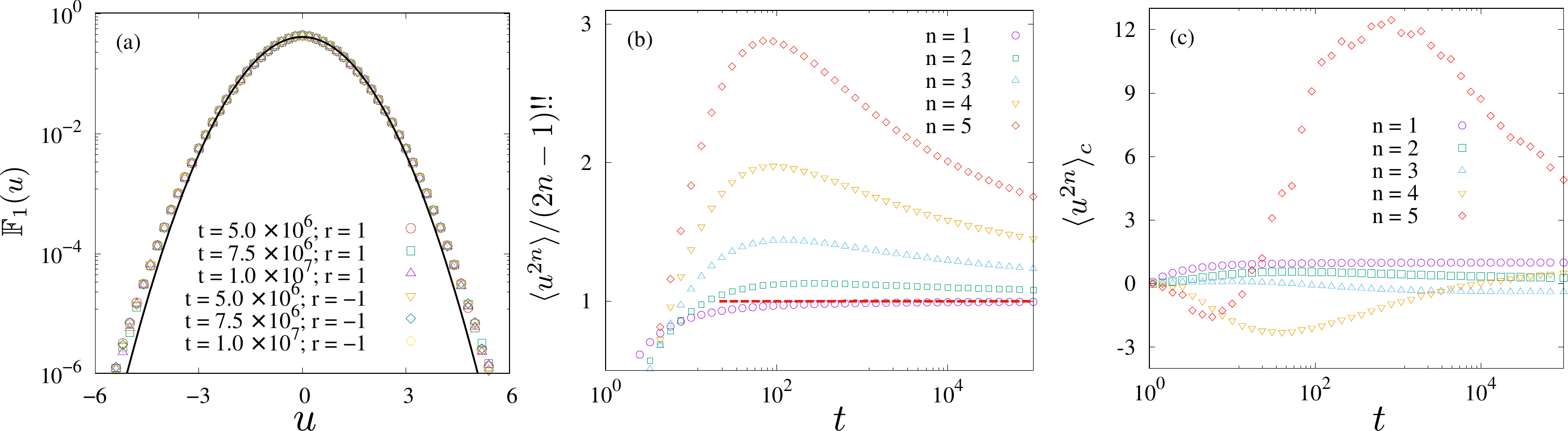}}
\caption{\label{pdf_vel_2} (Color online). (a) Probability distribution function of the scaled velocity $u = v/\sqrt{\sigma^2 t/\textrm{log}(t)}$ is plotted at different times for $r = \pm 1$ for $\alpha = 1$. In (b) and (c): we plot the simulation result of the first five (non-zero) even-order moments $\langle u^{2n}\rangle$ and cumulants $\langle u^{2n}\rangle_c$ of the scaled variable $u$ as function of time to demonstrate the slow convergence.} 
\end{figure*}

In the $\alpha \to 1$ limit the stretched exponential form of the scaling distribution $F_\alpha(y)$ in Eq.~\eref{F_alpha} approaches a Gaussian distribution form with mean zero and variance growing linearly with time. However, the small $s$ behaviour of $\tilde{\psi}(s)$ for $\alpha=1$  given in Eq.~\eref{psi_s->0} suggests us to expect a $\ln(t)$ dependence in the variance of $v$. To see this we use $\tilde{\psi}(s) \simeq 1 - (1 - \gamma_e - \ln(s)) s$ for small $s$ and $e^{k^2 \sigma^2/2} \simeq 1+ k^2 \sigma^2/2$ for small $k$ in Eq.~\eref{fl_r1} and expanding we get 
\bea
\widetilde{P_v}(k,s) \simeq \sum_{l=0}^{\infty} \frac{1}{s^{l+1}} \left( \frac{1}{\ln(s)} \left( e^{k^2 \sigma^2/2} -1\right) \right)^l.
\eea
We first perform the inverse Laplace transform with respect to $s$ and to do that we employ the Tauberian theorem (see \eref{ilt_6} in \ref{ILT_BR_TU}). We finally get the following approximate result in the $t \rightarrow \infty$ limit 
\bea
\widetilde{P_v}(k,t) \simeq \sum_{l=0}^{\infty} \frac{1}{l!} \left( - \frac{t~k^2 \sigma^2}{2~\ln(t)} \right)^l 
 \simeq \exp\left[-\frac{t~k^2 \sigma^2}{2~\ln(t) } \right].
\label{velkt_a10}
\eea
This suggests that for large $t$, the distribution $P(v,t)$ indeed has a mean zero Gaussian distribution but with variance growing with time as  $~\sigma^2 t/ \ln(t)$. This means the scaling variable $u = v/\sqrt{\sigma^2 t/\ln(t)}$ has a scaling distribution $\mathbb{F}_1(u)=G(u)$ where $G(u)$ is given in Eq.~\eref{Gaussian-SF}.
We attempt to illustrate this result numerically in fig.~\ref{pdf_vel_2}(a). In this figure, while we observe a nice data collapse in terms of the scaling variable $u$ (with y-axis appropriately scaled), the scaling distribution 
$\mathbb{F}_1(u)$ does not match properly with the Gaussian form at the tails. We believe this happens because of the slow convergence arising due to the absence of a time scale for $\alpha=1$. For  both $\alpha <1$ and $\alpha >1$, there exist a time scale $\tau^*$ over which one expects the distribution $P(v,t)$ to approach an appropriate scaling distribution. As can be seen from Eq.~\eref{ml_fn1} or Eq.~\eref{scl_dis_vel},  $\tau^* \sim \Gamma(1-\alpha)^{1/\alpha}$ for $\alpha <1$. On the other hand for $\alpha >1$, as will be shown in the next section,  $\tau^* \sim \alpha/(\alpha - 1)$. Observe from both these expressions of $\tau^*$ that it diverges as $\alpha$ approaches $1$ either from below or above. To investigate further about the slow convergence and approach to the Gaussian scaling form, we compute the moments and cumulants of the scaling variable $u = v/\sqrt{\sigma^2 t/\ln(t)}$  as functions of $t$ in numerical simulation and check if they converge to 
$\langle u^{2n} \rangle = (2n-1)!!$ and $\langle u^{2n} \rangle_c= \delta_{n,1}$ at large $t$, where $\delta_{i,j}$ is the Kronecker delta. Note all odd order moments and cumulants are identically zero by symmetry. Here $(2n-1)!!$ represents double factorial defined as $(2n-1)!!=1.3.5....(2n-1)$.  In figs.~\ref{pdf_vel_2}(b) and \ref{pdf_vel_2}(c) we plot the moments and cumulants till order $10$ ({\it i.e.} $n=5$) as functions of time. We observe that at large time the cumulants indeed approach zero. However,  the convergence time become larger and larger as the  order of the moments/cumulants increase as one expects.


\subsubsection{ For $\alpha > 1$\\\\}
\label{vel-dist-a>1}
\begin{figure*}[t!]
\centering{\includegraphics[width = 0.8\hsize]{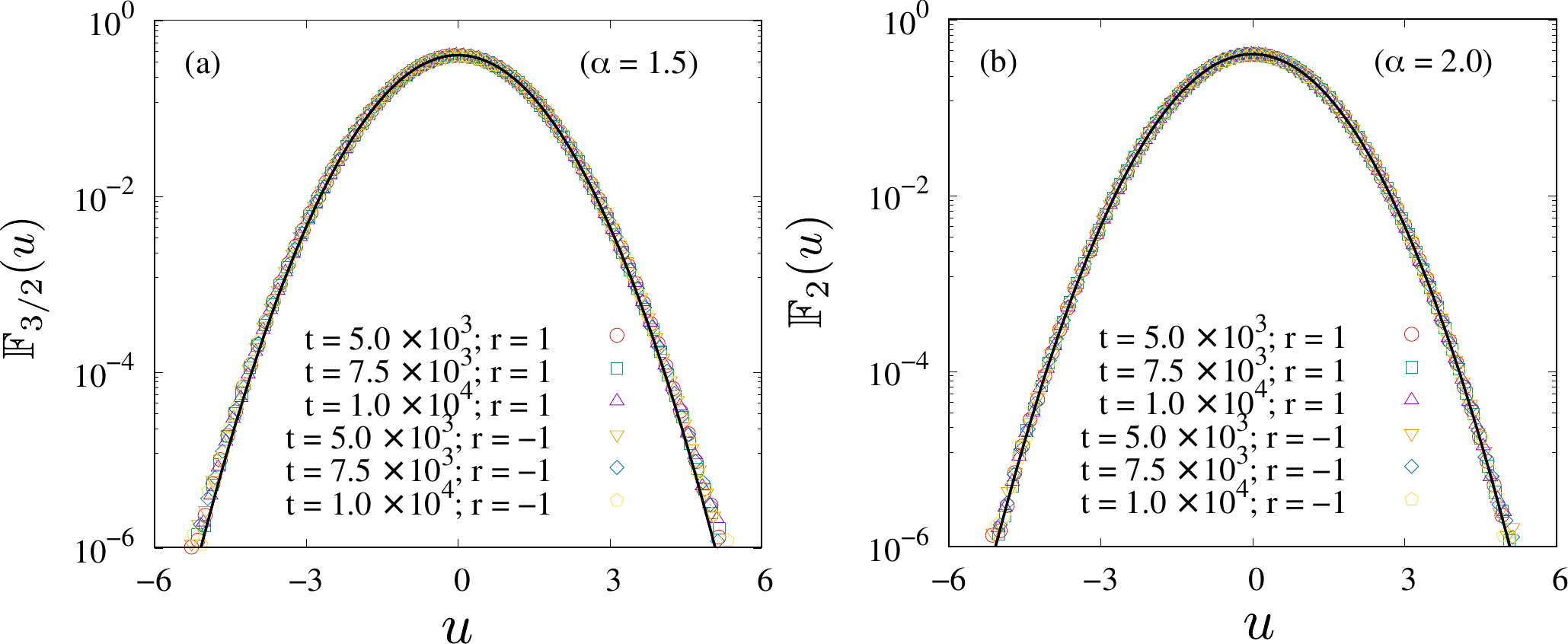}}
\caption{\label{pdf_vel_3} (Color online). Probability distribution function of the scaled variable $u = v/\sqrt{\sigma^2 t (\alpha -1)/\alpha}$ is plotted at different times for $r = \pm 1$ with (a) $\alpha = 1.5$ and (b) $\alpha = 2.0$. Symbols in each plot correspond to numerical simulation results which show excellent agreement with the normal Gaussian distribution $\mathbb{F}_\alpha(u) =G(u)$ (solid black lines) predicted theoretically (see sec.~\ref{vel-dist-a>1}). } 
\end{figure*}

In this case we can see from Eq.~\eref{psi_s->0} the function $\tilde{\psi}(s)$ in the $s \to 0$ has a linear order term {\it i.e.} $\tilde{\psi}(s) \simeq 1 - \alpha s/(\alpha -1)$. As a consequence, following a similar procedure and arguments like case II in sec.~\ref{C-2-vel-dist}, in this case also we find that the typical fluctuation around the mean is Gaussian with variance $\sigma^2 t (\alpha -1)/\alpha$.  This means that in this case also the scaling variable $u = v/\sqrt{\sigma^2 t(\alpha -1)/\alpha}$ has a scaling distribution $\mathbb{F}_\alpha(u)=G(u)$ where $G(u)$ is given in Eq.~\eref{Gaussian-SF}.
We demonstrate this result in fig.~\ref{pdf_vel_3} in terms of the scaled variable $u$.

\section{Variance of the position}
\label{var_xt}
We now study  the properties of the position $x(t)$ of the particle in time $t$ when it starts at the origin {\it i.e.} $x(0)=0$. The distribution $P(x,t)$ of the 
position can be written as 
\begin{equation}
P(x,t) 
= \sum_{m=0}^{\infty} \left\langle \delta \left(x-\sum_{j=0}^{m-1} v_{j}\tau_{j} - v_m \tau^{*}_m \right) \delta \left(t-\sum_{j=0}^{m-1}\tau_{j} - \tau^{*}_m \right) \right\rangle, 
\label{pdf_1}
\end{equation}
where we have used Eqs.~\eref{modl_3} and \eref{modl_4}. 
Performing Fourier-Laplace transform of $P(x,t)$ we get
\bea
\nonumber
\widetilde{P_x}(k,s) = & \int_{x=-\infty}^{\infty} dx \hspace{0.2cm} e^{ikx} \int_{t=0}^{\infty} dt \hspace{0.2cm} e^{-st} P(x,t) \\ & =  \sum_{m=0}^{\infty} \left \langle \exp \left[{- \frac{1}{2} \sigma^2 k^{2}\bar{\tau}_{m}^{T}\Sigma_{m}\bar{\tau}_{m}- sI_{m}^{T}\bar{\tau}_{m}} \right] \right \rangle_{\{\tau_{j}\}}
\label{tildeP_k_s}
\eea
where $\bar{\tau}^T_{m} = \left(\tau_{0}\;\;\tau_{1}\;\;\tau_{2}\;\;.....\;\tau^{*}_{m}\right)_{1\times(m+1)}$ and 
$I^T_{m} = \left(1\;\;1\;\;1\;\;.......\;1\right)_{1\times(m+1)}$
are two $(m+1)$ dimensional row vectors and $\Sigma_m$ 
is a symmetric $(m+1)\times(m+1)$ dimensional matrix with elements given by 
\bea
\nonumber
\Sigma_{m}(l,j) =  & \underbrace{ \delta_{l,j} \sum _{p=0}^{l-1} r^{2 p}}_{\textrm{diagonal elements}} \;+\; \underbrace{  (1 - \delta_{l,j}) \; \mathbb{I}(j > l) \sum _{p=0}^{l-1} (-r)^{j - l + 2p}}_{\textrm{upper-half off-diagonal elements}} \\& \;+\; \underbrace{(1 - \delta_{l,j}) \; \mathbb{I}(l > j) \sum _{p=0}^{j-1} (-r)^{l - j + 2p}}_{\textrm{lower-half off-diagonal elements}},
\label{ele_m}
\eea
where $\mathbb{I}(\textrm{condition})$ is the indicator function and takes value $1$ if the condition in the argument is true and $0$ otherwise.
The angular bracket in Eq.~\eref{tildeP_k_s} represents average over the time intervals $\{\tau_{m}\}$ which are chosen independently from distribution $\rho(\tau)$. We recall, in this paper we consider three choices of $\rho(\tau)$ as stated  in Eq.~\eref{time_dist_1}. Since the $\tau$s are positive random variables, performing the average over them in Eq.~\eref{tildeP_k_s} for $r \neq 0$ is difficult. However, one can compute moments of different order by evaluating derivatives of $\tilde{P}_x(k,s)$ with respect to $k$ in the $k \to 0$ limit. It is easy to observe from Eq.~\eref{tildeP_k_s} that $P(x,t)=P(-x,t)$. Hence all odd order moments and the cumulants of the position are zero. The even order moments are non-zero. In this paper we discuss the variance $\sigma_x^2(t)=\langle x^{2}(t) \rangle_c=\langle x^{2}(t) \rangle$ in detail while making some comments on higher order moments/cumulants. The Laplace transform of the variance $\sigma_x^2(t)$ is given by [see \ref{var_cal} for details]
\bea
\hspace{-1.5cm}
\widetilde{\sigma}_x^2(s)&= - \left[\frac{d^{2}}{dk^{2}} \widetilde{P_x}(k,s)\right]_{k=0} 
= \frac{2 \sigma^2}{s^3 (1 - r^2 \tilde{\psi}(s))}  \left( 1 + \frac{s \;\tilde{\psi}'(s)}{1 - \tilde{\psi}(s)} + \frac{r s \;\tilde{\psi}'(s)}{1 + r \;\tilde{\psi}(s)} \right).
\label{var_all}
\eea
This is a general result valid for any $\rho(\tau)$ and $-1 \leq r \leq 1$. Next we discuss three choices of $\rho(\tau)$ separately.

\subsection{Case I: $\rho(\tau) = \delta(\tau -a)$}
\label{var_cs1}
The behaviour of the variance at large $t$ can be easily found by using the small $s$ asymptotic $\tilde{\psi}(s) = e^{-a s} \simeq 1 - a s + \mathcal{O}(s^2)$ in Eq.~\eref{var_all} and then performing the inverse Laplace transform. We get 
\bea
\sigma_x^2(t)=\langle x^2 (t)\rangle_c \simeq
\left \{\begin{array}{ccc} 
a \sigma^2 t + \mathcal{O}(1)\;&\; \textrm{for}\ r = 0 \\
\frac{1}{3 a} \sigma^2 t^3 + \mathcal{O}(t^2)\;&\; \textrm{for}\ r = -1\\
\frac{1}{3} a \sigma^2 t + \mathcal{O}(1)\;&\; \textrm{for}\ r = 1.
\end{array}
            \right. \label{var-x_case_1}
\eea
As mentioned earlier, in this case one can actually perform the inverse Laplace transform exactly for arbitrary $t$ and $r$. As shown in  \ref{var_case1}, we get the following explicit 
expression of $\sigma_x^2(t)=\langle x^2 (t)\rangle_c$ 
\begin{align}
\label{exact_varx_cs1}
 \sigma_x^2(t)= 
& \frac{\sigma ^2 (1-r)^{-1} }{(r+1)^3} \left[(r+1)^2 t^2 \left(1-r^{2 \left\lfloor \frac{t}{a}\right\rfloor +2}\right)  - 2 a (r+1) t \left \{ r +\left\lfloor \frac{t}{a}\right\rfloor (r + 1)  \right. \right. \nonumber\\
& \left. \left. 
-r^{2 \left\lfloor \frac{t}{a}\right\rfloor +2} \left( \left\lfloor \frac{t}{a}\right\rfloor (r + 1) +1\right)  -(-r)^{\left\lfloor \frac{t}{a}\right\rfloor }(1-r^2) \right\} \right.\\ 
& \left.  -a^2 \left\{ 
\left[ \left( 1+\left\lfloor \frac{t}{a}\right\rfloor (r + 1) \right) (-r)^{1+ \left\lfloor \frac{t}{a}\right\rfloor} + r-1\right]^2 
-(r+1)^2 \left\lfloor \frac{t}{a}\right\rfloor \left(\left\lfloor \frac{t}{a}\right\rfloor +1\right) -1 \right\} \right]. \nonumber
\end{align}
From this expression it is easy to see that for $-1<r<1$
\bea
\sigma_x^2(t) \simeq \frac{a \sigma^2 t}{(1+r)^2}~~{\textrm{for~large~}} t, 
\label{x^2(t)_t->oo}
\eea 
which for $r=0$ correctly reproduces the result in Eq.~\eref{var-x_case_1}. This result is easy to understand, because for $r=0$ the position $x(t)$ at large $t$ becomes a sum of many  (of the order $\lfloor t/a \rfloor$) Gaussian random variables each of mean zero and variance $a^2\sigma^2$. Hence the variance of the position $x(t)$ should be $\sim a \sigma^2 t$ for large $t$. In particular for $|r| <1$, one can see that the motion of the particle at large $t$ can be effectively described by a free particle with its velocity governed by a stochastic Ornstein-Ulhenbeck process characterised by dissipation strength $\gamma=(1+r)/a$ and noise strength $D=\sigma^2/a$. Such a particle is also known in the literature as active Ornstein-Uhlenbeck particle \cite{dolai2020universal}. For such a particle it is easy to show that the variance of its position for large $t$ is given by 
$\sigma_x^2(t) \simeq (D/\gamma^2)~t,$ which indeed is  equal to 
$a \sigma^2 t/(1+r)^2$. This result is not valid for $r =\pm 1$ for which we have to analyse the large $t$ behaviour of $\sigma_x^2(t)$ separately.

\begin{figure}[t!]
\centering{\includegraphics[width=1.\hsize]{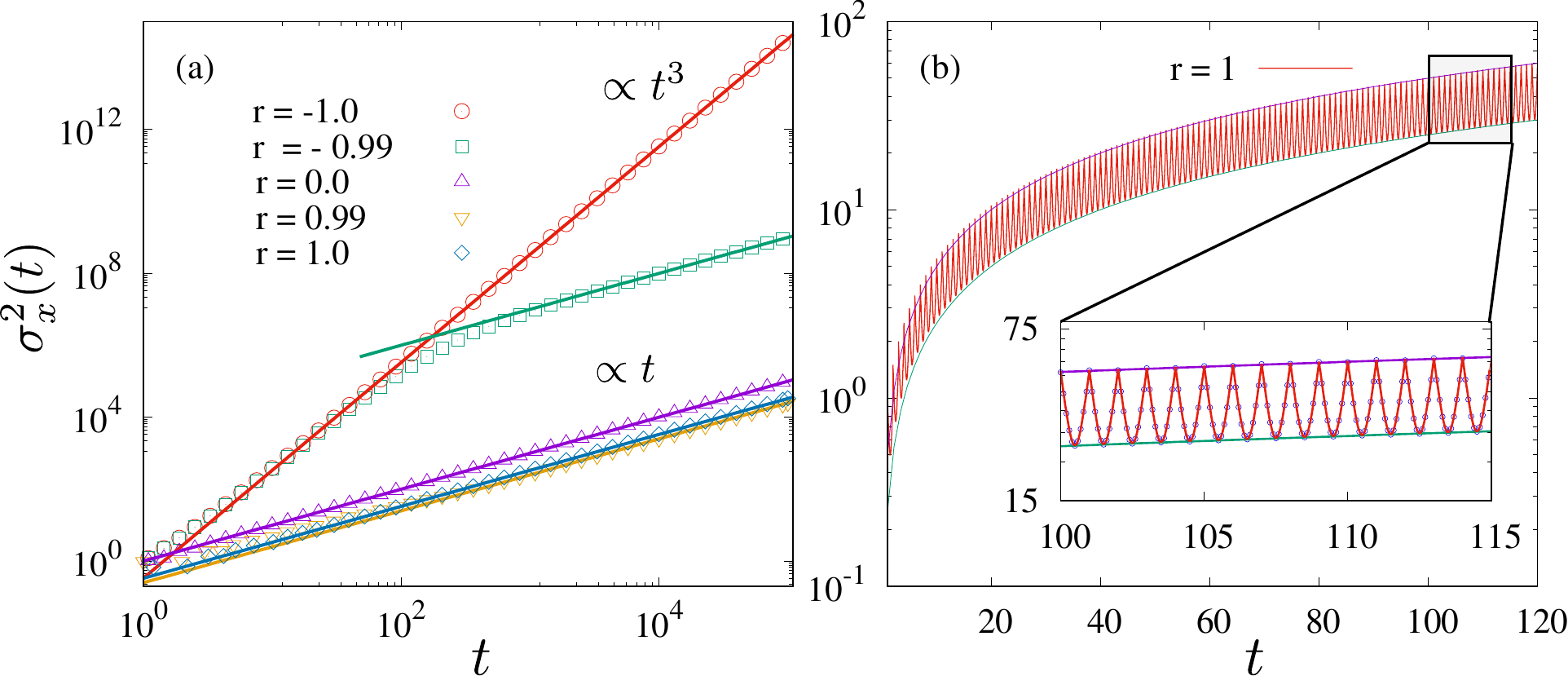}}
\caption{\label{variance_cs1} (Color online). Variance of the position is plotted as function of time for different $r$ in case I with $a = 1$. Discrete symbols in each plot are simulation results which are showing excellent agreements with the analytical results as shown by lines.} 
\end{figure}

For $r=1$, we see from Eq.~\eref{x^2(t)_t->oo} that $\sigma_x^2(t) \sim a \sigma^2 t/4$ for large $t$. On the other hand from the inverse Laplace transform calculation in Eq.~\eref{var-x_case_1} we get $\sigma_x^2(t) \sim a \sigma^2 t/3$ for large $t$.  However, in numerical simulation we do not observe  either of these two forms but we observe the variance to oscillate within an envelop which grows linearly with time (see fig.~\ref{variance_cs1}(b)).  From the exact expression in Eq.~\eref{exact_varx_cs1} we find
\bea
\sigma_x^2(t) \simeq \frac{1}{2} \sigma^2 a^2\left( 1 + \left\lfloor \frac{t}{a}\right\rfloor \right) (2 u^2 - 2 u + 1).
\label{x^2_t-os}
\eea
at large $t$ where $u = t/a - \lfloor t/a\rfloor$ with $u \in (0,1)$. Clearly, the slope depends on the value of $u$ at which one makes measurements in the simulation. For $u=0$ {\it i.e.} at times integer multiples of $a$, one finds $\sigma_x^2(t) \simeq  a \sigma^2 t/2$, whereas for $u = 1/2$ one finds $\sigma_x^2(t) \simeq a^2\sigma^2 \lfloor t/a\rfloor/4 \simeq a \sigma^2 t/4$ at large $t$.
In fact, the slope of $\sigma_x^2(t)= \langle x^2 (t)\rangle_c=\langle x^2 (t)\rangle$  oscillates between $a \sigma^2/2$ and $a \sigma^2/4$. The result in Eq.~\eref{var-x_case_1} corresponds to the average value
\bea
\hspace{-1.8cm}\sigma_x^2(t) \simeq \frac{\sigma^2a^2}{2} \left( 1 + \left\lfloor \frac{t}{a}\right\rfloor \right) \int_{0}^1 du \; (1 - 2 u + 2 u ^2)  \simeq \left( 1 + \left\lfloor \frac{t}{a}\right\rfloor \right) \frac{a^2 \sigma^2}{3} \simeq \frac{1}{3} a \sigma^2 t.
\eea
as stated in Eq.~\eref{x^2_t-os} and verified numerically in fig.~\ref{variance_cs1}(b).

Taking $r \to -1$ in Eq.~\eref{x^2(t)_t->oo} provides diverging result. One needs to take the $r \to -1$ limit first  and then take the large $t$ limit, following which we get $\sigma_x^2(t) \simeq \sigma^2 t^3/3a$. This result is verified numerically in fig.~\ref{variance_cs1}(a). In this figure we also observe that for $r$ very close to $-1$ one finds that  at moderately large $t$ the variance $\sigma_x^2(t)$ grows as $\sim t^3$ as for $r = -1$ but at really large times the growth crosses over to linear growth. A careful analysis shows that for $r \to -1$, the time dependence of the variance crosses over from $\sigma_x^2(t) \simeq (1-r)\sigma^2 t^3/6a$  to $a \sigma^2 t/(1+r)^2$ as $t$ increases and this crossover occurs at time scale 
$\tau_c \sim a/\sqrt{(1-r) (1+r)^2}$ which is indeed infinite at $r = -1$. This crossover behaviour is illustrated in fig.~\ref{variance_cs1}(a) for $r=-0.99$.

Before ending this section we  make an additional remark about the $r=-1$ case. The late time growth $\sigma_x^2(t) \simeq \sigma^2 t^3/3a$ can be understood by realising the fact that at large time the motion of the particle can be effectively described by a random acceleration process \cite{majumdar1999persistence,burkhardt2007random}  with noise strength being $D=\sigma^2/a$. For such a motion the variance of the position grows at large time as $\sim Dt^3/3 =\sigma^2 t^3/3a$. A small discussion  on the definition and properties of RAP is provided in \ref{appendix_RAP} for completeness.

\subsection{Case II: $\rho(\tau)=\beta e^{-\beta \tau}$}
\label{var_cs2}
\begin{figure}[t!]
\centering{\includegraphics[width=0.9\hsize]{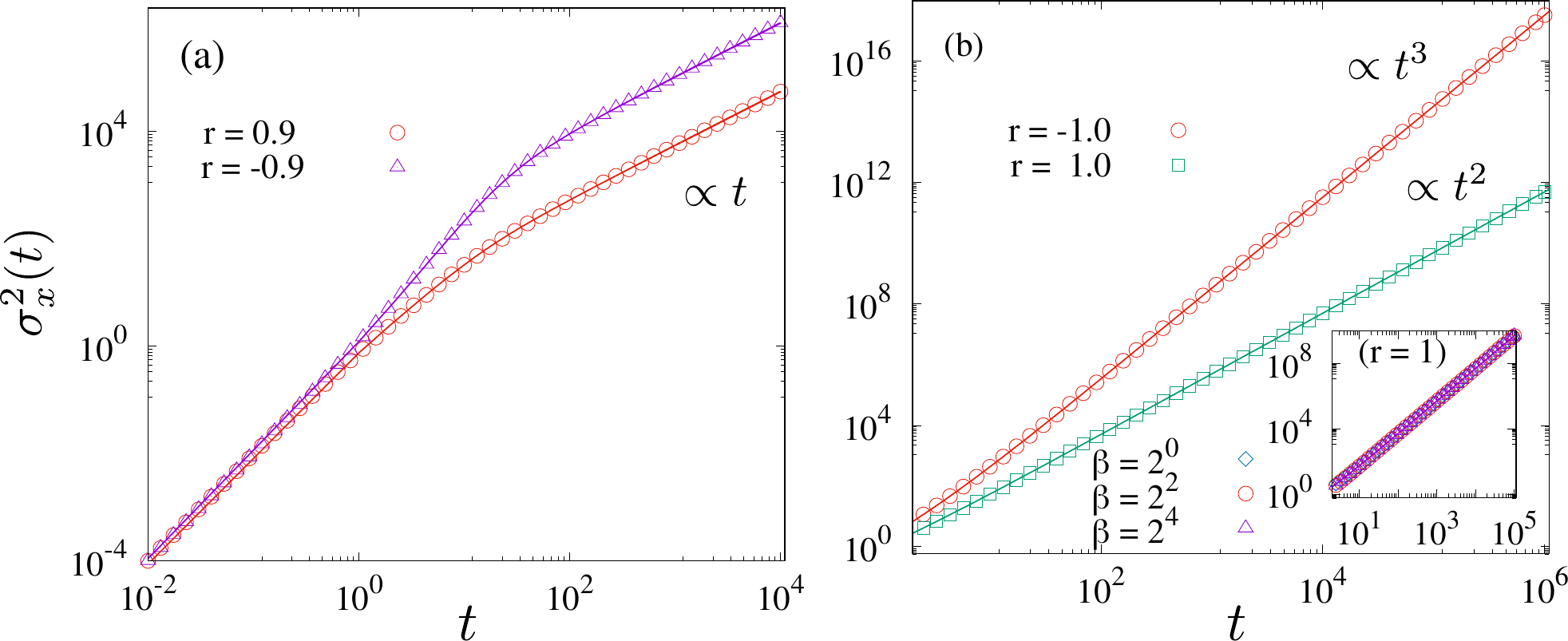}}
\caption{\label{variance_cs2} (Color online). The plots show $\sigma_x^2(t)$ vs. $t$ in case II for (a) $r = 0.9$ and $-0.9$ and for (b) $r =\pm 1$ with $\beta = 1$. Symbols represent the simulation results  and solid lines represent the analytical expression in Eq.~\eref{var_exp_r}. The inset of fig.~(b) verifies the $\beta$ independence of the variance for $r=1$, where we plot $\sigma_x^2(t)$ for different values of $\beta$. } 
\end{figure}

In this case also one can perform the inverse Laplace transform in Eq.~\eref{var_all} exactly for $-1\leq r \leq 1$ to get 
\bea
\nonumber
\sigma_x^2(t)=&\langle x^{2}(t) \rangle_c = \frac{2 \sigma^{2} t}{\beta (1+r)(1-r^{2})} - \frac{2 \sigma^{2} (1-r+r^{2})}{\beta^{2} (1+r)(1-r^{2})^{2}} \\& + \frac{2 \sigma^{2}}{\beta^{2} } \left( \frac{(1-r)^{2} e^{-\beta t(1+r)} + r e^{-\beta t(1- r^{2})}}{(1+r)(1-r^{2})^{2}} \right).
\label{var_exp_r} 
\eea
This result is verified numerically in fig.~\ref{variance_cs2}(a) for $r=\pm 0.9$. From this result we can see that there are two different time scales $\tau_{v} = (1 - r^{2})^{-1}/\beta$ and $\tau_{c} = (1 + r)^{-1}/\beta$ involved in the process. These times scales arise, respectively, from the variance of the velocity $\sigma^2_v(t)$ and the velocity auto-correlation function $C_{vv}(t) = \langle v(t_1) v(t_2) \rangle - \langle v(t_1)\rangle \langle v(t_2) \rangle$. Taking second derivative of the Fourier-Laplace transform $\tilde{P}_v(k,s)$ in Eq.~\eref{furlap_v} with respect to $k$ at $k \to 0$ and then performing inverse Laplace transform one can show that
\bea
\sigma_v^2(t)=\langle v^2(t)\rangle_c = \frac{\sigma^2 }{1-r^2} \left( 1 - r^2 e^{-\beta t(1-r^2)}\right).
\label{<v^2(t)>}
\eea 
It can be shown (see \ref{velo_corr}) that the velocity auto correlation function $C_{vv}(t_1,t_2)$ is given by 
\bea
C_{vv}(t_1, t_2) \simeq \langle v^2(\textrm{min} \{ t_1, t_2 \}) \rangle_c \; e^{-\beta(1+r)|t_1 - t_2|}.
\label{C_vv}
\eea 
for large $t_1$ and $t_2$.
From these expressions we observe that the variance approaches to a stationary value and the covariance  decreases to zero as $t$ increases but the time scales associated are given, respectively, by $\tau_{v} = (1 - r^{2})^{-1}/\beta$ and $\tau_{c} = (1 + r)^{-1}/\beta$. 
From Eq.~\eref{var_exp_r}, it is easy to see that for $r=-1$ 
 the variance behaves as  $\sigma_x^2(t)\sim  \beta \sigma^{2} t^{3}/3$ at late times which is similar to the variance of a RAP driven by white noise of strength $D=\beta \sigma^2$. This cubic growth of $\sigma_x^2(t)$  is verified numerically in 
fig.~\ref{variance_cs2}(b).
For $r=1$ it is interesting to observe from Eq.~\eref{var_exp_r} that $\sigma_x^2(t) \sim \sigma^{2} t^2/2$ at large times. Following the discussion presented for $r \to 1$ in the last sec.~\ref{var_cs1}, one may be tempted to assume that the motion of the particle can effectively be described by an active Ornstein-Ulhenbeck particle driven by white noise. But for such a particle the variance grows linearly  instead of quadratically with time as we see in our case. Remarkebly, we also observe that the variance is independent of $\beta$ (see the inset of fig.~\ref{variance_cs2}(b)). It is easy to see that in this case the variance of the position after a large number of `jumps', say $m$,  is $\sim \sigma ^2 m^2/2\beta^2$, whereas the typical number of `jump' events in time $t$ is $m \sim \beta t$ for large $t$. Hence the variance in the leading order for large $t$ becomes $\beta$ independent. This can also be easily seen from the small $s$ behaviour of $\widetilde{\sigma}_x^2(s) \simeq \sigma^2 s^{-3}$ [see Eq.~\eref{var_all}] which in the leading order is also $\beta$ independent.

Above large time behaviours of the variance of the position for different values of $r$, can also be obtained from the large time behaviours of velocity-velocity correlation $C_{vv}(t_1, t_2) $. From $x(t) = \int_{0}^{t} dt_1 \;v(t_1)$, one can easily write
\bea
\hspace{-1.5cm}
\sigma_x^2(t) & =\langle x^2(t) \rangle_c=  \left\langle \int_{0}^{t} dt_1 \;v(t_1) \int_{0}^{t} dt_2 \;v(t_2) \right\rangle_c 
= 2 \int_{0}^{t} dt_1 \int_{t_1}^{t} dt_2 \; C_{vv}(t_1, t_2). 
\label{x^2-C_vv-rela}
\eea

For $r=0$ it is easy to see from Eq.~\eref{fl_r0} that $\langle v^2(t) \rangle_c = \sigma^2$ using which in the above equation and performing the integral we get $\sigma_x^2(t) \simeq 2 \sigma^2 t /\beta$ for large $t$.

For $r = -1$, one can obtain the variance of the velocity from $\widetilde{P}_v(k,s)$ given in Eq.~\eref{fl_r1} as $\widetilde{\sigma}_v^2(s) = -\left[  \frac{d^2}{dk^2} \widetilde{P}_v(k,s)\right]_{k=0} = \frac{\sigma^2}{s(1 - \tilde{\psi}(s))}$ where for exponential distribution for the `jump' durations one has $\tilde{\psi}(s) =\beta/(\beta +s)$. Furthermore, using this and performing the inverse Laplace transform one can easily show that $\sigma^2_v(t) \simeq \sigma^2 \beta t$ for large $t$. Using this result in Eq.~\eref{C_vv} we get  $C_{vv}(t_1, t_2) \simeq \sigma^2 \beta \;\textrm{min} \{ t_1, t_2 \}$ putting which in Eq.~\eref{x^2-C_vv-rela} and performing the integrals we indeed get $\sigma_x^2(t) \simeq  \beta \sigma^2 t^3 /3$ for large $t$.
Note that the correlation function in this case comes out to be identical to the correlation function of the RAP \cite{majumdar1999persistence,burkhardt2007random} with the strength of the noise correlator $D = \sigma^2 \beta$. This similarity with RAP has been observed earlier in sec.~\ref{var_cs1}.

Now, for $r  = 1$, using $\sigma_v^2(t)\simeq \sigma^2 \beta t$, same as the case of $r = -1$, we find the correlation function as $C_{vv}(t_1, t_2) = \langle v^2(\textrm{min}\{t_1, t_2\}) \rangle_c \;e^{-2 \beta |t_1 - t_2|} \simeq \sigma^2 \beta \;\textrm{min}\{t_1, t_2\}$ $\;e^{-2 \beta |t_1 - t_2|}$ for $0 \le t_1, t_2 \le t$.
Using this expression in Eq.~\eref{x^2-C_vv-rela} we can easily see the dominant contribution of the integration $\int_{0}^{t-t_1} dz \; e^{-2\beta z} \propto 1/\beta$ which cancels out $\beta$ in the numerator.
It essentially makes $\sigma_x^2(t)=\langle x^2(t)\rangle_c \propto \sigma^2 \int_{0}^t dt_1 \; t_1 \propto \sigma^2 t^2$ to be independent of $\beta$.

\subsection{Case III: $\rho(\tau) \sim \tau^{-(1+\alpha)}$}
\label{var_case_3}
Unlike case I and II, in this case it is difficult to perform the inverse Laplace transform of $\widetilde{\sigma}_x^2(s)$ in Eq.~\eref{var_all} exactly for arbitrary $t$. However, one can find the variance at large $t$ at which we will mainly focus in the following. As mentioned earlier, we again note from Eqs.~\eref{modl_2} and \eref{modl_3} that the position $x_m$ of the particle after $m$ `collisions' or `jumps' can be written as a  sum of $m$ terms: $x_m= \sum_{j=1}^{m}\xi_j$ with $\xi=v_{j-1}\tau_{j-1}$. The random variable $\xi$s are correlated and additionally can have fat tails in their (marginal) probability distributions for power law `jump' time distributions. For $r=0$, they become independent and the distribution of $\xi$ has a power law tail of the form $\sim 1/|\xi|^{1+\alpha}$ (see sec.~\ref{pdf-x-r-0-case-III} for details). As a consequence, moments of order larger than $\lfloor \alpha \rfloor$ diverge. This suggests that the moments of $x_m$ of order larger than $\lfloor \alpha \rfloor$ should also diverge. This is not true for $x(t)$. Since for a given $t$, the `jump' duration (time between two successive `jumps') variables $\tau$-s can at maximum be $t$. Hence even for $\alpha <1$ we find, as we will see, the variance of $x(t)$ is finite.

At large $t$, the dominant contribution to the variance of $x(t)$ comes from the small $s$ properties of $\widetilde{\sigma}_x^2(s)$ given in Eq.~\eref{var_all}. In the following we present the computation of $\sigma_x^2(t)$ separately
for $r=0,~-1$ and $1$.

\subsubsection{For $r = 0$\\\\}
\label{pl_var_pdf_r0}
We start with $r=0$, for which we first obtain the small $s$ behaviour of $\widetilde{\sigma}_x^2(s)$ using the $s\to 0$ approximations of $\tilde{\psi}(s)$ given in Eq.~\eref{psi_s->0}. Then we perform the inverse Laplace transform using the Tauberian theorem (see \ref{ILT_BR_TU}) to get $\sigma_x^2(t)$ for large $t$. We get 
\begin{equation}
\widetilde{\sigma}_x^2(s) \underset{s \to 0}{\simeq}
\begin{cases}
\vspace{0.2cm}
\frac{2 \sigma^{2}  (1 - \alpha)}{s^3}   \\
\vspace{0.2cm}
\frac{2 \sigma^2}{s^3}  \frac{1}{1 - \gamma_e - \ln(s)}   \\
\vspace{0.2cm}
\frac{2 \sigma^2 (\alpha -1) \Gamma(2 - \alpha)}{\alpha \; s^{4-\alpha}}  \\
\vspace{0.2cm}
\frac{\sigma^2}{2 s^2} (1 - 2 \gamma_e - 2\;\ln(s)) \\
\frac{ \sigma^2 (\alpha -1)}{(\alpha - 2) s^{2}}   \\
\end{cases}
\xRightarrow{\mathcal{L}_t^{-1}} 
\sigma_x^2(t) \underset{t \to \infty}\simeq
\begin{cases}
\vspace{0.2cm}
 (1 - \alpha)\;\sigma^{2} t^{2} & \text{for}\ 0 < \alpha  < 1 \\
\vspace{0.2cm}
\frac{\sigma^2 t^2}{\ln(t)}  & \text{for}\ \alpha = 1\\
\vspace{0.2cm}
\frac{2(\alpha - 1) \Gamma(2 - \alpha)}{\alpha \Gamma(4 - \alpha)}  \sigma^2 t^{3-\alpha}  & \text{for}\ 1 < \alpha  < 2 \\
\vspace{0.2cm}
\sigma^2 t \; \ln(t)  &  \text{for}\ \alpha = 2\\
\frac{\alpha - 1}{\alpha -2 } \sigma^2 t & \text{for}\ \alpha  > 2. 
\end{cases}
\label{msd_pl_r0}
\end{equation}
Here, remember, $\mathcal{L}_t^{-1}[\tilde{f}(s)]$ represents the inverse Laplace transform of a function $\tilde{f}(s)$ to get $f(t)$ in the time domain. 
We verify these results numerically in fig.~\ref{var_pl_all}(a) for different values of $\alpha$ in each of the five regimes displayed in Eq.~\eref{msd_pl_r0}. For each cases we observe excellent agreement.
The time dependence of the variance are similar to that of one dimensional L\'evy walk  because, for $r=0$ the process can be thought of as a Levy walk in one dimension in which the velocity after each `jump' is chosen from a Gaussian distribution \cite{shlesinger1985comment, shlesinger1987levy, klafter1994levy, zaburdaev2015levy}.

In usual studies of L\'evy walks in one dimension,  the velocity distribution $p(v)$ is often taken of the form $p(v) = 1/2 ( \delta(v - v_0) + \delta(v + v_0) )$ \cite{klafter1994levy, zaburdaev2015levy} for some fixed magnitude $v_0$ of the velocity.
Other velocity distributions with variable magnitude and direction
 have also been investigated in the literature but mostly with $v_0 \gg \sigma$ where $\sigma$ is the variance of the velocity \cite{denisov2012levy, zaburdaev2015levy}. For distributions $p(v)$ with $v_0 \gg \sigma$, one observes  travelling peaks in the distribution of the position of the walker\cite{denisov2012levy,zaburdaev2011perturbation}. 
In contrast we here are working in the opposite regime $v_0 < \sigma$ because $v_0 = \langle |v| \rangle =\sigma/\sqrt{2\pi}$  in our case and consequently we do not observe such travelling peaks (see ~sec.~\ref{pdf-x-r-0-case-III}).
We expect the time dependence of the variance of the position  to remain same as in \cite{denisov2012levy, zaburdaev2015levy}, because for any symmetric velocity distribution $p(v)$ with zero mean and finite variance,  the variance of the position exhibits a quite general form as
$\widetilde{\sigma}_x^2(s)= \langle v^2\rangle_c\; \frac{\tilde{\psi}''(s) + s\; \tilde{\Psi}''(s)}{s \;(1 - \tilde{\psi}(s))} $
(see \ref{var_LW} for details).
The dependence of the velocity distribution $p(v)$ appears only through the time independent part of $\sigma^2_x(t)$ through $\sigma^2_v$.
For example, in our case, using $\tilde{\Psi}(s) = (1 - \tilde{\psi}(s))/s$ and $\langle v^2\rangle_c = \sigma^2$, we can correctly recover the exact expression of $\widetilde{\sigma}_x^2(s)$ for $r =0$ given in Eq.~\eref{var_all} which at large time provides the behaviour in Eq.~\eref{msd_pl_r0} for different values of $\alpha$.

To understand these late time behaviours of the variance $\sigma^2_x(t)$, we recall that for $r=0$, the position of the particle $x(t)$ at time $t$ can be written as sum of displacemsents $v \tau \equiv \eta \tau$ after each `collision' or `jump'. For $\alpha<1$, both the first and second moments of the `jump' time distribution $\rho(\tau)$ are divergent because the $\rho(\tau)$ has fat tail. 
In fact we have observed, as will be discussed in ~sec.~\ref{dist_x-0<a<1}, the position $x(t)$ for large $t$ gets the most dominant contribution from the maximum jump duration $\tau_{\max}(t)$ which is typically of the order of $t$. 
So for large $t$ we can approximate $x(t) \sim \eta \tau_{\max}$ which implies ballistic growth of the variance $\sigma_x^2(t) \sim \langle \eta^2 \tau_{\max}^2\rangle_c  \sim \sigma^2 t^2$ [see Eq.~\eref{P_taumax_t}]. In the $\alpha \to 1$ limit, the results in Eq.~\eref{msd_pl_r0} both from below ($1^-$) and above ($1^+$), provides $\sigma_x^2(t) \sim \sigma^2 t^2$. However, in our calculation we find an additional $\ln(t)$ modulation. 
Such a $\ln(t)$ correction appears through the distribution of $\tau_{\max}(t)$ [see ~\ref{P_tau_max_a=1}].

For $\alpha > 2$, all the terms in the sum of the individual jumps contribute to the final position $x(t)$ at time $t$ at the same order. 
Hence, for this case $\sigma_x^2(t) \sim $ [variance of the position in a single `jump'] $\times$ [number of `jump's in time $t$]. For $\alpha > 2$ both, the first and the second moments of the `jump' time distributions are finite and are given by $ \alpha/(\alpha-1)$ and $\alpha/(\alpha -2)$, respectively. Consequently, the number of `jump's in a large time interval is typically $m_t\sim t/\langle \tau \rangle = t(\alpha -1)/\alpha$ using which one gets diffusive growth of the variance 
$$\sigma_x^2(t)  \sim \langle \eta^2 \rangle \langle\tau^2 \rangle \times \left[t\frac{(\alpha -1)}{\alpha}\right] \sim \frac{(\alpha -1)}{(\alpha-2)}\sigma^2 t.$$

For $\alpha$ in the intermediate regime $1 < \alpha < 2$ the mean $\langle \tau \rangle = \alpha/(\alpha -1)$ is finite but all higher order moments are divergent.
These divergent higher order moments of the waiting time distribution give rise to the superdiffusive behaviour. For jump distributions with variance much smaller than the mean, it has been observed that the variance of the position gets the leading contributions from the last long ballistic jumps that has never changed till time $t$ \cite{denisov2012levy,zaburdaev2011perturbation}. In our case also we find that the leading contribution to the variance of the position $\sigma_x^2(t)$ comes from the tail of the distribution where trajectories with  `jumps'  of duration $\sim \mathcal{O}(t)$ contribute. As shown in the ~sec.~\ref{r0_alpha>1_pdf} later, the behaviour of the distribution $P(x,t)$  has different scaling forms in the central part (Eq.~\eref{P(x,t)_levy_a>1}) and at the tails (Eq.~\eref{P_{tail}(x,t)_a>1_r0_7}). 
Using these forms of the distribution, it is easy to show that the variance of the position for large $t$ grows as $\sigma_x^2(t)  \propto t^{3-\alpha}$.

Alternatively, this large $t$ growth can be obtained from the two-point velocity correlation function as well. For $1 < \alpha \le 2$, it is possible to show that the velocity-velocity correlation function decays slowly as  a power-law
\begin{equation} 
C_{vv}(t_1, t_2) \sim |t_1 - t_2|^{1-\alpha}, 
\end{equation} 
(see \ref{velo_corr_r0} for details).
Plugging this result into $\sigma_x^2(t) =\langle x^2(t) \rangle_c = 2 \int_{0}^{t} dt_1 \int_{t_1}^{t} dt_2 \; C_{vv}(t_1, t_2)$ and performing the integrations, we reproduce the behaviour of the variance $\propto t^{3-\alpha}$ for $1 < \alpha < 2$ and $\propto t \; \ln(t)$ for $\alpha = 2$ respectively.

\subsubsection{For $r = -1$\\\\}
\label{pl_pdf_rn1}
\begin{figure*}[t!]
\includegraphics[width=1.0\hsize]{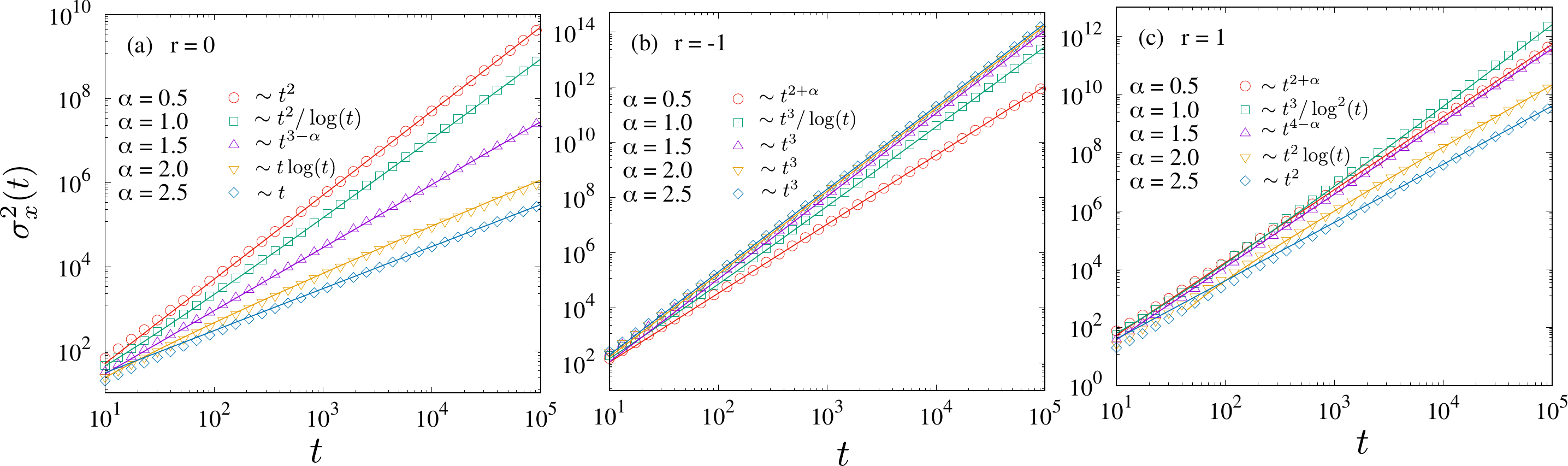}
\caption{\label{var_pl_all} (Color online). Variance of position $\sigma_x^2(t)$ are plotted for different $r$ and $\alpha$. Symbols are from numerical simulation and the solid lines are the corresponding analytical results from Eqs.~\eref{msd_pl_r0}, \eref{msd_pl_rn1} and \eref{msd_pl_r1} respectively.} 
\end{figure*}

For this case also we follow the same procedure as done for $r=0$. We find the large $t$ behaviour of $\sigma_x^2(t)$ from the small $s$ behaviour of $\widetilde{\sigma}_x^2(s)$. Performing inverse Laplace transform we get 
\begin{equation}
\widetilde{\sigma}_x^2(s) \underset{s \to 0}{\simeq}
\begin{cases}
\vspace{0.2cm}
\frac{2 \sigma^2}{\Gamma(1 - \alpha)} \frac{1}{s^{3+\alpha}}  \\
\vspace{0.2cm}
\frac{2 \sigma^2}{s^4}  \frac{1}{1 - \gamma_e - \ln(s)} \\
 \frac{2 \sigma^2}{s^4} \left( \frac{\alpha - 1}{\alpha} \right)  \\
\end{cases}
\qquad \xRightarrow{\mathcal{L}_t^{-1}} \qquad \sigma_x^2(t) \underset{t \to \infty}\simeq
\begin{cases}
\vspace{0.2cm}
\frac{2 \sigma^2 }{\Gamma(1-\alpha) \Gamma(3+\alpha) } t^{2+\alpha} 
&\; \text{for}\  \alpha  < 1 \\
\vspace{0.2cm}
\frac{\sigma^2 t^3}{3 \;\ln(t)}  &\; \text{for}\ \alpha = 1\\
\frac{\sigma^2 (\alpha - 1) }{3 \alpha} t^{3} &\; \text{for}\ \alpha  > 1.
\end{cases}
\label{msd_pl_rn1}
\end{equation}  
In this case the variance of the position shows super-diffusive behaviour for all ranges of $\alpha$.
We observe that the  exponent of super-diffusive growths depends on  $\alpha$ for $0  < \alpha < 1$, whereas it becomes independent of it for $\alpha > 1$ with a logarithmic correction to $\propto t^3$ behaviour for $\alpha = 1$.  These late time asymptotic behaviours of $\sigma_x^2(t)$ are verified numerically in fig.~\ref{var_pl_all}(b) for different values of $\alpha$ from the three regimes mentioned in Eq.~\eref{msd_pl_rn1} and we observe excellent agreement.

As can be observed from Eq.~\eref{modl_1} that, for $r=-1$  the velocity after $i^{\textrm{th}}$ `jump' event is given by 
$v_i= \sum_{l=0}^i \eta_l$ and the velocity after $j^{\textrm{th}} (> i)$ `jump' event is given by $v_j=v_i+\sum_{l=j+1}^i\eta_l$ for $j>i>0$. Hence the two point velocity correlation $C_{vv}(t_1,t_2)$ is given by $C_{vv}(t_1,t_2) = \langle v_{n(\textrm{min}(t_1,t_2))}^2 \rangle_c\; Q_{-1}(|t_1-t_2|)$, 
where $n(\textrm{min}(t_1,t_2))$ is the number of `jump' events till time $\textrm{min}(t_1,t_2)$. This can  be seen from Eq.~\eref{corrv_6} of \ref{velo_corr} where we note that $Q_{-1}(t) = 1$.  From Eq.~\eref{x^2-C_vv-rela} it is easy to see that  $\langle v_{n(\textrm{min}(t_1,t_2))}^2 \rangle_c=\sigma^2 \langle n(\textrm{min}(t_1,t_2))\rangle$.  The average number of events $\langle n(t)\rangle$ till time $t$ can be computed from its Laplace transform $\widetilde{\langle n \rangle}(s) = \tilde{\Psi}(s) \sum_{m=0}^{\infty} m \tilde{\psi}^m(s) = \frac{\tilde{\Psi}(s)\tilde{\psi}(s)}{(1 - \tilde{\psi}(s))^2}$ where $\tilde{\psi}(s)$ is the Laplace transform of the distribution $\rho(\tau)$ of the `jump' duration and $\tilde{\Psi}(s) = \frac{1-\tilde{\psi}(s)}{s}$. Hence $\widetilde{\langle n \rangle}(s) =\frac{\tilde{\psi}(s)}{s(1-\tilde{\psi}(s))}$. Since we are interested in the large $t$ behaviour of $\sigma^2_x(t)$, we require the large $t$ behaviour of $\langle n(t) \rangle$ as well and for that we focus on the small $s$ behaviour of $\tilde{\psi}(s)$ which are given in Eq.~\eref{psi_s->0}. Using these relations and performing the inverse Laplace transforms we get 
\be
\langle n(t) \rangle \simeq
\begin{cases}
 \left(\frac{\text{Sin}(\pi \alpha)}{\pi \alpha } \right) t^{\alpha}  & \text{for}\ \alpha  < 1 \\
  t \;\text{Log}^{-1}(t)  & \text{for}\ \alpha = 1\\
 \left(\frac{\alpha - 1}{\alpha} \right) t & \text{for}\ \alpha  > 1 
\end{cases}
~~~\text{for~large}~t.
\label{av-n_t-case-II-rm1}
\ee
Thus we get explicit expressions of the two-point velocity correlation $C_{vv}(t_1,t_2) = \sigma^2\langle n(\min(t_1,t_2)) \rangle $ for different values of $\alpha$. We now use these expressions in the relation between $\sigma^2_x(t)$ and $C_{vv}(t_1, t_2)$ given in Eq.~\eref{x^2-C_vv-rela} and, performing the integrations we reproduce the superdiffusive growths of the variance of the position for different values of $\alpha$ as announced in Eq.~\eref{msd_pl_rn1}. Note that for $\alpha>1$ we once again observe agreement with RAP {\it i.e.} $t^3$ scaling for the variance of the position (see \ref{appendix_RAP} for definition and properties of RAP).

\subsubsection{For $r = 1$\\\\}
\label{pl_pdf_r1}
Following a similar procedure like the previous two cases we, in this case, find the following large $t$ asymptotic behaviour of the variance 
$\sigma_x^2(t)$: 
\begin{equation}
\widetilde{\sigma}_x^2(s) \underset{s \to 0}{\simeq}
\begin{cases}
\vspace{0.2cm}
\frac{(1-\alpha)}{\Gamma(1-\alpha)} \frac{2 \sigma^2}{s^{3+\alpha}}  \\
\vspace{0.2cm}
\frac{2 \sigma ^2}{s^4(1- \gamma_e - \ln(s))^{2} }   \\
\vspace{0.2cm}
\frac{(\alpha -1)^2 \; \Gamma(2-\alpha)}{\alpha^2} \frac{2 \sigma^2}{s^{5 - \alpha}}   \\
\vspace{0.2cm}
 - \frac{\sigma^2 }{4 s^3} (3 + 2 \gamma_e + 2 \; \ln(s))   \\
\frac{\sigma^2}{\alpha (\alpha - 2) \;s^3}   \\
\end{cases}
\hspace{-1.0cm}
\xRightarrow{\mathcal{L}_t^{-1}} ~ \sigma_x^2(t) \underset{t \to \infty}\simeq
\begin{cases}
\vspace{0.2cm}
\frac{2 \sigma^2 (1-\alpha)}{\Gamma(1-\alpha) \Gamma(3 + \alpha)} \; t^{2+\alpha}  &\; \text{for}\  0 < \alpha  < 1 \\
\vspace{0.2cm}
\frac{\sigma^2 t^{3}}{3 \;\ln^2(t)}  &\; \text{for}\  \alpha  = 1 \\
\vspace{0.2cm}
\frac{2 \sigma^2 \Gamma(2-\alpha)}{\Gamma(5-\alpha)} \left( \frac{\alpha -1}{\alpha}\right)^2  t^{4-\alpha}   
 &\; \text{for}\  1< \alpha  < 2 \\
 \vspace{0.2cm}
\frac{1}{4}\;\sigma^2 t^2 \;(\ln(t) -3)  & \; \text{for}\ \alpha = 2\\
\frac{\sigma^2 t^2}{2 \alpha (\alpha - 2)} &\; \text{for}\ \alpha  > 2.
\end{cases}
\label{msd_pl_r1}
\end{equation}  
Once again these results are verified numerically in fig.~\ref{var_pl_all}(c) for $\alpha$ values lying in different regimes in Eq.\eref{msd_pl_r1}.

Although the variance of the velocity $\sigma_v^2(t)$ is same for $r = \pm 1$, from Eqs.~\eref{msd_pl_rn1} and \eref{msd_pl_r1} we see the behaviour of $\sigma_x^2(t)$ are different for them.
The growth of the variance $\sigma_x^2(t)$  increases with time with a $\alpha$ dependent exponent within the range $0  < \alpha < 1$. 
The exponent becomes maximum at $\alpha = 1$ with a logarithmic correction to $\sim t^3$ behaviour.
Within the range $1 < \alpha < 2$ this exponent decreases as $\sim({4-\alpha})$ and becomes $\alpha$ independent for $\alpha \geq 2$  with another logarithmic correction for $\alpha = 2$. In the following we try to understand the qualitative behaviour of the variance of the position in different regimes of $\alpha$.

As done for $r=-1$ case in the previous sec.~\ref{pl_pdf_rn1}, in this case also starting from Eq.~\eref{modl_1}, one can easily see that the two point velocity correlation is given by $C_{vv}(t_1,t_2) \simeq \langle v^2_{n(\min(t_1,t_2))}\rangle_c \;Q_{1}(|t_1-t_2|)$ as derived in Eq.~\eref{corrv_5} of \ref{velo_corr} for general $r$. Unlike the $r=-1$ case in this case $Q_{1}(t)$ is not equal to one in general. However, for $0 < \alpha < 1$ the dominant contribution to $Q_{1}(|t_1-t_2|)$ comes from the event in which there are typically no `jump' events in the time interval $|t_1-t_2|$. This happens becaues for large $t_1$ and $t_2$, the interval $|t_1-t_2|$ typically falls in the last incomplete step which is usually the largest. Hence in this case  also we approximate $Q_1(|t_1-t_2|) \simeq 1$. Thus the velocity correlation function $C_{vv}(t_1,t_2) \simeq  \sigma^2\langle n(\min(t_1,t_2)\rangle \simeq \sigma^2 \min(t_1,t_2)^\alpha$ as found in the previous sec.~\ref{pl_pdf_rn1} [see after Eq.~\eref{av-n_t-case-II-rm1}]. An alternative derivation of this result for $C_{vv}(t_1,t_2)$ is given in \ref{velo_corr_r1}. Using this expression in Eq.~\eref{x^2-C_vv-rela} and, performing the integrations we reproduce the superdiffusive  behaviour $\sim t^{2+\alpha}$ as announced in Eq.~\eref{msd_pl_r1} which obviously is  same as for $r=-1$ in the  $0<\alpha<1$ range.

A little more rigorous argument can be presented for  the superdiffusive  behaviour $\sim t^{2+\alpha}$ of $\sigma^2_x(t)$ for $r=1$ in the regime $0<\alpha<1$. As has been encountered, in this case the dominant contribution to the position comes from the largest jump duration $\tau_{\max}$ within time $t$. Evidence of this fact will be provided in sec.~\ref{dist_x_rpm1-0<a<1} [see fig.~\ref{pdf_pl_rpm1_a<1} bottom panel] where we study the distribution $P(x,t)$ of the position. 
So writing $x(t) \simeq v_l \tau_{\max}$ where $\tau_{\max}$ is the duration of the largest jump and $v_l$ is the velocity of the particle during this jump event. If in a particular trajectory, the particle makes $m$ jumps then this longest jump could happen at any of the steps or at the last incomplete step. Using this information one can compute the variance $\sigma_x^2(t)$. The Laplace transform of this variance is given by 
\bea
\nonumber
\hspace{-1.5cm}
\widetilde{\sigma_x^2}(s) \simeq \sum_{m=0}^\infty \left[ \left \langle \left( \mathbb{I}(m\ge 1) \sum_{l=0}^{m-1}\langle v_l^2 \rangle\tau_l^2 + \langle v_{l}^2\rangle {\tau^*_m}^2\right) e^{-s(\tau_0+\tau_1+...+\tau_{m-1}+\tau_m^*)}\right \rangle\right]\\ = \frac{\sigma^2}{s \;(1 - \tilde{\psi}(s))^2} \left( \tilde{\psi}''(s) + s \; \tilde{\Psi}''(s) \right),
\label{tilde_sig_x_r1_0a1}
\eea
where we have used $\langle v_l^2 \rangle = \sigma^2 (l+1)$ for $r=1$. Given that there are $m$ number of jump events within time $t$, the first term inside the first parenthesis corresponds to the case when the particle makes the longest jump at the $l$th step and the second term corresponds to the case when  longest jump occurs at the last incomplete step.

Recalling $\tilde{\Psi}(s) = (1-\tilde{\psi}(s))/s$ and using $\tilde{\psi}(s) \simeq 1 - \Gamma(1-\alpha) s^{\alpha}$ in Eq.~\eref{tilde_sig_x_r1_0a1} we get $\widetilde{\sigma_x^2}(s) \simeq \frac{2 \sigma^2 (1-\alpha)}{\Gamma(1-\alpha) \; s^{3+\alpha}}$ (for small $s) $ which after performing inverse Laplace transform would provide $\sigma_x^2(t)\simeq \frac{2 \sigma^2 (1-\alpha)}{\Gamma(1-\alpha)\; \Gamma(3+\alpha)} t^{2+\alpha}$ for large $t$.

We now focus for $1 < \alpha \le 2$, in which regime it is clear that one can not approximate $Q_1(|t_1-t_2|) \simeq 1$. In fact  it is possible to  show that the two-point velocity correlation function behaves for large $t_1,~t_2$ as 
\bea
C_{vv}(t_1, t_2) \sim \textrm{min}\{t_1, t_2\} \;|t_1 - t_2|^{1-\alpha}
\label{vel_corrv_r11} 
\eea
(see Eq.~\eref{equn13} in \ref{velo_corr_r1}).
Using this result in Eq.~\eref{x^2-C_vv-rela} and performing the integrals, one can easily recover large $t$ behaviours $\sigma_x^2(t)  \propto t^{4-\alpha}$ for $1 < \alpha < 2$ and $\propto t^2 \;\ln(t)$  and for $\alpha = 2$.

On the other hand, for $\alpha > 2$, the behaviour of $\sigma_x^2(t)  \propto t^2$ dominates over the $\sigma_x^2(t)  \propto t^{4-\alpha}$, {\it i.e.}, the variance shows ballistic growth with time as in case II. The explanation of this behaviour is similar to that of the exponential case discussed in sec.~\ref{var_cs2}.

\section{PDF of the position}
\label{pdf_dis}

We now study the distribution of the position. Like earlier two sections, here also we discuss three different cases of $\rho(\tau)$ separately for three limiting values of $r = 0, -1$ and $1$. We start with $r=0$ case.

\subsection{For $r = 0$}
This case is relatively simpler than $r=\pm1$ because for the later case the velocity of the particle at different time gets correlated as we have seen earlier. From Eq.~\eref{modl_1}, it possible to see that the position after $m$th jump event  can be described by a simple random walk of independent steps {\it i.e}
$x_m = \sum_{j=0}^{m-1}\xi_j$ where $\xi_j=\eta_j\tau_j$. Clearly, $\xi$ is a random variable which is a multiplication of two independent random variables $\tau$ and $\eta$. The distribution of $\xi=\eta \tau$ can be easily computed as 
\bea
\varrho(\xi) = \int_0^{\infty} d\tau \;\rho(\tau)  \int_{-\infty}^{\infty} d\eta \;p(\eta) \; \delta(\xi -\eta\tau),
\label{eqp_1p}
\eea
where recall $p(\eta)$ is Gaussian given in Eq.~\eref{p(eta)} and $\rho(\tau)$ for the three cases are given in Eq.~\eref{time_dist_1}. Since  $\tau$ is a positive random variable and the distribution of $\eta$ is symmetric about zero, the distribution of $\xi$ is also symmetric about zero. The characteristic function $\lambda(k)$ of $\varrho(\xi)$ is defined as
\bea
\lambda(k) =\left \langle e^{ik\xi}  \right \rangle_{\varrho(\xi)}= \int_{-\infty}^{\infty} d\xi \;e^{ik\xi} \;\varrho(\xi),
\label{lambda(k)} 
\eea
which will be used later.

\subsubsection{Case I: $\rho(\tau) = \delta(\tau - a)$\\\\}

In this case the number of complete steps in time $t$ is $\lfloor t/a \rfloor$  where $\lfloor z \rfloor$ represents the largest interger but not larger than $z$.
It is easy to see from Eq.~\eref{eqp_1p} that the position in each step is a Gaussian random variable (RV) with the variance $a^2 \sigma^2$.
The position made by the particle in the last incomplete step is also a Gaussian RV with variance $u^2 \sigma^2$ where $u = t - a \lfloor t/a \rfloor$.
Hence, the distribution of the position at time $t$ is a Gaussian distribution with variance $(\lfloor t/a \rfloor a^2 + u^2) \sigma^2 \simeq a \sigma^2 t$ for large~$t$ which describes typical fluctuations. The tails of the distribution should be described by an appropriate Large deviation function 

\subsubsection{Case II: $\rho(\tau) = \beta \e^{-\beta \tau}$\\\\}
\label{pdf_exp_r0}

\begin{figure}[t!]
\centering{\includegraphics[width=0.45\hsize]{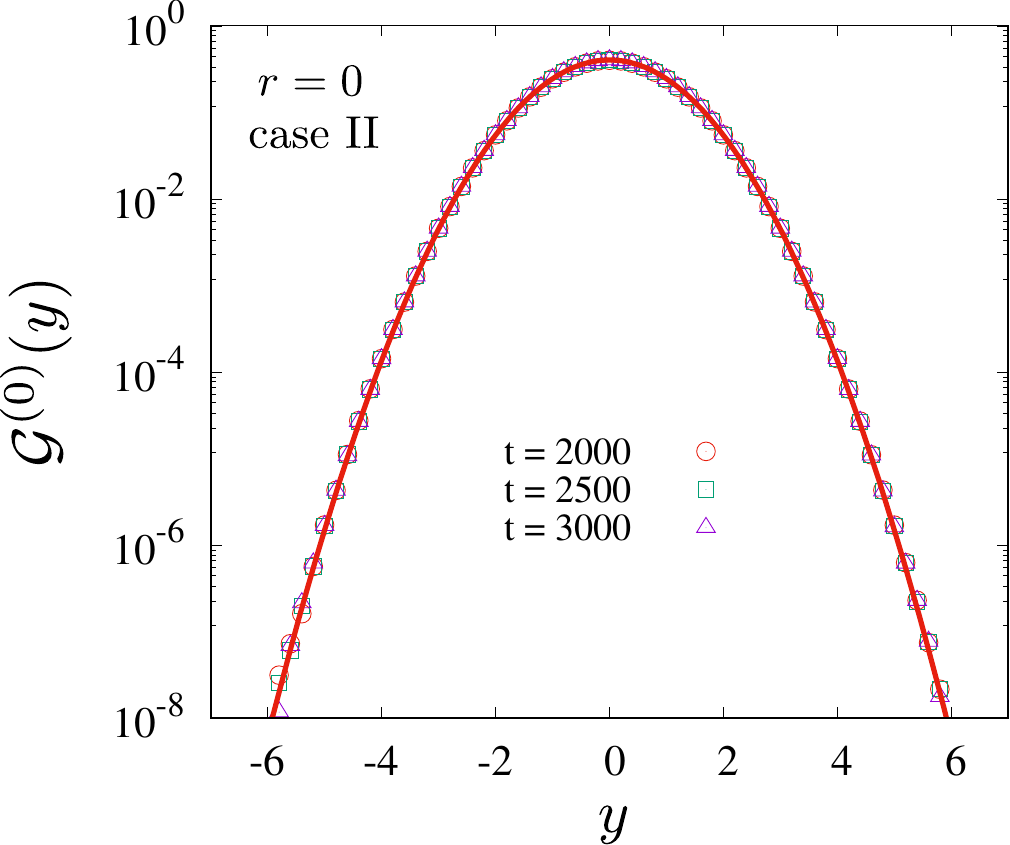}}
\caption{\label{pdf_r0_exp} (Color online). Probability distribution function of scaled position $y = x/\sqrt{2\sigma^2 t/\beta}$ are plotted for $r = 0$ with exponential waiting time distribution. Symbols are from numerical simulation which show good agreement with the analytical prediction (red solid line) given in Eq.~\eref{eq5}.} 
\end{figure}

We recall that in this case $\rho(\tau) = \beta \e^{-\beta \tau}$, inserting which in Eq.~\eref{eqp_1p} one finds that the distribution of $\xi$ is given by 
 \bea
 \nonumber
\varrho_1(u) = \int_0^{\infty} d\tau \;\rho(\tau)  \int_{-\infty}^{\infty} d\eta \;p(\eta) \; \delta(\xi -\eta\tau) \\
= \frac{\beta}{\sqrt{2\pi \sigma^2}} \; \int_0^{\infty} \frac{d\tau}{\tau}\; \exp\left[{-\beta \tau - \frac{\xi^2}{2 \sigma^2 \tau^2}}\right] 
\label{pdf_exp_gus1}
\eea
which is a symmetric distribution and decays for large $|\xi|$ as $\varrho(\xi) \propto |\xi|^{-1/3}\;\exp[-3/2 \;(|\xi|\beta/\sigma)^{2/3}]$. 
Hence, by virtue of central limit theorem, the fluctuation of the position $x_m$ after $m$ `jump' events  is Gaussian with variance $\langle x_m^2\rangle = 2 m \left( \sigma/\beta \right)^2$ and mean zero.
On the other hand, since each $\tau$ is chosen from exponential distribution, the average
time duration between two successive steps is $\beta^{-1}$ and the number of steps taken by the particle till time $t$ is typically of the order of $m \sim \beta t$ for large $t$.
As a result for large $t$, the distribution of the position $x(t)$ made by the particle till time $t$  is a Gaussian distribution with zero mean and variance  $\simeq \langle x_{\lfloor \beta t \rfloor}^2 \rangle_c \simeq 2 \sigma^2 t/\beta$ {\it i.e.}
\bea
\hspace{-1cm}
P(x,t)  \simeq  \sqrt{\frac{\beta}{2  t \sigma^2}}~\mathcal{G}^{(0)}\left( \frac{x}{\sqrt{2 \sigma^2 t/\beta}}\right),~\textrm{where},~~\mathcal{G}^{(0)}(y)= \frac{1}{\sqrt{2\pi}}\exp\left[-\frac{ y^2}{2} \right].
\label{eq5}
\eea
This result is verified numerically  in fig.~\ref{pdf_r0_exp}. For a more detailed calculation of $P(x,t)$ in this case, see \ref{pdf_single_stp_r0}. It is known that for $x$ really large, the distribution $P(x,t)$ can not be described by the above Gaussian form but in terms of a Large deviation function $\Phi_{II}(x/t)$  such that $P(x,t) \asymp e^{-t \Phi_{II}(x/t)}$ with $\Phi_{II}(v) = 
\underset {k}
{\min} \left[ikv -\beta \ln \lambda(k)\right]$ \cite{touchette2009large}.

\subsubsection{Case III: $\rho(\tau)=\alpha \tau^{-(1+\alpha)}$\\\\}
\label{pdf-x-r-0-case-III}

In this case the distribution of jump time durations is given by $\rho(\tau)=\alpha \tau^{-(1+\alpha)}$ (see Eq.~\eref{time_dist_1}). Inserting this distribution in Eq.~\eref{eqp_1p} we get 
\bea
\varrho(\xi) 
&= \frac{\alpha}{\sqrt{2\pi \sigma^2}} \;\int_1^{\infty} \frac{d\tau}{\tau^{1+\alpha}} \int_{-\infty}^{\infty} 
dv \;e^{-v^2/2 \sigma^2} \delta(\xi -v\tau)
\\&
= \frac{\alpha}{2\sqrt{2 \pi \sigma^2}}\; \left( \frac{2 \sigma^2}{\xi^2}\right)^{\frac{1+\alpha}{2}}\;\gamma \left( \frac{1+\alpha}{2}, \frac{\xi^2}{2 \sigma^2}\right)
\label{eqp_1}
\eea
where $\gamma(s, z)$ is the incomplete gamma function defined by $\gamma(s, z) = \int_{0}^z dw \;e^{-w}\; w^{s-1}$.
The distribution $\varrho(\xi) $ at the tails $|\xi|$ decays as 
\be
\varrho(\xi \rightarrow \pm \infty) \simeq \frac{\alpha}{2 \sqrt{2 \pi \sigma^2}} \;\Gamma \left(\frac{\alpha +1}{2}\right) \left(\frac{2 \sigma^2}{\xi^2}\right)^{\frac{\alpha +1}{2}}.
\label{eqp_3}
\ee
As realised earlier, in this case the position can be described by a L\'evy walk with velocity chosen from a Gaussian distribution of variance $\sigma^2$ \cite{shlesinger1985comment, shlesinger1987levy, klafter1994levy, zaburdaev2015levy}. 
As also mentioned earlier, in our case the velocity distribution has zero mean and non-zero variance $\sigma^2$.
The  characteristic function of the distribution $\varrho(\xi)$ in Eq.~\eref{eqp_1} is given by 

\bea
\label{lambda(k)_r0_a>1}
\nonumber
\lambda(k) &= \int_{-\infty}^{\infty} d\xi \;e^{ik\xi} \;\varrho(\xi) 
=\frac{\alpha}{2} \left(\frac{k^2\sigma^2}{2}\right)^{\frac{\alpha}{2}} \Gamma\left(-\frac{\alpha}{2},\frac{k^2\sigma^2}{2}\right) \\
&
\underset{k \rightarrow 0}
{\simeq}
\left \{\begin{array}{ccc}
1 - \frac{1}{2}\mathcal{D}_{\alpha}  \sigma^{\alpha} |k|^{\alpha} + \mathcal{O}(k^2) & \textrm{for}\;  \alpha < 2\\
1 + \frac{1}{2}\sigma^2 k^2 \ln(|k|^{2}) +\mathcal{O}(k^2) & \textrm{for}\; \alpha = 2, \\
1 - \frac{1}{2} \mathcal{D}_\alpha \sigma^2 k^2 + \mathcal{O}(|k|^{\alpha}) & \textrm{for}\; \alpha > 2
\end{array}
            \right.
~~~~\textrm{with}~~~~\\& \nonumber
\mathcal{D}_{\alpha} =
\left \{\begin{array}{ccc}
|\frac{ \alpha}{2^{\alpha/2}} \;\Gamma\left( -\frac{\alpha}{2}\right)|,~~&\textrm{for}~1<\alpha<2\\ 
\frac{\alpha}{\alpha -2},~~&\textrm{for}~\alpha>2.
\end{array}
            \right. 
\eea
Here, $\Gamma(s, z)$ represents the upper incomplete gamma function $\Gamma(s, z)  =\Gamma(s)-\gamma(s,z)$.

\subsubsubsection{For $0 < \alpha < 1$:}
\label{dist_x-0<a<1}


In this domain of $\alpha$, all the moments of the waiting time distribution diverge. We expect that in this case, for a given large but finite $t$, the number of jump events are not proportional to $t$ (see Eq.~\eref{av-n_t-case-II-rm1}). The final position $x(t)$ of the 
walker gets the most dominant contribution from the displacement made in the largest jump duration $\tau_{\max}(< t)$.  Consequently, we expect the distribution of the position $P(x,t)$ of the walker would be given by the distribution of the position $\Delta_{\max}=\eta \tau_{\max}$ made in the largest jump duration in the interval $[0,t]$.  Since for large $t$, $\tau_{\max}$ is typically of the order of $t$, it is expected to have a ballistic scaling for the position distribution $P(x,t)$  {\it i.e.} 
\be
P(x,t) \simeq \frac{1}{ t} \mathbb{G}^{(0)}_{\alpha<1}\left( \frac{x}{ t}\right). \label{P-ballis-scaling-0<a<1}
\ee
Such ballistic scaling have been discussed in previous studies of L\'evy walks \cite{klafter2011first,zaburdaev2015levy}.  In fact for arbitrary velocity distribution $p(\eta)$, a general exact but implicit expression for the scaling distribution $\mathbb{G}^{(0)}_{\alpha<1}$ has been obtained as \cite{froemberg2015asymptotic} 
\be 
\mathbb{G}^{(0)}_{\alpha<1}\left(y = x/t\right) = -\frac{1}{\pi } \;
\underset{\epsilon \rightarrow 0}
{\textrm{lim}} \; \textrm{Im} \; \frac{\int_{-\infty}^{\infty} d\eta \;p(\eta)\;(y + i \epsilon -\eta)^{\alpha-1}}{\int_{-\infty}^{\infty} d\eta\;p(\eta)\;(y + i \epsilon -\eta)^{\alpha}}.
\ee
Recall in this paper we consider $p(\eta)$ to be a zero mean  Gaussian with variance $\sigma^2$. From this expression one can in principle compute the scaling distribution but it is difficult to find an explicit form. However, from the distribution of $\Delta_{\max}$ it is possible to obtain an approximate but more explicit expression of the distribution $P(x,t)$.

\begin{figure}[t!]
\includegraphics[width=1.\hsize]{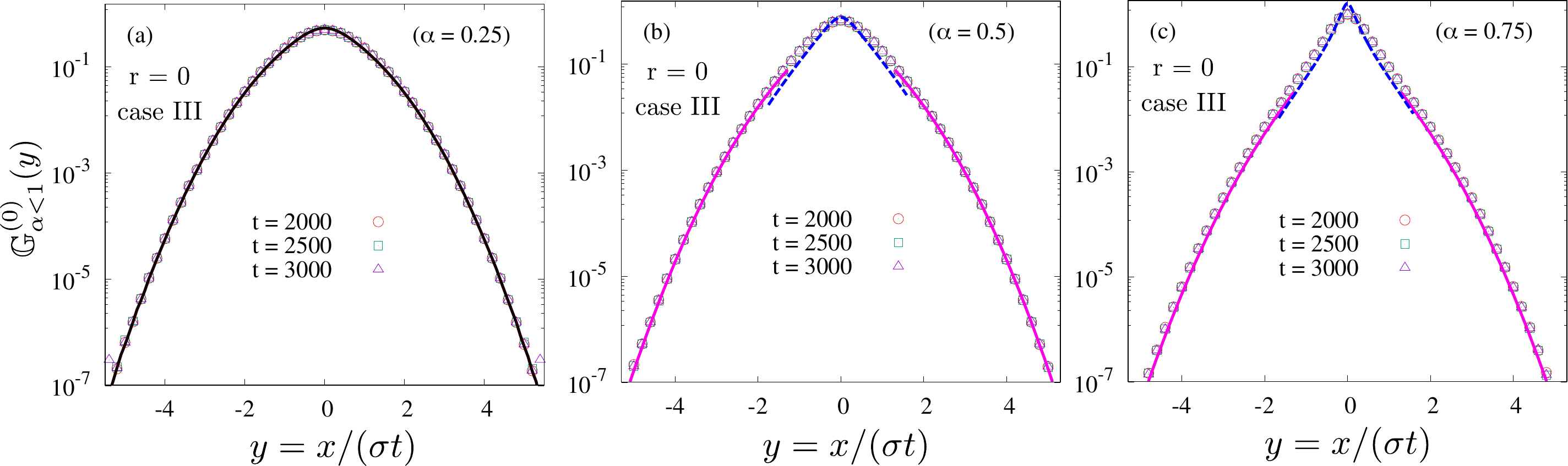}
\caption{(Color online). PDF of the scaled position $y = x/(\sigma t)$ are plotted for $r = 0$ and for (a) $\alpha = 0.25$, (b) $\alpha = 0.5$ and (c) $\alpha = 0.75$. In panel (a), we compare the distribution of $x(t)$ (symbols) and the distribution $\Delta_{\max}(t)$ (black solid line). The excellent agreement justifies the dominant contribution of the displacement $\Delta_{\max}$ made in the longest jump duration to the final position $x(t)$ for large $t$. In panels (b) and (c), we verify the our theoretical expressions of the distribution $P(x,t)$ in Eq.~\eref{G_alpha(y)} with numerical simulation. In both plots symbols correspond to simulation data whereas  the blue dashed line around the central part and the magenta solid lines  at the tails of the distribution, respectively, correspond to analytical expressions given in Eq.~\eref{G_alpha(y)} along with Eq.~\eref{G_alpha(y)}. }
\label{pdfpl_r0_a<1} 
\end{figure}

A numerical verification of the fact that the distribution of $\Delta_{\max}=\eta \tau_{\max}(t)$ describes the distribution of the final position $x(t)$ well, is presented in 
fig.~\ref{pdfpl_r0_a<1}(a) where for $\alpha=0.25$ the scaling  distribution  of the position is compared with the scaling distribution of $\Delta_{\max}(t)$. The symbols are obtained from the direct numerical simulation of the dynamics in Eqs.~(\ref{modl_2}-\ref{modl_3}) and corresponds to the (scaled) distribution of the position $x(t)$ whereas the solid (red) lines correspond to the (scaled) distribution of $\Delta_{\max}=\eta \tau_{\max}(t)$ also obtained numerically. We observe excellent match between the two which justifies the arguments given in the previous paragraph.

For given realisation of the trajectory of length $t$, the duration $\tau_{\max}$ depends on $t$ as can be clearly seen from the definition $\tau_{\max}(t) = \max[\tau_0, \tau_1,......., \tau_{m(t)-1}, \tau^{*}_{m(t)}]$ where $m(t)$ is the number of jumps occurred within time $t$ in that particular trajectory and $ \tau^{*}_{m(t)}$ is the duration of the last incomplete jump.  If $\mathcal{P}(\tau_{\max}|t)$ denotes the distribution of $\tau_{\max}$ then, using the above arguments we write 
\bea
\nonumber
P(x,t) \simeq &\int_{-\infty}^{\infty} d\eta\; p(\eta) \int_{1}^{t} d\tau_{\max} \; \mathcal{P}(\tau_{\max}|t) \; \delta(x - \eta \tau_{\max}) 
\\& \simeq \int_{1}^{t} \frac{d\tau_{\max}}{\tau_{\max}} \; \mathcal{P}(\tau_{\max}|t) \;p\left(\frac{x}{\tau_{\max}} \right).
\label{P_{tails}(x,t)_r0}
\eea
What is the distribution $\mathcal{P}(\tau_{\max}|t)$ of $\tau_{\max}$ for given $t$? This quantity has recently been studied in detail in Ref:~\cite{godreche2015statistics} where it has been shown that for large $t$ this distribution satisfy the following scaling form 
\be 
\mathcal{P}(\tau_{\max}|t) = \frac{1}{t} \;f_{\alpha}\left(\frac{\tau_{\max}}{t} \right),
\label{P_taumax_t}
\ee
with $f_\alpha(w)$ having the following asymptotic forms \cite{godreche2015statistics} 
\bea
f_\alpha(w) \simeq
\left \{\begin{array}{ccc} 
 \frac{c_{\alpha}}{\alpha w^2} \;e^{-c_{\alpha}/w},~& \textrm{for}~w \to 0^+ \\
 & \\
\frac{\textrm{Sin}(\pi \alpha)}{\pi} \; (1 - w)^{\alpha -1}\;\Theta(1-w),~& \textrm{for}~w \to 1^-.
\end{array}
            \right. \label{f_alpha}
\eea
Here $\Theta(x)$ is Heaviside theta function and $c_\alpha$ is a constant obtained from the solution of $\, _1F_1(1;1-\alpha ;-c_\alpha)=0$. Here $\, _1F_1(a;b;x)$ is hypergeometric function. Inserting the expression of $\mathcal{P}(\tau_{\max}|t)$ from Eq.~\eref{P_{tails}(x,t)_r0} along with Eq.~\eref{f_alpha} and using the explicit form of $p(\eta)$ from Eq.~\eref{p(eta)}, we get
\bea
\nonumber
P(x,t) \simeq & \frac{1}{ t}\;\mathbb{G}^{(0)}_{\alpha<1}\left( \frac{x}{ t} \right) \quad \textrm{where} \\ & \mathbb{G}^{(0)}_{\alpha<1}(y) \simeq \frac{1}{\sqrt{2 \pi {\sigma^2}}} \int_{0}^{1} \frac{dw}{w}\; f_{\alpha}(w) \;\exp\left[- \frac{y^2}{2 \sigma^2w^2} \right].
\label{P_{tails}(x,t)_m1}
\eea
for large $t$.
Using the asymptotic forms of $f_\alpha(w)$ from Eq.~\eref{f_alpha} in the above equation we get the approximate forms of the scaling function $\mathbb{G}^{(0)}_{\alpha<1}(y)$ in different asymptotic regimes. The $w \to 0$ asymptotic of $f_\alpha(w)$ provide us the central part of the scaling distribution $\mathbb{G}^{(0)}_{\alpha<1}(y)$ valid for small $|y|$  whereas  the asymptotic form of $f_\alpha(w)$ for $w \to 1$ provides $\mathbb{G}^{(0)}_{\alpha<1}(y)$ for large $|y|$. We get,
\bea
\mathbb{G}^{(0)}_{\alpha<1}(y) \simeq 
\left \{\begin{array}{ccc} 
\frac{c_{\alpha}}{\alpha\sqrt{2\pi {\sigma^2}}} \int_{0}^{1} \frac{dw}{w^3} \;\exp \left[ -\frac{c_{\alpha}}{w} - \frac{y^2}{2 \sigma^2 w^2}\right],~~&\textrm{for}~~|y|~\textrm{small} \\ 
 & \\
 \frac{\Gamma(\alpha) }{\sqrt{2 \pi {\sigma^2}}} \left(\frac{\textrm{Sin}(\pi \alpha)}{\pi}\right)\;\frac{\sigma^{2\alpha}\exp\left({-\frac{y^2}{2\sigma^2}}\right)}{|y| ^{2\alpha}},~~&
\textrm{for}~~|y|~\textrm{large}.
\end{array}
            \right. \label{G_alpha(y)}
\eea
For $\alpha \rightarrow 0$, employing $\Gamma(\alpha)\; \textrm{Sin}(\pi \alpha) \rightarrow \pi$ in the above equation, one can easily show $\mathbb{G}^{(0)}_{0}(y) 
\underset{y \rightarrow \pm \infty}
{\simeq} e^{-\frac{y^2}{2\sigma^2}}/\sqrt{2\pi \sigma^2} \simeq p(y)$ (see Eq.~\eref{Gaussian-SF}). This behaviour is intuitively expected because for $(\alpha \rightarrow 0)$ the first stem remains incomplete up to time $t $ in almost all realizations which effectively makes $\mathcal{P}(\tau_{\max}|t) \simeq \delta (\tau_{\max} -t)$ for large $t$ which can also be easily proved \cite{godreche2015statistics}. 
Using this result in Eq.~\eref{P_{tails}(x,t)_r0} immediately implies $\mathbb{G}^{(0)}_{\alpha<1}(y) \simeq p(y)$ for $\alpha \to0$.
The theoretical expression of $P(x,t)$ in Eq.~\eref{P_{tails}(x,t)_m1} along with Eq.~\eref{G_alpha(y)} is verified in fig~\ref{pdfpl_r0_a<1}(b) and fig.~\ref{pdfpl_r0_a<1}(c) where the magenta solid lines describe the tail behaviour and the blue dashed lines describe the central part. We observe nice agreement between theory and simulation. Note that the central regime becomes narrower as $\alpha$ decreases  and the tail behaviour in Eq.~\eref{G_alpha(y)} describes the distribution $P(x,t)$ over almost the entire region of $y$. As mentioned earlier, most of the earlier works of finding $P(x,t)$ for $0<\alpha<1$ have considered velocity distribution of the form $p(v) = 1/2 ( \delta(v - v_0) + \delta(v + v_0) )$ \cite{klafter1994levy, zaburdaev2015levy} with constant magnitude for which the distribution $P(x,t)$ is supported over finite range $-v_0t\leq x \leq v_0t$ and has a minimum at the center (U-shape) with integrable singularities (called ``chubchiks'') at the edges of the interval \cite{klafter2011first,zaburdaev2015levy}. This is in sharp contrast with what we obtain for Gaussian velocity distribution $p(v)=e^{-v^2/2\sigma^2}/\sqrt{2 \pi \sigma^2}$.  We get $P(x,t)$ defined over $-\infty \leq x \leq \infty$ with a peak at the center $x=0$ and decaying as power law for $|x| \to \infty$ as shown in fig.~\ref{pdfpl_r0_a<1}.

\subsubsubsection{For $\alpha > 1$}
\label{r0_alpha>1_pdf}

\begin{figure*}[t!]
\centering{\includegraphics[width=1.0\hsize]{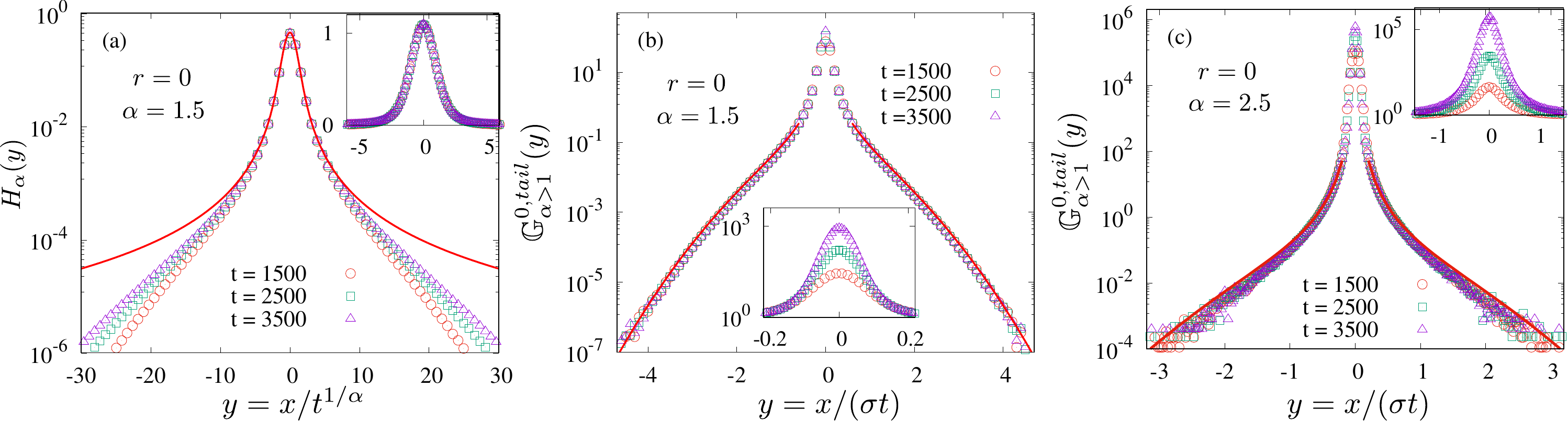}}
\caption{(Color online). In panel (a), PDF of scaled displacement $y \propto x t^{-1/\alpha}$, obtained from simulation (symbols), is compared with theoretical expression in Eq.~\eref{P(x,t)_levy_a>1} (red solid line) for $r =0$ and $\alpha = 1.5$. The excellent agreement (shown in the inset) verifies the scaling behaviour described by L\'evy stable function $H_{\alpha}(y)$ in the central part.  Discrete symbols in this plot are the simulation result and the solid (red) line is the numerical result of the L\'evy stable function $H_{\alpha}(y)$ defined in Eq.~\eref{P(x,t)_levy_a>1}.
In panels (b) and (c), we numerically verify the theoretical expressions for the tail behaviour of $P(x,t)$ given in Eq.~\eref{P_{tail}(x,t)_a>1_r0_7} (solid red lines) for $\alpha = 1.5$ and $\alpha = 2.5$ with $r =0$. Once again symbols in these plots are obtained from numerical simulation. In the inset of all the plots we have zoomed the distribution near the peak (in normal scale). } 
\label{pdf15_r0_cntr_tail} 
\end{figure*}

We first note that for $\alpha > 1$, the waiting time distribution $\rho(\tau)$ has finite mean $\langle \tau \rangle = \alpha/(\alpha-1)$. This implies that within a large time $t$ the  number of jump events  on an average  is $t/\langle \tau \rangle$.
Following a similar procedure as done for case II with $r = 0$ (see sec.~\ref{pdf_exp_r0}),  it is possible to write the following approximate equation for the distribution $P(x,t)$ for large $t$
\be
P(x,t) \simeq \frac{1}{2\pi} \int_{-\infty}^{\infty} dk\; \exp \left[ - i k x +  \frac{t}{\langle \tau \rangle}\ln \lambda(k)  \right].
\label{P(x,t)_r0_a>1} 
\ee
where $\lambda(k)$ is given in Eq.~\eref{lambda(k)_r0_a>1}. Note that small $k$ behaviour of $\lambda(k)$ is different for $\alpha<2$ and $\alpha >2$. So we need to perform the above integral separately for these cases. Executing this integral we find that for $\alpha>1$, the position distribution $P(x,t)$ satisfies the following scaling form  \cite{zaburdaev2015levy,zaburdaev2011perturbation,klafter2011first,denisov2012levy}
\bea
\nonumber
P(x,t) &\simeq \left( \frac{\langle \tau \rangle}{\mathcal{D}_{{\alpha}} \sigma^{{\alpha_0}} t}\right)^{\frac{1}{{\alpha_0}}} H_{{\alpha_0}} \left[ x \left( \frac{\langle \tau \rangle}{\mathcal{D}_{{\alpha_0}} \sigma^{{\alpha_0}} t}\right)^{\frac{1}{{\alpha_0}}} \right],~~\textrm{where}~~ \\ &
H_{{\alpha_0}}(y) = \frac{1}{2\pi} \int_{-\infty}^{\infty} dk \; e^{-i k y - \frac{|k|^{\alpha_0}}{2}},~~\textrm{and}~~\alpha_0=\min(\alpha,2),
\label{P(x,t)_levy_a>1} 
\eea
with $\mathcal{D}_\alpha$ is given in Eq.~\eref{lambda(k)_r0_a>1}.
This scaling from is valid for $\alpha>1$ except for $\alpha =2$ where we expect some $\ln(t)$ dependence in the distribution (see remarks later). Also this scaling function should describe the central part of the distribution $P(x,t)$ well {\it i.e.} for $|x|\lesssim \mathcal{O}(t^{1/{\min(\alpha,2)}})$. Note that for $\alpha <2$ the scaling function $H_\alpha(y)$ is L\'evy stable Law which has a power law tail $H_{\alpha}(y) \sim 1/|y|^{1+\alpha}$   \cite{montroll1984levy,bouchaud1990anomalous} which, as will see, smoothly connects to the tail of $P(x,t)$. On the other hand for $\alpha>2$ it is a Gaussian of mean zero and unit variance {\it i.e.} $H_2(y) =G(y)$ as in Eq.~\eref{Gaussian-SF}. The above scaling behaviour of $P(x,t)$ in the central part is verified numerically in fig.~\ref{pdf15_r0_cntr_tail}(a) for $\alpha=1.5$ where we observe excellent agreement. 

The behavior of $P(x,t)$ for L\'evy walks with constant speed has widely been studied for $1
< \alpha <2$ 
\cite{shlesinger1985comment, shlesinger1987levy, klafter1994levy, zaburdaev2015levy,klafter2011first,zaburdaev2011perturbation,denisov2012levy,dhar2019anomalous}. 
Cases with velocity distributions different from the ones with constant magnitude of speed have also been investigated in detail in Refs. \cite{zaburdaev2011perturbation,denisov2012levy}. It has been observed that the central part of the PDF $P(x,t)$ is universal across velocity distributions (with finite mean). 
However the behavior of the ballistic region at the tail depends strongly on the choice of the velocity distribution \cite{zaburdaev2011perturbation,denisov2012levy}.
As mentioned earlier,  most of the investigations are made in the limit $v_0 \gg \sigma$ where $v_0$ and $\sigma$ are the mean speed and standard deviation of velocity \cite{denisov2012levy}. In such cases, one observes  bumps  corresponding to those trajectories which have never changed velocity since the start. These bumps move ballistically with speed $v_0$ on both sides of the origin. Remember, in our case we consider Gaussian velocity distribution corresponding to the opposite limit $v_0 < \sigma$. Hence, we expect different behaviours at the tail which we explore in the next.\\

{\bf Tail behaviour:} Similar to $\alpha<1$ case, in this regime ($\alpha>1$) also we find  that the dominant contributions to the tail of $P(x,t)$ come from the largest jump duration $\tau_{\max}$ for large $t$. Numerical evidence of this fact is provided in fig.~\ref{pdf15_r0_cntr_tail}(b) and fig.~\ref{pdf15_r0_cntr_tail}(c) for $\alpha=1.5$ and $\alpha=2.5$ respectively. The distribution $\mathcal{P}_{\tau}(\tau_{\max}|t)$ of largest time interval $\tau_{\max}(t)$ within time $t$ for $\alpha > 1$ is given by \cite{godreche2015statistics} 
\be
\mathcal{P}_{\tau}(\tau_{\max}|t) \simeq \frac{t}{\langle \tau \rangle}\frac{\alpha}{\tau^{\alpha + 1}_{\textrm{max}}} 
\; \exp \left[ -\frac{t}{\langle \tau \rangle} \tau^{-\alpha}_{\textrm{max}} \right].
\label{P_{tail}(x,t)_a>1_r0_5}
\ee
Using this result in Eq.~\eref{P_{tails}(x,t)_r0} along with $\langle \tau \rangle = \alpha/(\alpha -1)$ and the explicit form of Gaussian velocity distribution we get following scaling form valid for $|x| \gtrsim \sigma t$
\bea
\nonumber
P(x,t) & \simeq  \frac{1}{\sigma t^{\alpha}} \mathbb{G}^{(0),tail}_{\alpha>1}\left( \frac{x}{\sigma t}\right) 
\quad \textrm{where} \\&
\mathbb{G}^{(0),tail}_{\alpha>1}(y) \simeq \frac{(\alpha -1)}{2\sqrt{2\pi}} \; \left(\frac{2}{y^2} \right)^{\frac{1+\alpha}{2}}\; \Gamma \left(\frac{1 + \alpha}{2},\frac{y^2}{2}\right).
\label{P_{tail}(x,t)_a>1_r0_7} 
\eea
This result is verified numerically for $\alpha=1.5$ and $\alpha=2.5$ in figs.~\ref{pdf15_r0_cntr_tail}(b) and (c) respectively, where red solid lines correspond to the above analytical expression and the symbols are obtained from numerical simulation.
It is interesting to observe that for $1<\alpha<2$ the form of the distribution $P(x,t)$ at the tails smoothly connects to the form of the distribution in the central part. This can be seen by comparing the $y \to \infty$ asymptotic of the central part in Eq.~\eref{P(x,t)_levy_a>1} with the $y \to 0$ asymptotic of the tail part in Eq.~\eref{P_{tail}(x,t)_a>1_r0_7}. To do so, we define 
$y_c = (\langle \tau \rangle/\mathcal{D}_{\alpha})^{1/\alpha} \times\;x/(\sigma t^{1/\alpha})$ to denote the scaling variable in the central part and $y_t = x/(\sigma t)$ for the tail part.
Note that these two variables are related via $y_c = y_t \times t^{1-1/\alpha} (\langle \tau\rangle/\mathcal{D}_{\alpha})^{1/\alpha}$.
Now, using the asymptotic form of L\'evy stable function \cite{montroll1984levy} 
\bea 
H_{\alpha}(y_c) &
\underset{y_c \rightarrow \pm \infty} 
{\sim} \frac{\Gamma(1+\alpha)}{\pi} \frac{\textrm{Sin}(\pi \alpha/2)}{|y_c|^{1+\alpha}},
\eea
in Eq.~\eref{P(x,t)_levy_a>1} we express it in terms of $y_t$ as 
\bea 
P(y_t) &  
\underset{y_t \rightarrow \pm \infty}
{\sim} 
\frac{(\alpha -1)}{2\sqrt{\pi}} \Gamma \left(\frac{1 + \alpha}{2}\right)2^{\alpha/2} \frac{t^{1-\alpha}}{|y_t|^{1+\alpha}}.
\label{P(y_t)_1a2}
\eea
On the other hand from Eq.~\eref{P_{tail}(x,t)_a>1_r0_7} we find
\bea 
P(y_t) & 
\underset{y_t \rightarrow 0}
{\simeq} 
\frac{(\alpha -1)}{2\sqrt{\pi}} \Gamma \left(\frac{1 + \alpha}{2}\right)2^{\alpha/2} \frac{t^{1-\alpha}}{|y_t|^{1+\alpha}},
\label{P_tail(y_t)}
\eea
which is exactly same as Eq.~\eref{P(y_t)_1a2}. Hence the central behaviour of $P(x,t)$ smoothly connects to the behaviour at the tails. Such matching does not happen for $\alpha>2$, possibly indicating the existence of an intermediate regime which seems difficult to find exactly. 

\subsubsubsection{For $\alpha = 1$}

\begin{figure}[t!]
\centering{\includegraphics[width=0.9\hsize]{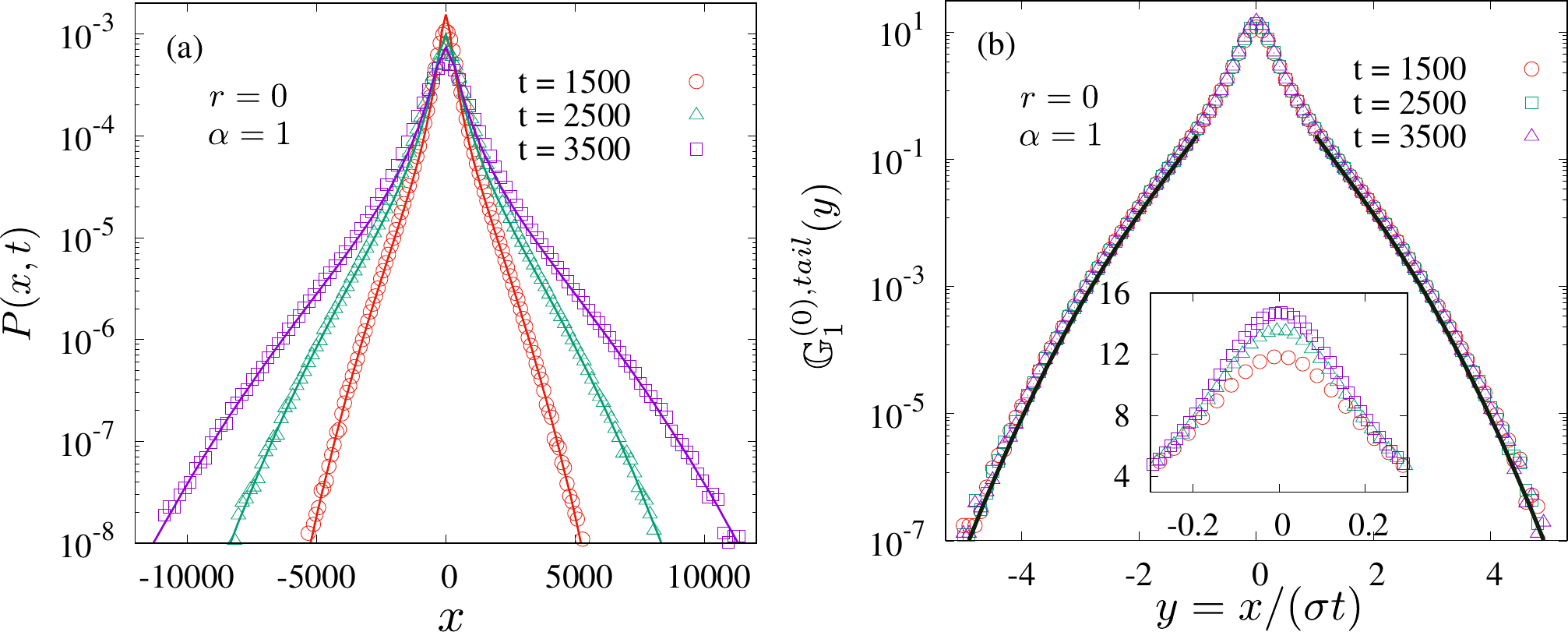}}
\caption{(Color online).{\bf Panel (a)}: Distributions of the {position} $x(t)$ (symbols) and the displacement $\Delta_{\max}(t)$ in the longest jump (solid lines) are plotted at different times for $\alpha = 1$ with $r=0$. The good agreement verifies the approximation $x(t) \simeq \Delta_{\max}(t)$ for typical trajectories. {\bf Panel (b):} The scaling behaviour distribution $P(x,t)$ at the tail described by  $\mathbb{G}_{1}^{0,tail}(y)$ in Eq.~\eref{P_{tails}(x,t)_a1}, is verified numerically.  Symbols are from numerical simulation and the solid black line represents analytical result. Inset shows zoomed version of the central part (in normal scale), exhibiting $x(t)$ does not scale ballistically there.} 
\label{pdfpl_r0_a=1}
\end{figure}

For $\alpha = 1$ the mean and all the higher order moments of the waiting time distribution diverges.
Like $\alpha < 1$, here also we numerically observe in fig.~\ref{pdfpl_r0_a=1}(a)  that the dominant contribution of the total position $x(t)$ within time $t$ is coming from the displacement associated with the largest jump time interval $\tau_{\max}(t)$ within time $t$ (solid lines).
The distribution of $\tau_{\max}(t)$ for $\alpha=1$ can be written as 
\be 
\mathcal{P}(\tau_{\max}=\tau|t) \simeq  \frac{ t (\ln (\tau) + 1)}{\tau^2 (\ln \tau)^2} \textrm{Exp}\left[-\frac{t}{\tau  \ln (\tau )}\right]
\label{Ptau_max-a-1}
\ee
for large $t$ (see \ref{P_tau_max_a=1} for details).

Although it seems difficult to find a scaling behavior of $\mathcal{P}(\tau_{\max}|t)$ as we got for $\alpha<1$. However, it is possible to show from Eq.~\eref{Ptau_max-a-1} that the distribution $P(x,t)$ at the tails (large $|x|$) poses a scaling form:
\bea 
\nonumber
P(x,t) \simeq & \frac{1}{\sigma t \ln(t)}\;\mathbb{G}^{(0),tail}_{1}\left( \frac{x}{\sigma t } \right) \quad \textrm{where} \\& \mathbb{G}^{(0),tail}_{1}(y) = \frac{1}{\sqrt{2 \pi}} \int_{0}^{1} \frac{dw}{w^3} \;\textrm{Exp}\left[- \frac{y^2}{2 w^2} \right]
= \frac{e^{-y^2/2}}{\sqrt{2 \pi}\; y^2}
\label{P_{tails}(x,t)_a1}
\eea
for  large $t$ at large $y$. This scaling result at the tails is plotted in fig.~\ref{pdfpl_r0_a=1}(b) where we once again observe  good agreement.

\subsubsubsection{For $\alpha = 2$}

\begin{figure}[t!]
\centering{\includegraphics[width=0.45\hsize]{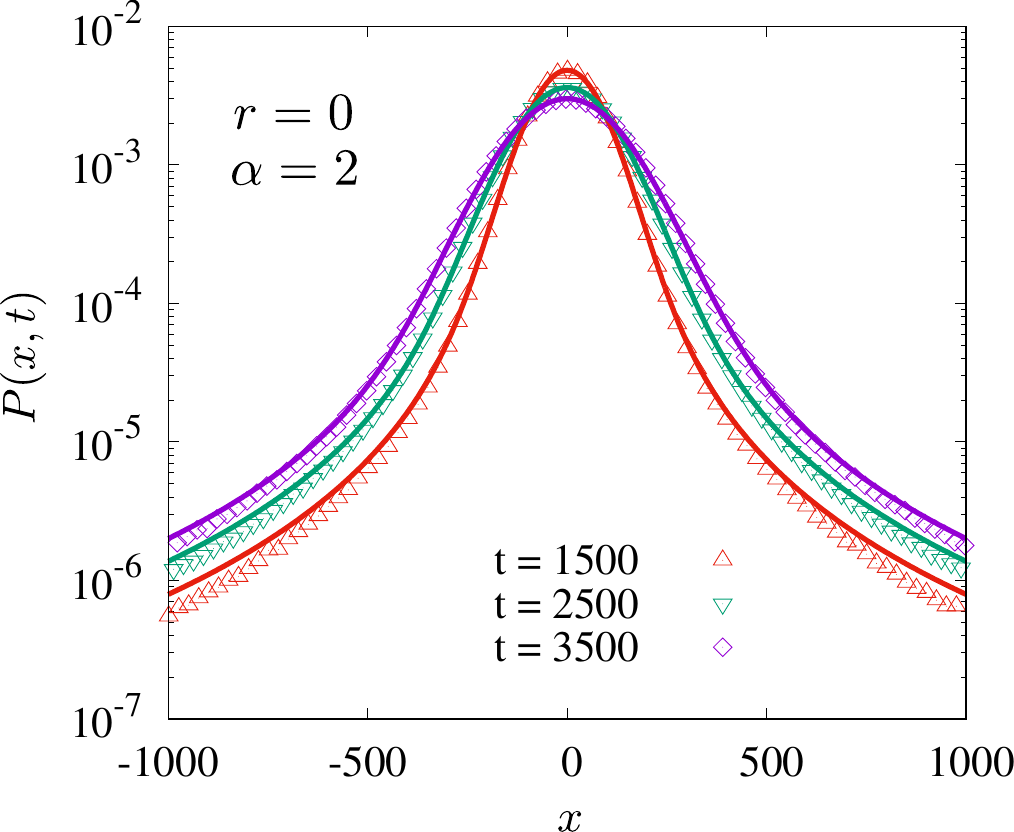}}
\caption{(Color online). PDF of the position $x(t)$ are plotted at different times for $r = 0$ with $\alpha = 2$. Symbols correspond to simulation results, whereas the solid lines are obtained performing the inverse Fourier transform in Eq.~\eref{P(x,t)_r0_a>1} numerically with $\alpha=2$. } 
\label{pdfpl_r0_a=2}
\end{figure}

For $\alpha = 2$ one can compute the distribution $P(x,t)$ from Eq.~\eref{P(x,t)_r0_a>1}  where $\lambda(k)$ is given in Eq.~\eref{lambda(k)_r0_a>1}. From this equation we expect a $\ln(t)$ dependence in the variance $\sigma^2_x(t)$ as also observed in Eq.~\eref{msd_pl_r0}. In this case it turns out difficult to find any scaling form of the distribution evidently. However, one can compute the distribution by evaluating the integral in Eq.~\eref{P(x,t)_r0_a>1} numerically which is plotted in fig.~\ref{pdfpl_r0_a=2} where we observe excellent agreement with the distribution  obtained from simulation (symbols).

\subsection{For $r = \pm 1$}
We now study the distribution $P(x,t)$ of the position for $r = \pm 1$. We discuss the three cases of jump time distribution separately.

\subsubsection{Case I: $\rho(\tau) = \delta(\tau - a)$\\\\}

First, in case of $\rho(\tau) = \delta (\tau -a)$ we can recall that the number of complete steps up to time $t$ is $\lfloor t/a \rfloor$.
It allows us to express $x(t)$ for an arbitrary $r$ as
\be
x(t) = \sum_{j=0}^{\lfloor t/a \rfloor -1} \tau_j v_j + \underbrace{ \left( t - a\left\lfloor \frac{t}{a} \right\rfloor \right) }_{\tau^*_{\lfloor t/a \rfloor}} v_{\lfloor t/a \rfloor}.
\label{xt_cs1_rpm1}
\ee
Note that the second term on the right hand side of Eq.~\eref{xt_cs1_rpm1} is denoting the contribution of the last incomplete step.
Using $v_j = \sum_{l=0}^{j} ( \pm 1)^{j-l}\;\eta_l$ for $r = \mp 1$ from Eq.~\eref{modl_2} we can further simplify and express $x(t)$ in Eq.~\eref{xt_cs1_rpm1} as
\bea
\nonumber
x(t) =  \sum_{j=0}^{\lfloor t/a \rfloor} \eta_{i} \chi_j(\mp 1,a,t) 
\qquad \textrm{with} \\
\chi_j(\mp 1,a,t) = a \left( \frac{1 - (\pm 1)^{\lfloor t/a \rfloor -j}}{1 - (\pm 1)} \right) + \left( t - a\left\lfloor \frac{t}{a} \right\rfloor \right) \;(\pm 1)^{\lfloor t/a \rfloor -j}.
\label{x(t)_cs3_1}
\eea
This $x(t)$ Eq.~\eref{x(t)_cs3_1} represents a weighted sum of i.i.d. Gaussian random variables. 
Hence, by employing the central limit theorem we find the distribution of $x(t)$ is also a Gaussian distribution with the variance $\propto t$. A detailed discussion and explicit expression of the variance are provided in sec.~\ref{var_cs1}.

\subsubsection{Case II: $\rho(\tau) = \beta e^{-\beta \tau}$\\\\}
\label{pdf_cs2_rmp1}
\begin{figure*}[t]
\centering{\includegraphics[width=0.9\hsize]{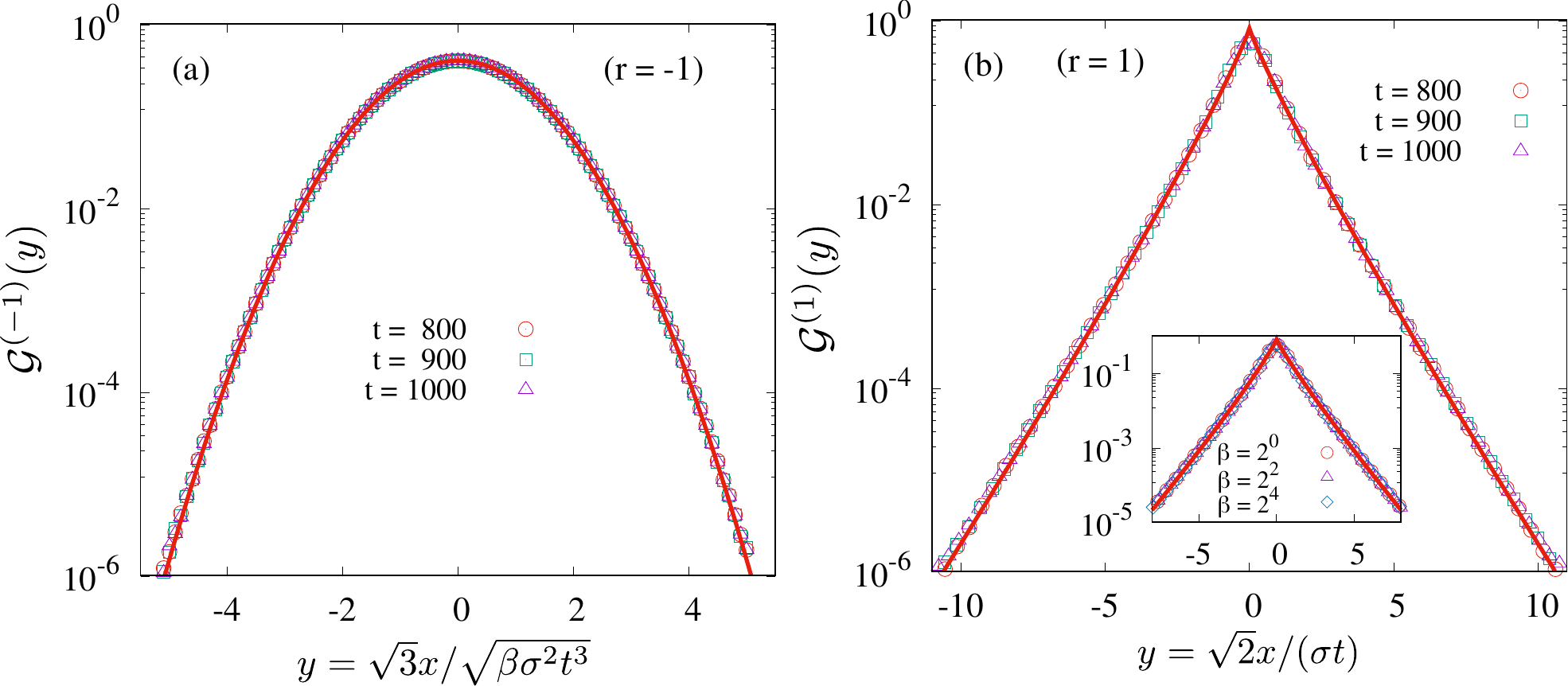}}
\caption{(Color online).  {\bf Panel (a):} PDF of the scaled position $y = \sqrt{3} x/\sqrt{\beta \sigma^2 t^3}$ are plotted at different times for $r = -1$ and $\beta=1$ in case II and compared with theoretical expression in Eq.~\eref{pdf-case-II-r_-1}. {\bf Panel (b):} PDF of the scaled position $y = \sqrt{2} x/{(\sigma t)}$  are plotted at different times for $r = -1$ and $\beta=1$ and compared with approximate theoretical result in Eq.~\eref{st_exp_r1_1} with $\lambda=1.109$. In both plots symbols correspond to simulation results and solid red lined correspond to the respective theoretical results.  In the inset of (b) numerically verify the $\beta$ independence of the distribution of $x(t)$ at $t=900$. }
\label{pdf_exp_rmp1} 
\end{figure*}

First we discuss the $r = -1$ case. We have observed earlier that in this case, for large time $t$, the dynamics of the particle can be effectively described by a random acceleration process with noise strength $D = \sigma^2 \beta$. 
Hence we expect the distribution $P(x,t)$ is a Gaussian with zero mean and  variance $D t^3/3 \simeq \beta \sigma^2 t^3/3$ for $t \gg \beta^{-1}$ {\it i.e.}
\bea
\hspace{-1.5cm}
P(x,t) \simeq 	\frac{\sqrt{3}}{\sqrt{\beta \sigma^2 t^3}}~\mathcal{G}^{(-1)}\left( \frac{\sqrt{3} x}{\sqrt{\beta \sigma^2 t^3}}\right),~~~\textrm{with}~~\mathcal{G}^{(-1)}(y)=\frac{1}{\sqrt{2\pi}} \exp \left( -\frac{y^2}{2} \right).
\label{pdf-case-II-r_-1}
\eea
 We verify this result using numerical simulation in fig.~\ref{pdf_exp_rmp1}(a).

Computing the distribution $P(x,t)$ for $r=1$ turns out to be difficult. 
To compute the distribution $P(x,t)$ one can in principle perform the inverse Fourier-Laplace transform in Eq.~\eref{tildeP_k_s}. Since in this case the jump time distribution is exponentially distributed,  all the jump durations in a trajectory of length $t$ are typically of same order {\it i.e.} $\mathcal{O}(\beta^{-1})$. Consequently, all jump durations contribute to same order in $x(t)$ and they are strongly correlated which makes it difficult compute $P(x,t)$. From the microscopic dynamics in Eqs.~\eref{modl_1} and \eref{modl_2} with $r=1$, it is easy to see that the motion of the particle should have a ballistic scaling {\it i.e.} $x \sim t$. In fig. \ref{pdf_exp_rmp1}(b) we plot the distribution of the scaled variable $y=\sqrt{2} x/(\sigma t)$ obtained in numerical simulation at times $t=800,~900$ and $t=1000$ and observe excellent data collapse. Since computing this scaling distribution exactly seems difficult we, instead, perform an approximate calculation below which proposes a stretched exponential form for the distribution. 

As has been observed earlier, the second moment $\langle x_m^2\rangle$ of the position after $m^{\textrm{th}}$ jump
grows quadratically as 
\be
\langle x_m^2 \rangle 
\simeq
\frac{\sigma^2 m^2}{2 \beta^2},
\label{x_m^2_cs2_rpm1}
\ee
for $r=1$. Similarly, after a lengthy and tedious computation one can show that $\langle x_m^4\rangle \simeq 7 (m \sigma/\beta)^4/4$ in the limit of large $m$ (see Eq.~\eref{x_m^4_app_rpm1} in \ref{4th_mnt_rpm1_app}). Since for large $t$, typical number of jump events is $m \sim \beta t$, we get 
 $\langle x^2(t)\rangle \simeq \sigma^2 t^2/2$ and $\langle x^4(t)\rangle \simeq 7 (\sigma t)^4/4 $.
In terms of the scaled variable $y = \sqrt{2} x/(\sigma t)$ we get $\langle y^2\rangle \simeq 1$ and $\langle y^4\rangle \simeq 7$. Similarly, one can compute higher order moments but the computation quickly becomes too involved.
However from the first two moments and the plots in fig.~\ref{pdf_exp_rmp1}(b) we make a guess for $P(x,t)$. We assume a stretched-exponential of the form  
\bea
\nonumber
P(x,t) \simeq	\frac{\sqrt{2} }{\sigma t}~\mathcal{G}^{(1)}\left( \frac{\sqrt{2} x}{\sigma t}\right),~~~\textrm{with}~~
\\ \mathcal{G}^{(1)}(y) 
= \frac{1}{2 \;\Gamma(1+\lambda)} \left( \frac{\Gamma(3 \lambda)}{\Gamma(\lambda)}\right)^{1/2}\;
\exp\left[- \left( \frac{\Gamma(3 \lambda)}{\Gamma(\lambda)} y^2 \right)^{1/2\lambda}  \right].
\label{st_exp_r1_1}
\eea
For an arbitrary $\lambda\in (0:\infty)$ this distribution is normalised and has $\langle y^{2} \rangle = 1$ and $\langle y^{4} \rangle 
= \frac{\Gamma(\lambda) \;\Gamma(5\lambda)}{(\Gamma(3\lambda))^2}$. 
Note that for $\lambda = 1/2$, $\mathcal{G}^{(1)}(y)$ is Gaussian distribution with $\langle y^2\rangle = 1$ and $\langle y^4\rangle = 3$.
For $\lambda = 1$, the distribution in Eq.~\eref{st_exp_r1_1} is a double-sided exponential distribution with $\langle y^2\rangle = 1$ and $\langle y^4\rangle = 6$. This means in our case we should expect $\lambda>1$. 
To find the value of $\lambda$ we numerically solve $\langle y^{4} \rangle = \frac{\Gamma(\lambda) \;\Gamma(5\lambda)}{(\Gamma(3\lambda))^2}\simeq 7$ and get $\lambda \simeq 1.109$.
Using this value for $\lambda$ in Eq.~\eref{st_exp_r1_1} we find that $\mathcal{G}^{(1)}(y)$ matches remarkably well with the numerical simulation result as displayed in fig.~\ref{pdf_exp_rmp1}(b). In the inset of this figure we show that, not only the variance (see fig.~\ref{variance_cs2}(c)) but the full scaling distribution is also independent of $\beta$.

\subsubsection{Case III: $\rho(\tau) \sim \tau^{-(1+\alpha)}$\\\\}

Like $r = 0$ case discussed in sec.~\ref{pdf-x-r-0-case-III}, here also we discuss different regimes of $\alpha$ separately.

\subsubsubsection{For $0 < \alpha < 1$}
\label{dist_x_rpm1-0<a<1}

\begin{figure*}[t!]
\includegraphics[width=1.0\hsize]{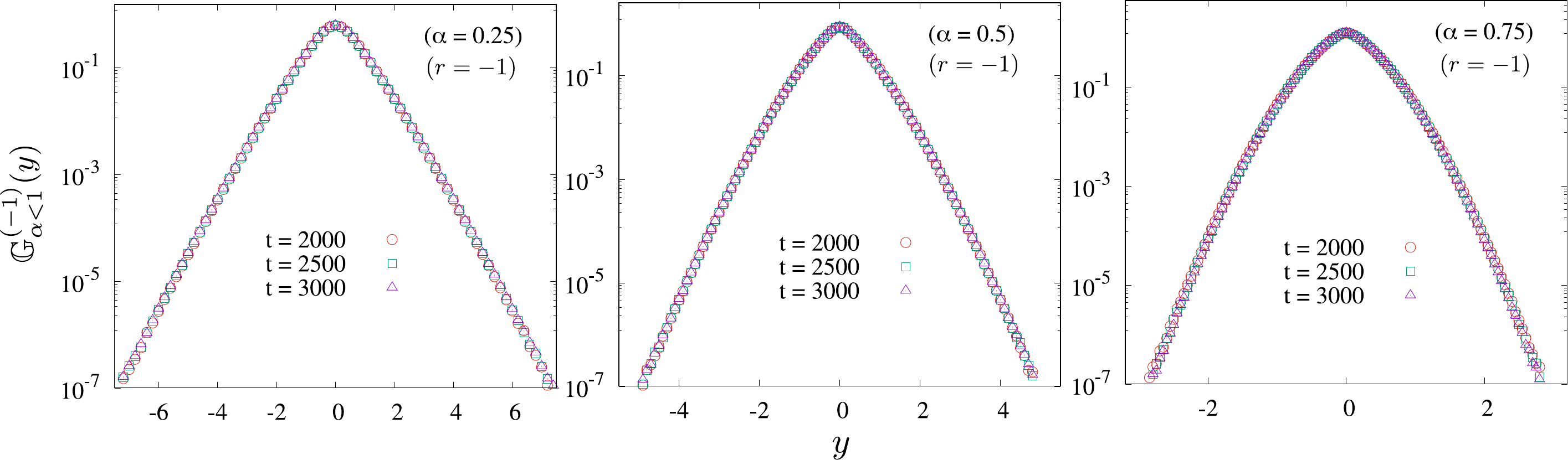}
\includegraphics[width=1.0\hsize]{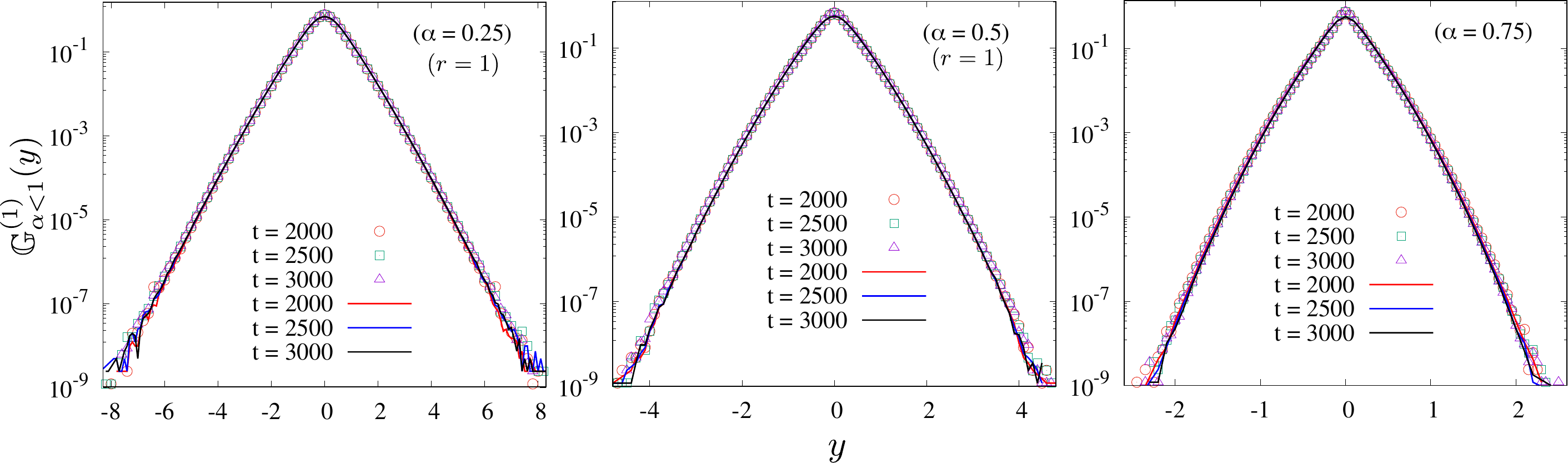}
\caption{\label{pdf_pl_rpm1_a<1} (Color online). Probability distribution function of the scaled position $y = x(t)/t^{1+\alpha/2}$ are plotted at different times for $r = -1$ (first row) and $r = 1$ (second row) with different values of $\alpha$.
Solid lines in each plot in the second row (for $r = 1$) represent the (scaled) distribution of displacement $\Delta_{\max}(t)$ made in the longest jump duration $\tau_{\max}$. These distributions are obtained numerically  and plotted in terms of the scaling variable $y=\Delta_{\max}/t^{1+\alpha/2}$. We observe excellent agreement between the (scaled) distributions of $x(t)$ and $\Delta_{\max}(t)$  for all values of $\alpha$. } 
\end{figure*}

Let us first consider the $r=-1$ case. Computing the distribution $P(x,t)$ for power law `jump' time distribution with $r \neq 0$ is even harder than previous casess because the velocity of the particle at different times are strongly correlated in addition to the fact that all the moments of the jump time distribution diverge. 
Instead of computing an analytical expression for the distribution, we focus on the scaling behaviour of $P(x,t)$ if there exists any. Since $r \neq 0$, we expect the velocity to appear as a function of time $t$ is a smoother function (upto first derivative)  over a long time duration than the velocity in $r = 0$ case. Hence, we write the position as $x(t) = \int_{0}^t dw \;(t -w) \frac{dv}{dw}$ (recall, $x(0)=0$) where $v(w)$ is the velocity at time $t$.  It is easy to see from Eq.~\eref{sc-P(v,t)-0<a<1} that for large $t$, 
$v(w) \sim w^{\alpha/2}$, using which in the above equation we find that typically $x(t) \sim t^{1+\alpha/2}$. This suggests the following scaling form for the distribution 
\be 
P(x,t) \simeq \frac{1}{t^{1+\alpha/2}} \; \mathbb{G}_{\alpha<1}^{(1)} \left( \frac{x}{t^{1+\alpha/2}} \right).
\label{scl_pdf_a<1_rpm1}
\ee
where $\mathbb{G}_{\alpha<1}^{(1)}(y)$ is the scaling distribution of  scaled variable $y =x/t^{1+\alpha/2}$. In fig.~\ref{pdf_pl_rpm1_a<1} (top row) we numerically verify this scaling form where we plot the distribution of the scaled variable $y$ obtained in simulation at three times $t=2000,~2500$ and $t=3000$ and for three values of $\alpha$. In all cases we observe excellent data collapse.

The above argument should also work for $r=1$ and hence the scaling form in Eq.~\eref{scl_pdf_a<1_rpm1} should also hold as can be seen in the bottom row of fig.~\ref{pdf_pl_rpm1_a<1} where we plot the scaling distribution obtained in numerical simulation (symbols). In this case, we however note that for large $t$ the dominant contribution to the position comes from the jumps with longest duration $\tau_{\max}$. This fact is verified in fig.~\ref{pdf_pl_rpm1_a<1} (bottom row) where solid lines correspond to the distribution of $\Delta_{\max}(t)=v_l \tau_{\max}$ with $v_l$ being the velocity in the longest jump duration. We observe that for large $t$ the distributions of $\Delta_{\max}(t)$ and $x(t)$ match  very well. 
Using the fact $x(t) \simeq \Delta_{\max}(t)$ it is possible to argue that $x(t) \sim  t^{1+\alpha/2}$ in an alternative way. This idea has been used in Eq.~\eref{tilde_sig_x_r1_0a1} where we show that 
$\widetilde{\sigma_x^2}(s) \simeq  s^{-(3+\alpha)}$  yielding $\sigma_x^2(t) = \langle x^2(t) \rangle \sim t^{2+\alpha}$. Following a similar calculation, one can show that 
$\langle x^{2n}(t)\rangle \sim t^{2n+\alpha}$ for $n=1,2,3,...$ (see \aref{HOM-r-1-case-III}). This behaviour is consistent with the scaling form in Eq.~\eref{scl_pdf_a<1_rpm1}. We end this section by making a remark that even though the scaling behaviour of $P(x,t)$ is same for both $r=1$ and $r=-1$ in this range of $\alpha$, the argument based on dominant contribution from longest jump duration does not work for $r=-1$ at least over the range of position $x$ accessible in numerical simulation.

\begin{figure*}[t]
\centering{\includegraphics[width=1.0\hsize]{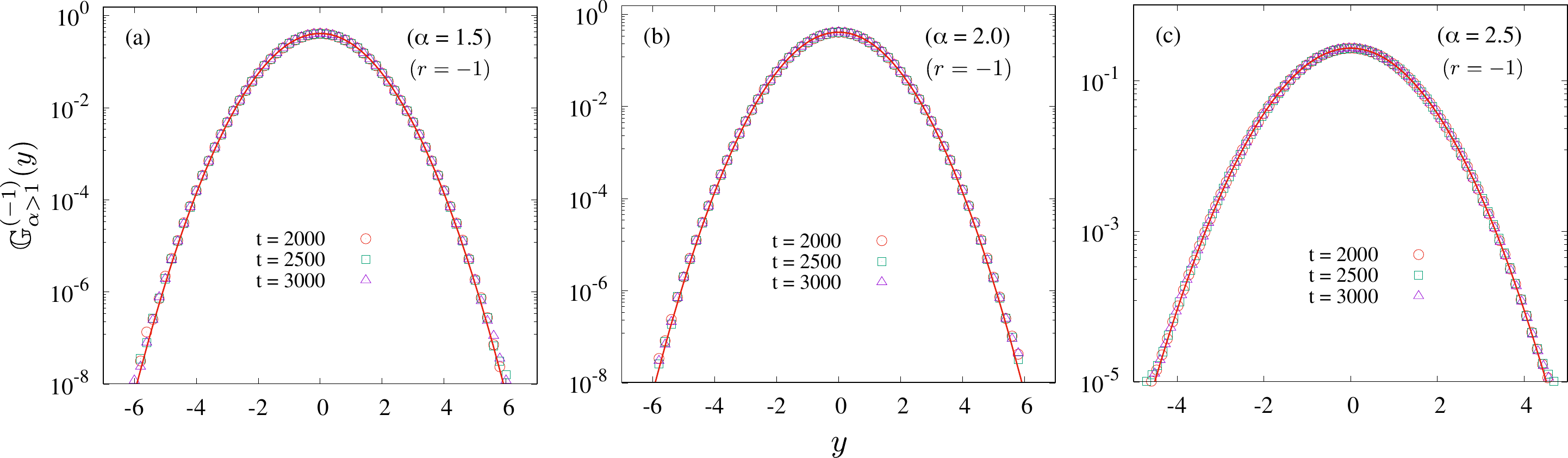}}
\caption{\label{pdf_pl_rn1_agt1} (Color online). Probability distribution functions of the scaled position $y = x(t)\sqrt{3\alpha/((\alpha-1)\sigma^2 t^3)}$ are plotted for $r = -1$ and different values $\alpha$. Symbols are obtained from simulation results which show good agreement with the analytical distribution $\mathbb{G}^{(-1)}_{\alpha>1}(y)$ (red solid lines) discussed in sec.~\ref{dist_x_rpm1-agt1}.} 
\end{figure*}

\subsubsubsection{For $1<\alpha <2$}
\label{dist_x_rpm1-agt1}
Recall that for $\alpha$ in this range, the mean jump duration is finite $\langle \tau \rangle = \frac{\alpha}{\alpha -1}$. As in the exponential case studied previously in  sec.~\ref{pdf_cs2_rmp1}, in this case also for $r=-1$ we expect the distribution of the position $x(t)$ after a long time $t$ is given by  Gaussian with the variance $D t^3/3 \simeq (\alpha -1) \sigma^2 t^3/(3\alpha)$ as in a random acceleration process driven by a zero mean white noise having strength $D = \sigma^2 (\alpha -1)/\alpha$.  Although in this section we are focusing in the range $1<\alpha<2$, this Gaussian scaling form is also valid for $\alpha>2$. Hence it is more appropriate to denote the distribution of the scaling variable $y = x(t)\sqrt{3\alpha/((\alpha-1)\sigma^2 t^3)}$ by $\mathbb{G}^{(-1)}_{\alpha>1}(y)$ which  is essentially $\mathbb{G}^{(-1)}_{\alpha>1}(y)=G(y)$  where $G(y)$ is given in Eq.~\eref{Gaussian-SF}. This distribution is verified numerically for $\alpha= 1.5$, $2.0$ and $2.5$ in fig.~\ref{pdf_pl_rn1_agt1}, where the red solid lines represent theoretical result  and the symbols are from numerical simulation.

\begin{figure*}[t]
\centering{
\includegraphics[width=1.\hsize]{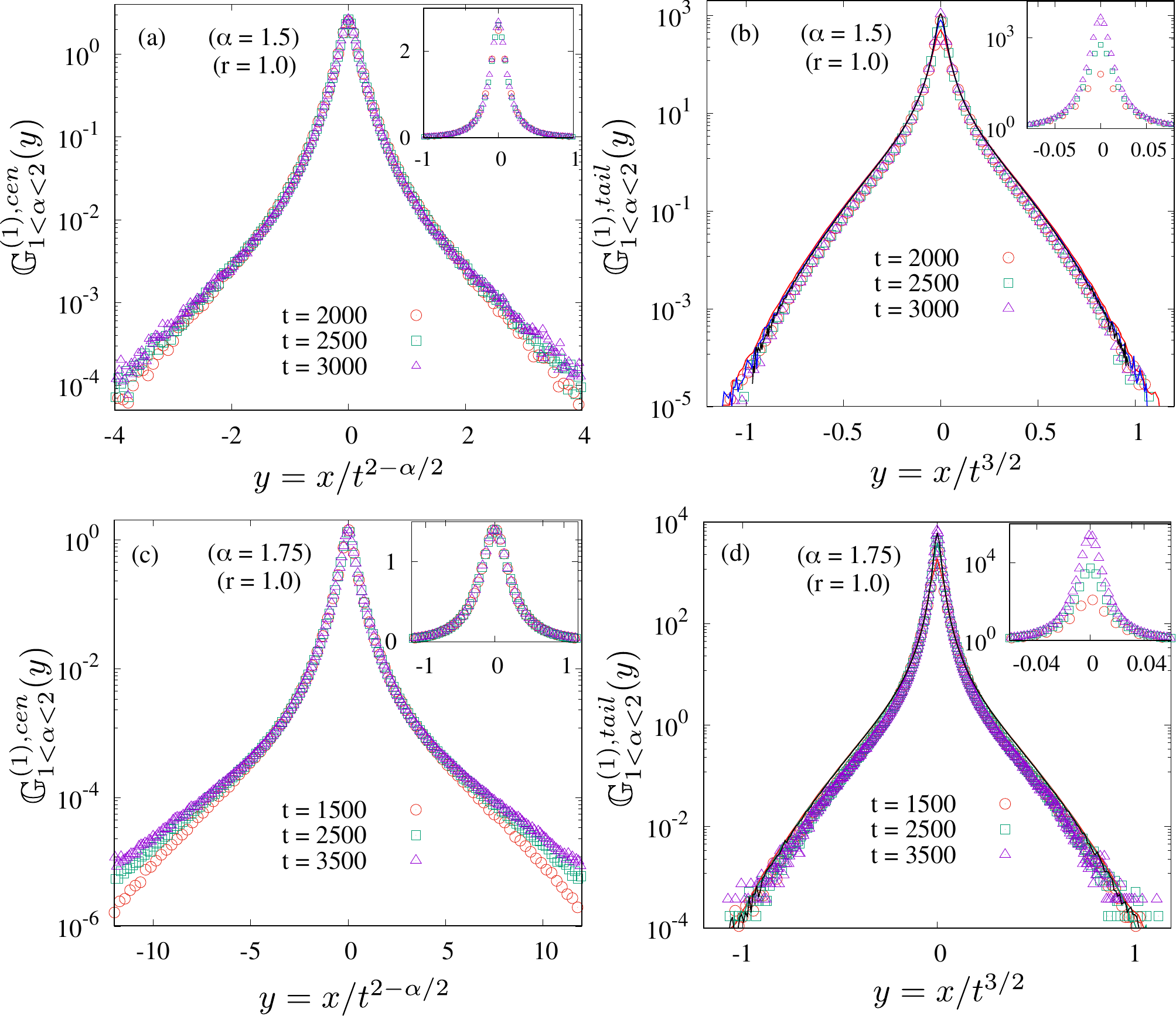}}
\caption{(Color online). Distributions $\mathbb{G}_{1<\alpha<2}^{1,cen}\left( y\right)$ and $\mathbb{G}_{1<\alpha<2}^{1,tail}\left( y\right)$ of the scaled positions $y = x/t^{2-\alpha/2}$ and $y = x/t^{3/2}$ are  plotted, respectively, in 
the left (a,c) and right (b,d) columns for $r=1$  [see Eqs.~(\ref{sc-form-central-r_1-1<a<2}) and (\ref{sc-form-tail-r_1-1<a<2})]. The top row corresponds plots for $\alpha=1.5$ and the bottom row corresponds plots for $\alpha=1.75$
. Solid lines in figures (b) and (d)  plot the numerical data for the distribution of scaled variable $\Delta_{\max}/t^{3/2}$ associated to the displacement  $\Delta_{\max}(t)$ made in the longest jump duration for $\alpha=1.5$ and $\alpha=1.75$ respectively. In the inset of figures (b) and (d) we highlight that $t^{3/2}$ scaling of the position $x(t)$ does not work in the central part as can be seen from the data collapse  failure whereas scaling with $t^{2-\alpha/2}$ works very well in the central part. Note that the distributions in the insets of all the plotted in normal scale.}
\label{pdf_pl_r1_agt1} 
\end{figure*}

We now discuss the behaviour of $P(x,t)$ for $r = 1$ with $1<\alpha <2$. As shown in Eq.~\eref{<x^2n>-vs-t-r_1} of \ref{HOM-r-1-case-III}, in this case the variance grows as $\sigma_x^2(t) = \langle x^2(t) \rangle_c \sim t^{4-\alpha}$. This may naively suggest a scaling form once $x(t)$ is scaled by $\sigma_x(t)$ {\it i.e.} 
\bea
P(x,t) \approx \frac{1}{t^{2-\alpha/2}} \mathbb{G}_{1<\alpha<2}^{(1),cen}\left( \frac{|x|}{t^{2-\alpha/2}}\right).
\label{sc-form-central-r_1-1<a<2}
\eea
In figs.~\ref{pdf_pl_r1_agt1}(a) and \ref{pdf_pl_r1_agt1}(c) we try to numerically verify this for $\alpha=1.5$ and $1.75$ where we plot data of $P(x,t)$ obtained from simulation in this scaling form. We observe excellent data collapse but only over the central part ($|x| \lesssim \sigma_x(t)$) of the distribution implying that such a scaling form is not valid outside this region {\it i.e.} at the tails of the distribution. This can also be seen from the large $t$ growth of the higher order moments: $\langle x^{2n}(t) \rangle \sim t^{3n+1-\alpha}$ (see \ref{HOM-r-1-case-III}) which, one will not be able to obtain from the scaling form in Eq.~\eref{sc-form-central-r_1-1<a<2}. This suggests a different scaling form at the tails of the distribution $P(x,t)$ as in the $r=0$ case discussed in sec.~\ref{r0_alpha>1_pdf}.

The late time behaviour of the higher moments suggests us to guess the following scaling form for the distribution at the tail ($|x| \gg \sigma_x(t)$ possibly $|x| \gtrsim t^{3/2}$)
\bea
P(x,t) \approx \frac{1}{t^{\alpha+1/2}} \mathbb{G}_{1<\alpha<2}^{(1),tail}\left( \frac{|x|}{t^{3/2}}\right),
\label{sc-form-tail-r_1-1<a<2}
\eea
which is verified numerically in figs.~\ref{pdf_pl_r1_agt1}(b) and  \ref{pdf_pl_r1_agt1}(d)  for $\alpha=1.5$ and $1.75$, respectively. The symbols are from simulation. Note in these figures that the data collapse is excellent at the tails but do not collapse at the central part as shown in the insets. 
The fact that the distribution at the tail has a different scaling form stems from the fact that here also for large $t$, the dominant contribution at the tails comes the displacememt $\Delta_{\max}(t)$ in the longest jump duration $\tau_{\max}$ which indeed leads to  $\langle x^{2n}(t) \rangle \sim t^{3n+1-\alpha}$ (see \ref{HOM-r-1-case-III}).

Like $\alpha<1$ case (discussed in the previous section), in this range of $\alpha$ also, the fact of dominant contribution from $\Delta_{\max}$ is also true as verified in figs.~\ref{pdf_pl_r1_agt1}(b) and \ref{pdf_pl_r1_agt1}(d) where we observe excellent match between the distributions of $x(t)$ (symbols) and $\Delta_{\max}(t)$ (solid lines) both obtained numerically. Note these distributions are plotted in terms of the scaling variables $y=x/t^{3/2}$ and $y=\Delta_{\max}/t^{3/2}$. 

We close this section by mentioning that unlike the $r=0$ case, for $r=\pm1$ it seems difficult to find analytical expressions of the scaling functions mainly because the velocity at different steps are highly correlated which makes it harder to compute the joint probability of the velocity at $l^{\textrm{th}}$ jump and this jump being the longest one.

\begin{figure*}[t]
\centering{\includegraphics[width=0.55\hsize]{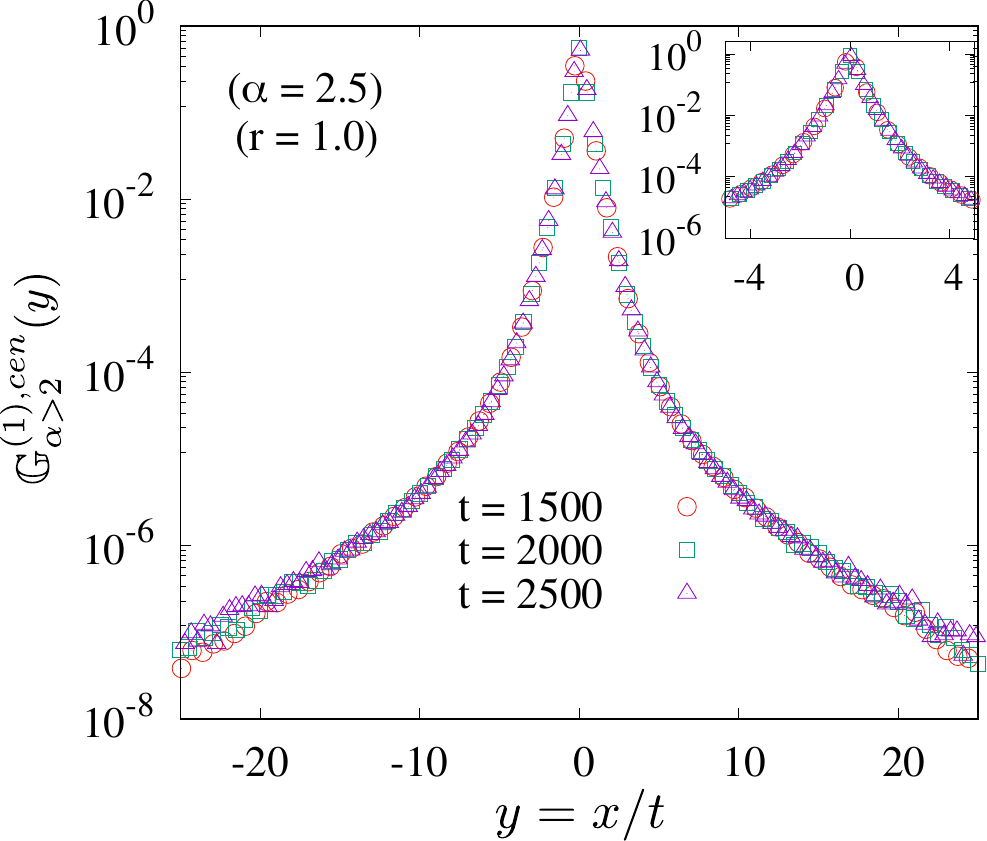}}
\caption{\label{pdf_pl_r1_agt2} (Color online). Distribution  $\mathbb{G}_{\alpha>2}^{(1),cen}\left( y\right)$ of the scaled variable position $y = x/t$ is plotted for $\alpha=2.5$ and $r=1$ at different times. The excellent data collapse validates the scaling form in Eq.~\eref{dist-sc-alpha>2-r-1}. Inset shows a zoomed version of the distribution near the peak. The inset in figure (b) shows the distribution in normal scale.}
\end{figure*}

\subsubsubsection{For $\alpha >2$}
\label{dist_x_rpm1-agt2}
For this range of $\alpha$, since both mean and variance of the jump time distribution are finite, all jump steps contribute more or less equally to the total position. Unlike the $0<\alpha<1$ case, the single jump of longest duration does not dominate in this case. Hence for large $t$ the  distribution can not be described by the position made in the longest jump. However, such events in fact should provide dominant contribution at the tail for large $t$. The central part of the distribution would be described by a scaling distribution when $x(t)$ is scaled by the standard  deviation $\sigma_x(t)$ which in this case is given by $\sigma_x^2(t) \simeq \frac{\sigma^2 t^2}{2 \alpha (\alpha - 2)}$ (see Eq.~\eref{msd_pl_r1}). Hence the scaling form for the distribution $P(x,t)$ seems to be
\bea
P(x,t) \approx \frac{1}{t} \mathbb{G}_{\alpha>2}^{(1),cen}\left( \frac{|x|}{t}\right),
\label{dist-sc-alpha>2-r-1}
\eea
which is verified numerically in fig.~\ref{pdf_pl_r1_agt2} for $\alpha=2.5$. 

\begin{figure*}[t!]
\centering{\includegraphics[width=1.\hsize]{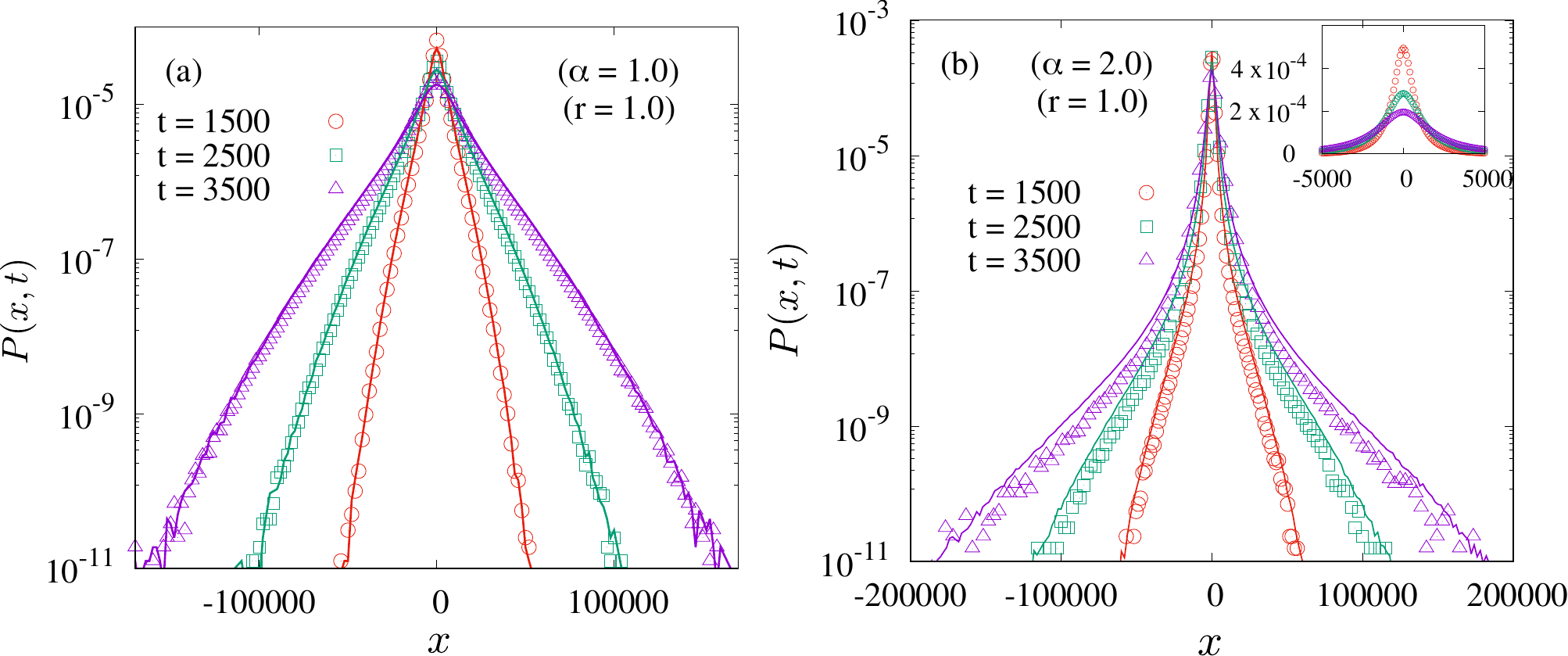}}
\caption{\label{pdf_pl_r1_1020} (Color online). Distributions of the position $x(t)$ (symbols) and the displacement $\Delta_{\max}(t)$ made in the longest `jump' duration $\tau_{\max}$ are plotted  for $\alpha = 1.0$ (a) and $2.0$ (b), respectively, at different times. Both distributions are obtained numerically and we observe they match quite well.}
\end{figure*}

\subsubsubsection{For $\alpha =1$ and $\alpha=2$}
\label{dist_x_rpm1-a-1_a-2}
Finally we make some remarks about $\alpha=1$ and $\alpha=2$ cases. These cases represent marginal cases in which one expects to have some $\ln(t)$ dependences in the moments (see \ref{HOM-r-1-case-III}) as well as in the distributions. It turns out that for these two cases also the position $x(t)$ can be well approximated the displacement $\Delta_{\max}(t)$ in the longest jump duration $\tau_{\max}(t)$ within a large time interval $t$. This fact is numerically verified in fig.~\ref{pdf_pl_r1_1020} where we compare the distributions of $x(t)$ and $\Delta_{\max}(t)$ for both $\alpha=1$ and $\alpha=2$. The good agreement indeed justifies the approximation  $x(t) \simeq \Delta_{\max}(t)$ for large~$t$.





\section{Summary and conclusion}
\label{discussion}

In this paper we have studied a simple extension of the LW walk model by introducing correlation among the velocities of the walker at different steps. This correlation has been introduced by relating the velocities of two successive `jump' steps such that the velocity at any step retains some part of the velocity in the previous step plus a random noise coming from the surrounding medium as written in Eq.~\eref{modl_1}. The parameter $r$ controls the degree of correlation which becomes largest in the limit $r \to \pm 1$ and zero for $r=0$. Also the process discussed in this paper can describe the dynamics of a granular particle being driven and dissipating energies from the collision with a heavy wall at thermal equilibrium. The parameter $r$ in this context serves as a coefficient of restitution. 

The particle in a `jump' step moves with a fixed velocity for a random interval of time drawn from a distribution $\rho(\tau)$. For different choices of $\rho(\tau)$ and values of $r$ (with more focus on $r=\pm 1$ and $0$) we have studied the statistical properties of the velocity $v(t)$ and position $x(t)$ of the particle at time $t$. We have shown that for different values of $r$ with thee choices of $\rho(\tau)$ one finds interesting ballistic, super-diffusive, diffusive and sub-diffusive scaling of these quantities. In particular, we find that in case of power law `jump' time distribution (case III) with $\alpha\leq 2$ the distribution of the position $x(t)$ is dominated by the the displacement $\Delta_{\max}(t)$ made in the longest `jump' interval at the leading order for large $t$. This allowed us to find explicit asymptotic form for the distribution $P(x,t)$ at the tails for $r=0$ and enabled us to make guess on the scaling behaviour, again  at the tails of the distributions for $r= \pm 1$. 

We believe that our study provides another simple yet nontrivial model of non-Markovian process for which many things can be computed analytically and understood theoretically. Several important future questions can be asked as well as extensions can be made. While the observation that the $\Delta_{\max}(t)$ provides dominating contribution to $x(t)$ for large $t$ allows us to guess the scaling behaviour for position distribution $P(x,t)$ at the tails, it does not immediately tell us the form at the tails. It would be interesting to explore in this direction by studying the joint distribution of the velocity at a step and that step being the longest one. Another  direction to explore would be to look at survival problem of the walker in our model and ask how does the persistent properties depend on $r$ and $\rho(\tau)$. Also studying, extreme statistics and path functionals would lead to interesting directions.


\section{Acknowledgement}
\label{acknowledge}
 We would like to thank Urna Basu, Sanjib Sabhapandit and Abhishek Dhar for their comments and suggestions. We thank Prashant Singh for a careful reading of the manuscript and interesting suggestions. SD and AK would like to thank financial support from DST, Government of India grant under project No. ECR/2017/000634. SD would also like to acknowledge the hospitality of the International Centre for Theoretical Sciences, Bangalore. AK would like to acknowledge support of the Department of Atomic Energy, Government of India, under project no.12-R\&D-TFR- 5.10-1100.

\appendix
\section*{Appendix}
\label{appendix}

\section{Calculation of $\widetilde{P}_v(k,s)$ in Eq.~\eref{furlap_v}}
\label{furlap_v_app}

We can write and simplify the Fourier-Laplace tranform of Eq.~\eref{dist_vt} as 
\bea
\hspace{-1cm}  \widetilde{P_v}(k,s) 
\nonumber
= \sum_{m=0}^{\infty} \int_{v=-\infty}^{\infty} dv \; e^{ikv} \int_{t=0}^{\infty} dt \; e^{-st} \left\langle \delta \left(v - v_m\right) \right\rangle \left\langle \delta \left(t-\sum_{j=0}^{m-1}\tau_{j} - \tau^{*}_m \right) \right\rangle \\
= \sum_{m=0}^{\infty}  \left\langle e^{i k v_m}\right\rangle   \left\langle e^{- s \sum_{j=0}^{m-1}\tau_j -s \tau_m^*}\right\rangle.
\label{fl_v1}
\eea
Since the time intervals are independent of each others we, using Eqs.~\eref{psi(s)} and \eref{Psi(s)} we can simplify Eq.~\eref{fl_v1} as
\bea
\widetilde{P_v}(k,s) = \sum_{m=0}^{\infty}  \left\langle e^{i k \sum_{j=0}^m \eta_{m-j} (-r)^j}\right\rangle \; \tilde{\psi}^m(s) \;\tilde{\Psi}(s).
\label{fl_v2}
\eea
where we have used  $v_m = \sum_{j=0}^m \eta_{m-j} (-r)^j$ from Eq.~\eref{modl_2}.
Also since, all the $\eta$'s are independent Gaussian noise with mean zero and variance $\sigma^2$, we can further 
simplify Eq.~\eref{fl_v2} as
\bea
\nonumber
\widetilde{P_v}(k,s) = \sum_{m=0}^{\infty} \exp \left[-\frac{1}{2} k^2 \sigma^2 \sum_{j=0}^m (-r)^{2j} \right] \tilde{\psi}^m(s) \;\tilde{\Psi}(s) \\
= \sum_{m=0}^{\infty} \exp \left[-\frac{1}{2} k^2 \sigma^2 \frac{1 - r^{2m+2}}{1-r^2} \right] \tilde{\psi}^m(s) \;\tilde{\Psi}(s),
\eea
which is what we have written in Eq.~\eref{furlap_v} in the main text.

\section{Saddle point integration of Eq.~\eref{scl_dis_vel} for $0 < \alpha < 1$}
\label{saddle_point_alpha}

For $0 < \alpha < 1$, the asymptotic form of the Mittag-Leffler function $E_\alpha(u)$ is $E_\alpha(u) \simeq \exp[u^{1/\alpha}]/\alpha$ for $|u| \rightarrow \infty$\cite{haubold2011mittag}. It allows us to write
\bea
E_{\alpha}(- \widetilde{k}^2) \simeq \frac{1}{\alpha} \exp\left[(- \widetilde{k}^2)^{1/\alpha} \right],
\eea
for large $|\tilde{k}|$, inserting which in Eq.~\eref{scl_dis_vel} and using the $\tilde{k} \to -\tilde{k}$ symmetry we  get
\bea
\mathbb{F}_\alpha(y) \simeq \frac{1}{2\pi \alpha} \int_{-\infty}^{\infty} d\widetilde{k} \;e^{|y| \varphi(\widetilde{k}, y)},~~~\textrm{where},~~~
 \varphi (\widetilde{k},y) = (i\widetilde{k})^{2/\alpha}/|y| - i \widetilde{k}.
\eea
We now evaluate this integral using saddle point method. It is easy to see that the saddle point is $\widetilde{k}^* = (\frac{\alpha |y|}{2})^{\frac{\alpha}{(2 - \alpha)}} e^{-i\frac{\pi}{2} }$, at which 
\bea
\hspace{-2.3cm} \varphi (\widetilde{k}^*) = -\left( \frac{2}{\alpha} - 1\right) \left( \frac{\alpha |y|}{2} \right)^{\frac{2}{2 - \alpha}},~~~\textrm{and}~~~
\varphi''(\widetilde{k}^*)  = -\frac{1}{|y|} \;\frac{2}{\alpha} \left( \frac{2}{\alpha} - 1\right) \left( \frac{\alpha |y|}{2} \right)^{\frac{2(1-\alpha)}{2 - \alpha}}.
\eea
Performing the saddle point integration and using the above expressions we finally get  
\bea
\hspace{-2.3cm} \mathbb{F}_\alpha(y) \simeq \frac{e^{y \varphi (\widetilde{k}^*)}}{\alpha \;\sqrt{2 \pi y \;|\varphi''(\widetilde{k}^*)|}} \simeq  \frac{1}{\sqrt{4 \pi (2 - \alpha)}} \left( \frac{\alpha y}{2} \right)^{-\frac{1-\alpha}{2 - \alpha}} \; \exp\left[- \left( \frac{2}{\alpha} - 1 \right) \left( \frac{\alpha y}{2}\right)^{\frac{2}{2-\alpha}} \right].
\label{spi_alp}
\eea
Note that $\mathbb{F}_\alpha(y)$ has the symmetry $\mathbb{F}_\alpha(y)=\mathbb{F}_\alpha(-y)$. 

\section{Inverse Laplace transform: Bromwich integral and Tauberian theorems}
\label{ILT_BR_TU}

In this section we discuss few inverse Laplace transform results which we have used in the main text. The well known methods to compute inverse Laplace transform is to evaluate the Bromwich integral (BI)
~\cite{spiegel1965laplace}. 
Using the formalism of Bromwich integral we first find out the inverse Laplace transform of $\ln(s)/s^{n}$ for any $n \ge 1$ \cite{spiegel1965laplace} 
as
\bea
\footnotesize
\mathcal{L}^{-1}_t\left( \frac{\ln(s)}{s^n}\right) = \left( 1 + \frac{1}{2} +..+ \frac{1}{n-1} - \gamma - \ln(t)  \right) \frac{t^{n-1}}{(n-1)!}.
\label{ilt_1}
\eea
Often we are interested in the large $t$ asymptotic for which one can use the Tauberian theorem~\cite{feller2008introduction} instead of performing the BI. 
This theorem states that if $L(t)$ is a slowly varying function at infinity and $0 \leq \nu < \infty$, then each of the relation
\bea
f(t) &\simeq \frac{t^{\nu} L(t)}{\Gamma(\nu + 1)} ~~~~~~~~~~~~~~~~~~~~~~~\textrm{for}\ t \rightarrow \infty \\
\tilde{f}(s) &=\mathcal{L}[f(t)]\simeq s^{-(\nu+1)} L(1/s) \;\;\; \; \textrm{for}\ s \rightarrow 0
\label{ilt_4}
\eea
implies the others~\cite{feller2008introduction}.  
For an arbitrary constant $c$ the slowly varying function $L$ implies $L(cu)/L(u) \rightarrow 1$ in the limit of $u \rightarrow \infty$.
For example, if $L(u)= \ln(u)$, then it satisfy the limiting condition.
Hence, by using this theorem we can find out an approximate result 
\bea
\mathcal{L}^{-1}_t\left( \frac{\ln(s)}{s^n} \right) \simeq - \frac{t^{n-1}}{(n-1)!} \; \ln(t)
\label{ilt_5}
\eea
for any integer $n \geq 1$ in the limit $t \rightarrow \infty$.
Clearly, this result agrees with the large $t$ asymptotic of the result in Eq.~\eref{ilt_1}.

Similarly, we can consider $L(u) = \ln^{-1}(u)$ which also satisfies the above condition of a slowly varying function.
It enable us to find an approximate, asymptotic result of the inverse Laplace transform of $s^{-n} \;\ln^{-m}(s)$ as 
\bea
\mathcal{L}^{-1}_t\left(\frac{1}{{s^{n} \;\ln^{m}(s)}}\right) \simeq (-1)^m \frac{t^{n-1}}{(n-1)!} \; \ln^{-m}(t)
\label{ilt_6}
\eea
for any integer $n, m \geq 1$ in the limit $t \rightarrow \infty$.

\section{Calculation of $\widetilde{\sigma}_x^2(s)$ for $-1 \le r \le 1$}
\label{var_cal}

Here we show the details of calculation of $\widetilde{\sigma}_x^2(s)$ in Eq.~\eref{var_all}. 
\bea
\nonumber
\hspace{-2cm}\widetilde{\sigma}_x^2(s) =\widetilde{ \langle x^{2} \rangle}(s) = - \left[\frac{d^{2}}{dk^{2}} \widetilde{P_x}(k,s)\right]_{k=0} = \sigma^{2}\; \sum_{m=0}^{\infty} \left \langle \left( \bar{\tau}_{m}^{T}\Sigma_{m}\bar{\tau}_{m}\right) e^{- sI_{m}^{T}\bar{\tau}_{m}} \right \rangle_{\{\tau_{j}\}} \\
\nonumber
\hspace{-1cm} = \sigma^2 \sum_{m=0}^{\infty} \left \langle \left( \mathbb{I}(m \ge 1) \sum_{l=0}^{m-1} \tau^{2}_{l} \sum_{j=0}^{l} r^{2 j}  + {\tau^{*}_m}^{2} \sum_{l=0}^{m} r^{2 l}  + 2\;\mathbb{I}(m \ge 2) \sum_{l=0}^{m-1}  \tau_l\; \right. \right. \\ \nonumber \hspace{-2.3cm} \times \sum_{j=l+1}^{m-1} \tau_{j} \; (-r)^{j-l} \sum_{p=0}^{l} r^{2 p}  + \left.\left. 2 \; \mathbb{I}(m \ge 1) \; \tau^{*}_m \sum_{l=0}^{m-1}  \tau_l\; (-r)^{m-l} \sum_{p=0}^{l} r^{2 p} \right) e^{-s(\tau_0 + \tau_1 +....+\tau^{*}_m)} \right\rangle_{\{\tau_{j}\}} \\ \nonumber
\hspace{-2.2cm} = \sigma^2 \sum_{m=0}^{\infty} \left( \frac{m (1 - r^2) - r^2 (1 -r^{2 m}) }{(1 - r^2)^2} \tilde{\Psi}(s)\;\tilde{\psi}^{m-1}(s)\;\tilde{\psi}''(s) \right. + \frac{1- (r^2)^{1+m}}{1-r^2} \tilde{\psi}^m(s) \tilde{\Psi}''(s)  \\ \nonumber \hspace{-1.2cm} \;+\; 2 \frac{(1-r)^2 (-r)^{m+1} -r^{2 +2m} + r (1-m-r+r^2+m r^2)}{(1- r)^2 (1 + r)^3} \tilde{\Psi}(s)\;\tilde{\psi}^{m-2}(s)\;\tilde{\psi}'^2(s) \\ \left. \right. \left.  \;-\; 2 \;\frac{(-r)^{1+m}(1-r) + r - r^{2(1+m)} }{(1-r) (1+r)^2} \;\tilde{\psi}^{m-1}(s)\;\tilde{\psi}'(s)\;\tilde{\Psi}'(s) \right) 
\eea
Executing all the summation over $m$ 
and using $\tilde{\psi}(s) = \langle e^{-s \tau}\rangle$, $\tilde{\psi}'(s) = -\langle \tau e^{-s \tau}\rangle$, $\tilde{\psi}''(s) = \langle \tau^2 e^{-s \tau}\rangle$, $\tilde{\Psi}(s) = \langle e^{-s \tau^*}\rangle$, $\tilde{\Psi}'(s) = -\langle \tau^{*} e^{-s \tau^*}\rangle$, $\tilde{\Psi}''(s) = \langle {\tau^*}^2 e^{-s \tau^*}\rangle$ (here the averages are over $\rho(\tau)$) we find 
\bea
\nonumber
\hspace{-1cm}\widetilde{\sigma}_x^2(s) 
= \sigma^2 \underbrace{ \left( \frac{\tilde{\Psi}(s) \;\tilde{\psi}''(s)}{(1-\tilde{\psi}(s))^2 \;(1 - r^2 \;\tilde{\psi}(s))}\right.}_{\textrm{I}} \;+\; \underbrace{ \frac{\tilde{\Psi}''(s)}{(1-\tilde{\psi}(s))\; (1 - r^2 \;\tilde{\psi}(s))}}_{\textrm{II}} \\ \hspace{-2.9cm}
\;-\; \underbrace{ \frac{2 r \;\tilde{\Psi}(s) \;\tilde{\psi}'^2(s)}{(1-\tilde{\psi}(s))^2 \; (1 + r\;\tilde{\psi}(s)) (1 - r^2 \;\tilde{\psi}(s))}}_{\textrm{III}} \;-\; \underbrace{ \left. \frac{2 r \;\tilde{\Psi}'(s) \;\tilde{\psi}'(s)}{(1-\tilde{\psi}(s)) \; (1 + r\;\tilde{\psi}(s)) \; \; (1 - r^2 \;\tilde{\psi}(s))} \right)}_{\textrm{IV}}. 
\label{def_momentl}
\eea
Simplifying this equation further by using $\tilde{\Psi}(s)=(1-\tilde{\psi}(s))/s$ we get Eq.~\eref{var_all}.

\section{Exact result of $\sigma_x^2(t)$ in case I {\it i.e.} for $\rho(\tau)=\delta(\tau -a)$}
\label{var_case1}

In case of $\rho(\tau) = \delta(\tau -a)$, using $\tilde{\psi}(s) = e^{-a s}$ in Eq.~\eref{var_all} we can write 
\bea
\widetilde{\sigma}_x^2(s) = \frac{2 \sigma^2}{s^3 (1 - r^2 e^{-as})} \left( 1 - \frac{as \;e^{-as}}{1 - e^{-as}} - \frac{r as \; e^{-as}}{1 + r\; e^{-as}} \right).
\eea
Let us first consider $r = -1$ case, for which the above equation simplifies to 
\bea
\widetilde{\sigma}_x^2(s) = \frac{2 \sigma^2}{s^3 (1 - e^{-as})} = 2 \sigma^2 s^{-3} \sum_{n=0}^{\infty} e^{-nas}.
\eea
Performing inverse Laplace transform yields
\bea
\nonumber
\sigma_x^2(t) = \sigma^2 \sum_{n=0}^{\lfloor t/a \rfloor} (t - a n)^2 = \frac{\sigma^2 }{6} \left( (a^2 - 6 at +6 t^2)  \left\lfloor \frac{t}{a}\right\rfloor \right.\\ + \left.  (3 a^2 - 6 at) \left\lfloor \frac{t}{a}\right\rfloor^2 + 2 a^2 \left\lfloor \frac{t}{a}\right\rfloor^3 + 6 t^2 \right) \simeq \frac{\sigma^2 t^3}{3 a} + \mathcal{O}(t^2)
\eea
in the limit of large time assuming $\lfloor t/a \rfloor \simeq t/a$.
Following a similar procedure we can find out the exact result for $-1< r< 1$ as 
\bea
\nonumber \hspace{-2.5cm}
\sigma_x^2(t) = \frac{\sigma ^2 (r-1)^{-1} }{(r+1)^3} \left(a^2 \left(r^{2 \left\lfloor \frac{t}{a}\right\rfloor +2} \left( \left\lfloor \frac{t}{a}\right\rfloor (r + 1) +1\right)^2 -2 (r-1) r \left( \left\lfloor \frac{t}{a}\right\rfloor (r + 1) + 1\right) 
\right. \right. \\ \left. \left. 
\nonumber \hspace{-2.1cm} \times (-r)^{\left\lfloor \frac{t}{a}\right\rfloor }-(r+1)^2 \left\lfloor \frac{t}{a}\right\rfloor ^2 
-(r+1)^2 \left\lfloor \frac{t}{a}\right\rfloor +(r-2) r\right) + (r+1)^2 t^2 \left(r^{2 \left\lfloor \frac{t}{a}\right\rfloor +2}-1\right) 
\right. \\ \left.
\nonumber \hspace{-1.1cm} 
+ 2 a (r+1) t \left(-r^{2 \left\lfloor \frac{t}{a}\right\rfloor +2} \left( \left\lfloor \frac{t}{a}\right\rfloor (r + 1) +1\right)  +(-r)^{\left\lfloor \frac{t}{a}\right\rfloor +1} \right. \right. \\ 
\left. \left. + (-r)^{\left\lfloor \frac{t}{a}\right\rfloor +2} + \left\lfloor \frac{t}{a}\right\rfloor (r + 1) +r\right)\right)  
\simeq \frac{a \sigma^2}{(1+r)^2} t 
\label{exact_varx_cs1-app}
\eea
in the limit of large time $t$.
Similarly, for $r = 1$, we calculate
\bea
\nonumber
\sigma_x^2(t)  = \frac{\sigma^2}{2} \left(  (a^2 - 2 at + 2 t^2) + (3 a^2  - 6 at + 2 t^2) \left\lfloor \frac{t}{a}\right\rfloor \right. \\ \left. + 4 a (a - t) \left\lfloor \frac{t}{a}\right\rfloor^2 + 2 a^2 \left\lfloor \frac{t}{a}\right\rfloor^3 \right).
\eea 
For any real $t/a$ we can write $t/a = \lfloor t/a\rfloor + u$ with $u \in (0:a)$.
Using this in the above expression we get
\bea
\sigma_x^2(t)  = \frac{1}{2} \sigma^2 a^2\left( 1 + \left\lfloor \frac{t}{a}\right\rfloor \right) (2 u^2 - 2 u + 1),
\label{r1_apx}
\eea
as announced in Eq.~\eref{x^2_t-os}.

\section{Variance and the distribution of the position in random acceleration process (RAP)}
\label{appendix_RAP}

Random acceleration process is a simple non-Markovian stochastic process that yields several non-trivial results \cite{majumdar1999persistence,burkhardt2007random}.
In RAP, a point particle moving in one dimension is accelerated by white noise as 
\bea
\ddot{x}(t) = \zeta(t) 
\eea
where $\zeta(t)$ is a Gaussian white noise with the zero mean and  delta correlation of the strength $D$, {\it i.e.}, 
\bea
\langle \zeta(t)\rangle = 0, \;\; \textrm{and}\;\; 
\langle \zeta(t) \zeta(t')\rangle = D \delta(t - t').
\label{rap_1}
\eea
Notably, the velocity $v(t) = \dot{x}(t)$ in this process executes a Brownian motion and the position $x(t)$ is the area under the Brownian motion trajectory. 
Using Eq.~\eref{rap_1} we can calculate the velocity-velocity correlation function 
\bea
\hspace{-1cm}
C_{vv}(t', t'') = \langle v(t') v(t'') \rangle_c 
= \int_{0}^{t'} d\tau_1 \int_{0}^{t''} d\tau_2 \;\langle \zeta(\tau_1) \zeta(\tau_2) \rangle 
= D\; \textrm{min}(t', t'').
\label{vel_cor_rap}
\eea
where we have used $\langle v(t) \rangle=0$.
Writing the position $x(t)$ up to time $t$ by the integral $x(t) = \int_{0}^t dt' v(t')$.
Clearly, it provides $\langle x(t) \rangle = 0$. Hence the variance can be written in terms of the correlation as
\bea
\nonumber
\sigma_x^2(t)=\langle x^2(t) \rangle_c=\langle x^2(t) \rangle = \int_{0}^{t} dt' \int_{0}^{t} dt'' \; C_{vv}(t', t'') \\
= \underbrace{\int_{0}^{t} dt' \int_{0}^{t'} dt'' \; C_{vv}(t', t'')}_{t'> t''} + 
\underbrace{\int_{0}^{t} dt' \int_{t'}^{t} dt'' \; C_{vv}(t', t'')}_{t' < t''}
= \frac{1}{3}D t^3.
\eea
The joint probability distribution of position and velocity $(x, v)$ for a random accelerated particle at time $t$ starting initially from $x_0$ and $v_0$ reads \cite{burkhardt1993semiflexible,singh2020random} 
\bea
\hspace{-2.5cm}
\mathcal{P}(x,v,t|x_0,v_0) = \frac{\sqrt{3}}{\pi D t^2} \exp\left[ -\frac{6}{D t^3} \left( (x - x_0 -v_0 t)(x - x_0 -v t) + \frac{1}{3} (v - v_0)^2 t^2 \right) \right].
\eea
Integrating out $v_0$ and $v$ with the zero mean gaussian distribution with variance $D$, one gets 
\bea
P(x-x_0,t) \simeq \sqrt{\frac{3}{2\pi D t^3}} \exp\left[ - \frac{3 (x - x_0)^2}{2 D t^3} \right]
\eea
in the limit of large $t$.

\section{Velocity-velocity correlation function}
\label{velo_corr}
Consider our walker starts at $x(0) = 0$ and is moving with a velocity $v(t)$ at time $t$.
We can define the velocity-velocity correlation function in this case as $C_{vv}(t_1, t_2) = \langle v(t_1) v(t_2) \rangle - \langle v(t_1) \rangle \langle v(t_2) \rangle$.
Since the noise in our case has symmetric distribution (see from Eq.~\eref{modl_1}) it makes $\langle v(t) \rangle = 0$, and hence $C_{vv}(t_1, t_2) = \langle v(t_1) v(t_2) \rangle$.
Assumin $t_2 > t_1$ we write
\bea
\hspace{-1cm}
{C_{vv}(t_1, t_2)}_{t_2 > t_1} = \langle v(t_1) v(t_2) \rangle 
= \sum_{m=0}^{\infty} \sum_{n=0}^{\infty}  \mathcal{P}(m,n; t_1, t_2 - t_1) \; \langle v_m v_{m+n} \rangle_c 
\label{corrv_2}
\eea
where $\mathcal{P}(m,n; t_1, t_2 - t_1)$ is the joint  probability of $m$ and $n$ `jump' events to occur within time intervals $t_1$ and $(t_2 - t_1)$ respectively.
From the dynamical rule of the velocity in Eq.~\eref{modl_1} it is easy to write 
$v_{m+n} = (-r)^n v_m + \sum_{j=1}^n (-r)^{n-j} \;\eta_{m+j}$ for any $n\ge 0$. This suggests 
$ \langle v_m v_{m+n} \rangle_c = (-r)^n  \langle v_m^2 \rangle_c$.  Hence, from Eq.~\eref{corrv_2} we write, 
\bea
{C_{vv}(t_1, t_2)}_{t_2 > t_1} = \sum_{m=0}^{\infty} \langle v_m^2 \rangle_c \sum_{n=0}^{\infty} (-r)^n \mathbb{P}(n, t_2 - t_1|m, t_1) \; \mathcal{P}(m, t_1)\;  
\label{corrv_3}
\eea
where we have written
\be 
\mathcal{P}(m,n; t_1, t_2 - t_1) = \mathbb{P}(n, t_2 - t_1|m, t_1) \; \mathcal{P}(m, t_1).
\ee
Here $\mathcal{P}(m, t_1)$ is the probability of having $m$ number of `jump' events within time $t_1$ and $\mathbb{P}(n, t_2 - t_1|m, t_1)$ is the conditional probability to have $n$ `jump' events in duration $t_2-t_1$ given that there were $m$ `jump' events till time $t_1<t_2$. The correlation in Eq.~\eref{corrv_3} can now be written as 
\bea
{C_{vv}(t_1, t_2)}_{t_2 > t_1} = \sum_{m=0}^{\infty} \mathcal{P}(m, t_1) \; \langle v^2_m \rangle_c \sum_{n=0}^{\infty} (-r)^n \;\mathbb{P}(n, t_2 - t_1|m, t_1).  
\label{corrv_4}
\eea
{To compute $\mathbb{P}(n, t_2 - t_1|m, t_1)$ one needs to be careful because  the $m^{\textrm{th}}$ jump event may not occur exactly at time $t_1$. It can occur before time $t_1$ and the $(m+1)^{\textrm{th}}$ jump event occurs after time $t_1$. 
In this case the problem of computing $\mathbb{P}(n, t_2 - t_1|m, t_1)$ is equivalent to finding the probability $\bar{\mathcal{P}}(n, t;t_0)$ of having $n$ renewal events in a renewal process within a observation time interval $[t_0,t+t_0]$ where $t_0$ is not the starting time of the process. Like in the random incidence problem \cite{ibe2014fundamentals,feller2008introduction}, the incident time $t_0$ can fall at a random moment between  two consecutive renewal events (in our problem the `kick' events), one before $t_0$ and one after $t_0$. Following \cite{ibe2014fundamentals}, let the time interval between these two events is denoted by $w$ and the probability distribution of the length $w$ of the interval between two successive renewal events in which the incidence time $t_0$ falls, is denoted by $f_w(w)$. Assuming that the probability that a random incidence occurs in an interval (gap) of length between $w$ to $w+dw$ is proportional to the gap length itself, we write $f_w(w) \propto w \rho(w)$ implying 
\bea
f_w(w) = \frac{w \rho(w)}{\langle w \rangle} . \label{f_w(w)}
\eea

Now assume that, starting from $t_0$ the  $1^{\textrm{st}}$ event ( equivalently $(m+1)^{\textrm{th}}$ event in our problem) occurs after duration $r$ {\it i.e.} at time $t_0+r$. The duration $r$ is called the residual time in the context of renewal process. The joint probability distribution of the residual time $r$ and the length $w$ of the gap in which the random incidence time $t_0$ falls, is given by $f_{rw}(r,w) = f_{r|w}(r|w) f_w(w)$ where $f_{r|w}(r|w)$ is the conditional probability distribution of the residual time given that gap length is $w$. Since, the random incidence time $t_0$ falls uniformly within the gap, we should have $f_{r|w}(r|w)=\frac{1}{w}$ which implies 
$f_{rw}(r,w) = f_{r|w}(r|w) f_w(w) = \frac{\rho(w)}{\langle w \rangle}.$
Integrating this joint probability distribution over $w$ we get the distribution of the residual time $\tau$ (denoted by $f_r(r)$) as 
\bea
f_r(r) =  \frac{1}{\langle w \rangle} \int_r^\infty dw~ \rho(w).
\eea

We can now write the probability $\bar{\mathcal{P}}(n, t;t_0)$ of having $n$ renewal events within the interval $[t_0,t_0+t]$. Let the the first event occurs at time $t_1=t_0+r_1$, the second at $t_2=t_0+r_1+r_2$ and so on and the last ($n^{\textrm{th}}$) event occurs at time $t_n=t_0+r_1+r_2+...+r_n$. The joint probability distribution of having $n$ events  with the following time interval configuration $(r_1,r_2,...,r_n,r_l)$ is given by 
\bea
\nonumber
\bar{\mathcal{P}}(n, t;t_0) = \int \int ...\int dr_1dr_2...dr_ndr_l f_r(r_1) \rho(r_2)....\rho(r_n) \\ \times \Psi(r_l)\delta(t-r_1-r_2-...-r_n-r_l),~~~\textrm{for}~n>0,
\eea
where $r_l=t-t_n$ is the last incomplete interval and recall, $\Psi(t)=\int_t^\infty d\tau ~\rho(\tau)$. Similarly for $n=0$ we have 
\bea
\bar{\mathcal{P}}(n=0, t;t_0) = g_r(t),~~\textrm{where}~~g_r(t) = \int_t^\infty dr~f_r(r).
\eea
Since the events after $t_0$ are independent of events or the number of events before $t_0$, we should have 
\bea
\nonumber
\mathbb{P}(n, t_2 - t_1=t~ |~m, t_1) = \delta_{n,0}g_r(t) + (1-\delta_{n,0})\int \int ...\int dr_1dr_2...dr_ndr_l \\ \times f_r(r_1) \rho(r_2)....\rho(r_n) \Psi(r_l)\delta\left(t-\sum_{i=1}^nr_i-r_l\right).
\label{mbbP_big-exp}
\eea
Note that $\mathbb{P}(n, t_2 - t_1=t~ |~m, t_1)$ does not depend explicitly on $t_1$ and $m$. Hence we can omit their explicit appearance in this distribution and denote it by just $\mathbb{P}(n, t_2 - t_1=t)$.  The velocity correlation can now be written as 
\bea 
\hspace{-2.3cm}
{C_{vv}(t_1, t_2)}
= \underbrace{\sum_{m=0}^{\infty} \mathcal{P}(m, t_1) \;\langle v^2_m \rangle}_{\langle v^2 (t_1) \rangle} \times \underbrace{\sum_{n=0}^{\infty} (-r)^n \;\mathbb{P}(n, t_2 - t_1)}_{Q_r(t_2 - t_1)} \simeq \langle v^2 (t_1) \rangle_c \;Q_r(t_2 - t_1).
\label{corrv_5}
\eea
Hence for  large $t_1$ and $t_2$ we get 
\bea
\hspace{-2.3cm}
C_{vv}(t_1, t_2) &= \langle v^2 (\textrm{min}\{t_1, t_2\}) \rangle_c \;Q_r(|t_2 - t_1|),~~\textrm{with,}
~~Q_r(t) = \sum_{n=0}^{\infty} (-r)^n \;\mathbb{P}(n, t).
\label{corrv_6}
\eea
Taking Laplace transform of $Q_r(t) $ with respect to time we get 
\bea
\tilde{Q}_r(s)=  \sum_{n=0}^{\infty} (-r)^n \;\tilde{\mathbb{P}}(n, s),\label{tilde_Q(s)}
\eea
where from Eq.~\eref{mbbP_big-exp} (after taking Laplace transform with respect to $t$ on both sides)  
we get,
\bea 
\widetilde{\mathbb{P}}(n, s) = \widetilde{g}_r(s) \;\delta_{n,0} +  \widetilde{f}_r(s) \;\tilde{\psi}^{n-1}(s) \;\tilde{\Psi}(s)\; (1 - \delta_{n,0}),
\label{P_R(m, s)}
\eea 
where $\tilde{f}_r(s)$ and $\tilde{g}_r(s)$ are Laplace transforms of $f_r(r)$ and $g_r(r)$ respectively and, $\tilde{\psi}(s)$ and $\tilde{\Psi}(s)$ are defined in Eqs.~(\ref{psi(s)}-\ref{Psi(s)}). Using this expression of $\widetilde{\mathbb{P}}(n, s)  $ in Eq.~\eref{tilde_Q(s)}  we get 
\bea 
\widetilde{Q}_r(s) = \widetilde{g}_r(s) - \frac{r \;\tilde{\Psi}(s) \;\widetilde{f}_r(s)}{1 + r \tilde{\psi}(s)}.
\label{Q_R(m, s)}
\eea

}

\subsection{Velocity-velocity correlation function in case II}
\label{velo_corr_cs2}
In case II, {\it i.e.}, for exponential waiting time distribution $ \rho(\tau) = \beta e^{-\beta \tau}$, we have $f_r(\tau) = \beta \int_{\tau}^{\infty} dz \;\rho(z) = \beta e^{-\beta \tau} = \rho(\tau)$. Hence, in this case $\tilde{f}_r(s)=\tilde{\psi}(s)$ and $\tilde{g}_r(s)=\tilde{\Psi}(s)$, using which in Eq.~\eref{Q_R(m, s)} we get 
\bea 
\widetilde{Q}_r(s) = \frac{\tilde{\Psi}(s)}{1 + r \;\tilde{\psi}(s)}.
\label{equn6}
\eea
Remember for exponential waiting time distribution $\tilde{\psi}(s) = \beta/(\beta +s)$ and $\tilde{\Psi}(s) = 1/(\beta +s)$. Using these expressions in the above equation we get $\widetilde{Q}_r(s) = (s + (1+r)\beta)^{-1}$. Performing inverse Laplace transform provides $Q_r(|t_2 - t_1|) = e^{-\beta(1+r)|t_1 - t_2|}$, using which in Eq.~\eref{corrv_6} yields
\bea 
C_{vv}(t_1, t_2) \simeq \langle v^2 (\textrm{min}\{t_1, t_2\}) \rangle_c \; e^{-\beta (1+r) |t_1 - t_2|}.
\label{corrv_8}
\eea
Note for $r=0$, the variance of the velocity $\langle v^2 (\textrm{min}\{t_1, t_2\}) \rangle_c =\sigma^2$.

\subsection{Velocity-velocity correlation function in case III with $\alpha >1$}
\label{velo_corr_cs3}

\subsubsection{For $r = -1$\\\\}
\label{velo_corr_rn1}
Computing $C_{vv}(t_1,t_2)$ in this case is simpler because for $r=-1$, it is easy to see that $Q_{-1}(u) = \sum_{m=0}^{\infty} \mathbb{P}(m, u) = 1$ due to the normalization of $\mathbb{P}(m, u)$. Hence, in this case we simply get 
\be
C_{vv}(t_1, t_2) \simeq \langle v^2( \textrm{min}\{ t_1, t_2\})\rangle_c.
\label{C_vv_t1t2_rn1}
\ee

\subsubsection{For $r = 0$\\\\}
\label{velo_corr_r0}

For $r = 0$, we see from Eq.~\eref{Q_R(m, s)} that $\widetilde{Q}_0(s) = \widetilde{g}_r(s)$.  Performing inverse Laplace transform we get 
\bea
\hspace{-1.cm}
Q_0(u) = g_r(u) = \int_{u}^{\infty} dw \; f_r(w)
= \frac{1}{\langle \tau \rangle} \int_{u}^{\infty} dw \; \int_{w}^{\infty} dz \; \rho(z)
= \frac{1}{\alpha} u^{-\alpha + 1},
\label{equn11}
\eea
\bea
\hspace{-1.9cm}
\textrm{which~finally~provides~} \qquad C_{vv}(t_1, t_2) = \sigma^2 \; Q_0(|t_1 - t_2|) = \frac{\sigma^2}{\alpha} \; |t_1 - t_2|^{-\alpha+1} 
\eea
where we have used  $ \langle v^2(t)\rangle_c=\sigma^2$ for $r =0$.

\subsubsection{For $r = 1$\\\\}
\label{velo_corr_r1}

For $r = 1$, using $g_r(s) = (1 - f_r(s))/s$, $\tilde{\Psi}(s) = (1 - \tilde{\psi}(s))/s$, and $\tilde{\psi}(s) \simeq 1 - \alpha s/(\alpha -1) + \mathcal{O}(s^{\alpha})$ for $\alpha > 1$ in the limit $s \rightarrow 0$ we can simplify Eq.~\eref{Q_R(m, s)} and get 
\bea 
\widetilde{Q}_1(s) = \frac{1}{s} - \frac{2 \widetilde{f}_r(s)}{s (1 + \tilde{\psi}(s))}
\simeq \frac{1 - \widetilde{f}_r(s) }{s} \simeq \widetilde{g}_r(s).
\eea
It implies Eq.~\eref{equn11} is also valid for $r = 1$ in the limit of large $t$, but the correlation function is modified because  $\langle v^2(t) \rangle \simeq \sigma^2 t/\langle \tau \rangle_c$ in this case, instead of $\sigma^2$.
Hence we can get
\bea 
\hspace{-2.5cm}
C_{vv}(t_1, t_2) \simeq \langle v^2 \;\textrm{min}\{t_1, t_2 \} \rangle_c\; Q_1(|t_1 - t_2|) \simeq \frac{\sigma^2(\alpha-1)}{\alpha^2} \; \textrm{min}\{t_1, t_2 \}\; |t_1 - t_2|^{-\alpha+1}.
\label{equn13}
\eea

\section{Variance of the position of  L\'evy Walk}
\label{var_LW}

Let $P_{lw}(x,t)$ denote the position $x(t)$ made by a space-time coupled L\'evy walker after time $t$. Also let $\mathcal{Q}_{lw}(x,t)$ denote the probability distribution that the walker lands at $x$ exactly at time $t$. This probability distribution satisfies the following balance equation \cite{klafter2011first,zaburdaev2015levy} 
\bea
\mathcal{Q}_{lw}(x,t) = \int_{-\infty}^{\infty} dx' \int_{0}^{\infty} dt' \; \mathcal{Q}_{lw}(x',t') \; \psi_{lw}(x-x',t-t')  + \delta(x)\delta(t).
\label{eqn1}
\eea
Here, $\mathcal{Q}_{lw}(x',t')$ denotes the distribution of the position $x'(t')$ at the last completed step in time $t' < t$ with $\psi_{lw}(x,\tau)$ denoting the joint distribution of `jump' length $x$ and jump duration $\tau$.
The second term in Eq.~\eref{eqn1} arises from the initial condition $P_{lw}(x=0,t=0) = \delta(x) \delta(t)$. The evolution equationfor $P_{lw}(x,t)$ is related to the distribution $\mathcal{Q}_{lw}(x,t)$ by 
\bea
P_{lw}(x,t) = \int dx' \int dt' \; \mathcal{Q}_{lw}(x',t') \; \Psi_{lw}(x-x',t-t')
\label{eqn3}
\eea
where $\Psi_{lw}(x,t)$ denotes the joint distribution of `jump' length and time of the last incomplete step.  Given the distribution $p(v)$ of velocity and the distribution  $\rho(\tau)$ of the jump duration $\tau$, $\psi_{lw}(x,t)$ can be  written as 
$\psi_{lw}(x,\tau) = \rho(\tau)~ \int_{-\infty}^{\infty} dv \;p(v) \;\delta(x - v \tau)$ which simplifies to $\psi(x,\tau) = \frac{p(x/\tau)}{\tau}\rho(\tau)$. Similarly, $\Psi_{lw}(x,t)$ can be written as $\Psi_{lw}(x,\tau)  =  \frac{p(x/\tau)}{\tau}\int_{\tau}^{\infty} du \;\rho(u)$.
Performing joint Fourier-Laplace transform on both sides of Eqs.~\eref{eqn1} and \eref{eqn3} we get 
\bea
\widetilde{\mathcal{Q}}_{lw}(k,s) &= \frac{1}{1 - \widetilde{\psi}_{lw}(k,s)},~~\textrm{and,} \label{eqn2}\\ 
\widetilde{P}_{lw}(k,s) &= \widetilde{\Psi}_{lw}(k,s)\;\widetilde{\mathcal{Q}}_{lw}(k,s) = \frac{\widetilde{\Psi}_{lw}(k,s)}{1 - \widetilde{\psi}_{lw}(k,s)},
\label{eqn4}
\eea 
where $\widetilde{F}_{lw}(k,s) = \int_{-\infty}^\infty e^{i k x}e^{-s \tau}F_{lw}(x,\tau)$ is the Fourier-Laplace transform of the function $F_{lw}(x,\tau)$.

Note that $\widetilde{\psi}_{lw}(k = 0,s) = \tilde{\psi}(s)=\mathcal{L}[\rho(\tau)]$ and $\widetilde{\Psi}_{lw}(0,s) = \tilde{\Psi}(s)=\mathcal{L}[\int_{\tau}^{\infty} du \;\rho(u)]$ as given in Eqs.~(\ref{psi(s)}) and (\ref{Psi(s)}). Taking second derivative of $\widetilde{P}_{lw}(k,s) $ with respect to $k$  at $k=0$ provides the Laplace transform of the variance of the position of the L\'evy walker at time~$t$ 
\bea 
\hspace{-1.3cm}
\widetilde{\sigma}_x^2(s) &= - \frac{d^2}{dk^2} \left[ \widetilde{P}(k,s)\right]_{k=0} 
= - \frac{d^2}{dk^2} \left[ \frac{\widetilde{\Psi}_{lw}(k,s)}{1 - \widetilde{\psi}_{lw}(k,s)} \right]_{k=0} \nonumber \\
&= - \left( \frac{\widetilde{\Psi}_{lw}(0,s) \; \widetilde{\psi}_{lw}''(0,s)}{(1 - \widetilde{\psi}_{lw}(0,s))^2} + \frac{\widetilde{\Psi}_{lw}''(0,s)}{1 - \widetilde{\psi}_{lw}(0,s)} \right) = \langle v^2\rangle\; \frac{\tilde{\psi}''(s) + s\; \tilde{\Psi}''(s)}{s \;(1 - \tilde{\psi}(s))}.
\label{eqn6}
\eea
where we have used the following relations
$\widetilde{\psi}_{lw}'(0,s) = - i \langle v\rangle \tilde{\psi}(s)$, $\widetilde{\psi}_{lw}''(0,s) = - \langle v^2\rangle \tilde{\psi}(s)$, $\widetilde{\Psi}_{lw}'(0,s) = - i \langle v\rangle \tilde{\Psi}'(s)$ and $\widetilde{\Psi}_{lw}''(0,s) = - \langle v^2\rangle \tilde{\Psi}''(s)$ with$\langle v\rangle$ and $\langle v^2\rangle_c$ being the mean and variance of $p(v)$. Remember that in this paper we have takes $p(v)$ to be Gaussian with $\langle v\rangle=0$.

\section{Distribution of displacement made in a single step for $r = 0$ in case II}
\label{pdf_single_stp_r0}

The distribution $P(x,t)$ of the position $x$ after time $t$ is given approximately by 
\bea
P(x,t) \simeq \sum_{m=0}^{\infty} \mathcal{P}_x(x|m)\; \mathcal{P}(m,t),
\label{P(x,t)-exp}
\eea
for large $t$ where, remember, $\mathcal{P}(m,t)$ is the probability of taking $m$ steps within time $t$ and $ \mathcal{P}_x(x|m)$ is probability distribution of finding the particle at position $x$ after $m$ jump steps. The approximately equal sign in the above expression is because we can neglect the contribution of the position of the last incomplete step for large $t$ in the case of  $\rho(\tau)=\beta e^{-\beta \tau}$. Note for $r=0$ the position of the walker at $m^{\textrm{th}}$ step is given by $x_m=\sum_{i=0}^{m-1}\xi_i$ where the distribution of the position $\xi = \tau \eta$ in a single step can be calculated as follows
\bea
\nonumber
\varrho(\xi) = &\int_0^{\infty} d\tau \;\rho(\tau)  \int_{-\infty}^{\infty} d\eta \;p(\eta) \; \delta(\xi -\eta\tau) \\ &
= \frac{\beta}{\sqrt{2\pi \sigma^2}} \;\int_0^{\infty} d\tau \; e^{-\beta \tau} \int_{-\infty}^{\infty} 
d\eta \;e^{-\eta^2/2 \sigma^2} \delta(\xi -\eta\tau).
\label{eq1}
\eea
Using $\delta(\xi - \eta \tau) = \delta(\xi/\tau - \eta)/|\tau|$ we get 
\bea
\hspace{-2.5cm}
\varrho(\xi) =  \frac{\beta}{\sqrt{2\pi \sigma^2}} \; \int_0^{\infty} \frac{d\tau}{\tau}\; \exp\left[{-\beta \tau - \frac{\xi^2}{2 \sigma^2 \tau^2}}\right] 
\underset{\xi \to \infty} 
{\simeq} 
 \frac{2\sqrt{2}\sigma^{1/3}}{\beta^{1/3}}\frac{1}{\xi^{1/3}}
 \exp\left(-\frac{3 \beta^{2/3}}{2\sigma^{2/3}}\xi^{2/3}\right),
\label{pdf_exp_gus1}
\eea
which shows $\varrho(\xi)$ decays faster than a power law at large $\xi$.
The characteristic function of $\varrho(\xi)$ is given by 
\be
\lambda(k) = \int_{-\infty}^{\infty} d\xi \; e^{i k \xi} \varrho(\xi)
= \sqrt{\frac{\pi \beta^2}{2 k^2 \sigma^2}}\;\exp\left[\frac{\beta^2}{2 k^2 \sigma^2} \right]\;\textrm{Erfc}\left[\sqrt{\frac{\beta^2}{2 k^2 \sigma^2} } \right]
\ee
where $\textrm{Erfc}[z] = (2/\sqrt{\pi}) \int_{z}^{\infty} du \;e^{-u^2}$ is complimentary error function. 
Using this $\lambda(k)$ we write the distribution of the position $x_m=\sum_{i=0}^{m-1}\xi_i$ after $m$ jumps or steps as
\bea
\mathcal{P}_x(x|m) = \frac{1}{2\pi} \int_{-\infty}^{\infty} dk\; e^{-i k x}\; \lambda^m(k).
\label{eq2}
\eea

On the other hand, to compute $\mathcal{P}(m,t)$ in this case we note that the time interval between all successive events are taken from exponential distribution.  Hence, the probability of making $m$ steps within time $t$ is given by $p(m,t) = e^{-\beta t} (\beta t)^m/m!$.  Using this result and Eq.~\eref{eq2} in Eq.~\eref{P(x,t)-exp} we get
\bea
\hspace{-2cm}
P(x,t) \simeq \frac{1}{2\pi} \int_{-\infty}^{\infty} dk\; e^{- i k x} \;\sum_{m=0}^{\infty} \frac{e^{-\beta t}}{m!} (\beta t \lambda(k))^m
\simeq \frac{1}{2\pi} \int_{-\infty}^{\infty} dk\; e^{- i k x + \beta t (\lambda(k) - 1)}.
\label{eq3}
\eea
For large $t$ one can perform a saddle point calculation keeping the ratio $x/t$ fixed and get a large deviation form of the distribution $P(x,t)$. From this calculation it is easy to show that, this distribution behaves as a Gaussian around the mean with a variance $2 \sigma^2 t/\beta$.

\section{Calculation of $\mathcal{P}(\tau_{\max}|t)$ for $\alpha = 1$:}
\label{P_tau_max_a=1}

To obtain the behaviour of $\mathcal{P}(\tau_{\max}|t)$, we start with Laplace transform $\widetilde{\mathcal{H}}_{\max}(\tau,s)$ of the cumulative distribution function $\mathcal{H}_{\max}(\tau, t) = \textrm{Prob}(\tau_{\max}(t) \le \tau)$ of $\mathcal{P}(\tau_{\max}|t)$. 
Since for large $t$ we expect $\tau_{\max}$ would also be large. Hence we focus in the small $s$ behaviour the function $\widetilde{\mathcal{H}}_{\max}(\tau,s)$. This function is given in Ref.~ \cite{godreche2015statistics}, from which we write 
\bea
\hspace{-1.cm}
\frac{1}{s} - \widetilde{\mathcal{H}}_{\max}(\tau,s) \simeq \frac{\mathcal{G}_{\max}(\tau s)}{s} 
\qquad \textrm{where} \\
\hspace{-1.cm}
\mathcal{G}_{\max}(\tau s) 
\simeq \left( 1 + \tau s\; e^{\tau s} \int_{s}^{\tau s} du \; u^{-1}\;e^{-u}\right)^{-1} 
\simeq \left( 1 + \tau s\; \ln(\tau) + \mathcal{O}(s^2) \right)^{-1}.
\label{G_tau_s_alpha1}
\eea
Performing inverse Laplace transform, we get
\bea 
\mathcal{H}_{\max}(\tau,t) \simeq \; \textrm{Exp}\left[-\frac{t}{\tau \ln(\tau)} \right].
\eea
Now, using the relation $\mathcal{P}(\tau_{\max}|t) =  \left[ \frac{d}{d\tau} \mathcal{H}_{\max}(\tau, t)\right]_{\tau = \tau_{\max}} $ we get
\bea 
\mathcal{P}(\tau_{\max}=\tau|t) \simeq \frac{ t (\ln (\tau) + 1)}{\tau^2 (\ln \tau)^2 } \textrm{Exp}\left[-\frac{t}{\tau  \ln \tau}\right].
\eea

\section{Calculation of fourth moment of the position for $r = \mp 1$ in case II}
\label{4th_mnt_rpm1_app}
In this section we calculate the fourth moment of position exactly after $m$ `jump' steps. Note from Eq.~\eref{modl_3} that $x_m = \sum_{i =1}^{m-1} v_i \tau_i$.
Using $v_i = \sum_{j=0}^i \eta_i\;(-r)^{i-j}$ from Eq.~\eref{modl_2} we can write 
\bea
x_m = \sum_{i =1}^{m-1} \tau_i \sum_{j=0}^i \eta_i\;(-r)^{i-j}.
\eea 
Starting from this expression, the fourth moment of $x_m$ is written as 
\bea
\hspace{-2cm}
\langle x_m^4\rangle = \sum_{i = 0}^{m-1} \sum_{j = 0}^{m-1} \sum_{k = 0}^{m-1} \sum_{l = 0}^{m-1} \langle \tau_i \tau_j  \tau_k \tau_l\rangle \sum_{p=0}^{i} 
\sum_{q = 0}^{j} \sum_{w = 0}^{k} \sum_{s = 0}^{l} \langle \eta_p \eta_q  \eta_r \eta_s\rangle \;(-r)^{i+j+k+l-p-q-w-s}.
\label{4mnt}
\eea
To compute the averages of the noises, we use the Wick's theorem $\langle \eta_p \eta_q \eta_w \eta_s \rangle = \delta_{p,q} \delta_{w,s} + \delta_{p,w} \delta_{q,s} + \delta_{p,s} \delta_{q,w}$ and calculate different parts of the fourth moment one by one. 
First we consider the case $i=j=k=l$ in Eq.~\eref{4mnt} for which we  get 
\bea
\nonumber
\langle x_m^4\rangle_{i = j = k = l} = \sum_{i = 0}^{m-1} \langle \tau_i^4 \rangle \;(-r)^{4i} \sum_{p=0}^{i} 
\sum_{q = 0}^{i} \sum_{w = 0}^{i} \sum_{s = 0}^{i} \langle \eta_p \eta_q \eta_w \eta_s \rangle (-r)^{-p-q-w-s} 
\\ \nonumber
 = \langle \tau^4 \rangle \langle \eta^2 \rangle^2 \sum_{i = 0}^{m-1} \sum_{p=0}^{i} \sum_{q = 0}^{i} \sum_{w = 0}^{i} 
\sum_{s = 0}^{i}  \left[ \delta_{p,q} \delta_{w,s} + \delta_{p,w} \delta_{q,s} + \delta_{p,s} 
\delta_{q,w} \right] \\ = 3 \; \langle \tau^4 \rangle \langle \eta^2 \rangle^2 \sum_{i = 0}^{m-1} \sum_{p=0}^{i} 
\sum_{w = 0}^{i} 1 \simeq 4! \;\frac{\sigma^4 m^3}{\beta^4} + \mathcal{O}(m^2) \qquad \textrm{for}\;\; r = \mp 1.
\label{eq35}
\eea
Note that we have used $r = \mp 1$ in the second line in Eq.~\eref{eq35}.
We now follow the same method to compute the contribution to $\langle x_m^4\rangle$ if $i = j = k \ne l$,
\bea
\nonumber
\hspace{-2cm}
\langle x_m^4\rangle_{i = j = k \ne l} = \left( \sum_{i = 0}^{m-2} \sum_{l = i+1}^{m-1} + \sum_{l = 0}^{m-2} 
\sum_{i = l+1}^{m-1} \right) \langle \tau_i^3 \rangle \langle \tau_l \rangle  \;(-r)^{3i+l} \sum_{p=0}^{i} 
\sum_{q = 0}^{i} \sum_{w = 0}^{i} \sum_{s = 0}^{l} \langle \eta_p \eta_q \eta_r \eta_s \rangle \\ \nonumber
\hspace{-1cm} \times (-r)^{-p-q-w-s} 
 = \langle \tau \rangle \langle \tau^3 \rangle \langle \eta^2 \rangle^2 \left( \sum_{i = 0}^{m-2} \sum_{l = i+1}^{m-1} + \sum_{l = 0}^{m-2} 
\sum_{i = l+1}^{m-1} \right) \;(-1)^{3i+l}
\\ \nonumber
\hspace{-1.5cm} \times \sum_{p=0}^{i} \sum_{q = 0}^{i} \sum_{w = 0}^{i} \sum_{s = 0}^{l} 
\left[ \delta_{p,q} \delta_{w,s} + \delta_{p,w} \delta_{q,s} + \delta_{p,s} \delta_{q,w} \right] 
 = 3 \; \langle \tau \rangle \langle \tau^3 \rangle \langle \eta^2 \rangle^2 \left( \sum_{i = 0}^{m-2} \sum_{l = i+1}^{m-1} (i+1)^2  
 \right. \\ \left. \hspace{-2.3cm} \times  (\pm 1)^{3i+l} 
+ \sum_{l = 0}^{m-2} \sum_{i = l+1}^{m-1} (i+1) (l+1) (\pm 1)^{3i+l} \right) 
\simeq 
\left \{\begin{array}{ccc} 
\frac{5}{4} \;\frac{\sigma^4 m^4}{\beta^4} + \mathcal{O}(m^3) ~~~ \textrm{for}\; r = -1\\
- 3! \;\frac{\sigma^4 m^3}{\beta^4} + \mathcal{O}(m^2) ~~ \textrm{for}\; r = 1.
\end{array}
            \right. \label{eq36}
\eea
In the second line in Eq.~\eref{eq36} we have used $r = \mp 1$ and employed Wick's theorem.
Same method can be followed to compute the contribution from the remaining two cases $i = j \ne k = l$, $i = j \ne k \ne l$, and $i \ne j \ne k \ne l$ to $\langle x_m^4\rangle$ as following:
\bea
\hspace{-2.3cm}
\nonumber
\langle x_m^4\rangle_{i = j \ne k = l} = 3\times 2 \sum_{i=0}^{m-2}\sum_{k=i+1}^{m-1} \langle \tau_i^2 \rangle \langle \tau_k^2 \rangle (-r)^{2i+2k} \sum_{p=0}^{i} \sum_{q = 0}^{i} \sum_{ w = 0}^{k} \sum_{s = 0}^{k} \langle \eta_p \eta_q \eta_w \eta_s \rangle (-r)^{-p-q-w-s} 
\\ \hspace{-2.cm} \nonumber = 6 \;\langle \tau^2 \rangle^2 \langle \eta^2 \rangle^2 \sum_{i=0}^{m-2} \sum_{k=i+1}^{m-1} \sum_{p=0}^{i} \sum_{q = 0}^{i} \sum_{ w = 0}^{k} \sum_{s = 0}^{k} \left[ \delta_{p,q} \delta_{w,s} + \delta_{p,w} 
\delta_{q,s} + \delta_{p,s} \delta_{q,w} \right] 
\\ \hspace{-2.cm} \nonumber = 6 \;\langle \tau^2 \rangle^2 \langle \eta^2 \rangle^2 \sum_{i=0}^{m-2} \sum_{k=i+1}^{m-1} 
\left[(i+1)(k+1) + 2 (i+1)^2  \right] 
\\ \simeq 7 \left( \frac{\sigma m}{\beta}\right)^4 + \mathcal{O}(m^3) \qquad \textrm{for}\; r \mp 1.
\label{eq37}
\eea

\bea
\hspace{-2.cm} \nonumber 
\langle x_m^4\rangle_{i = j \ne k \ne l} = 6 \times 2 \left( \sum_{i=0}^{m-3} \sum_{k=i+1}^{m-2} \sum_{l=k+1}^{m-1} + \sum_{k=0}^{m-3} \sum_{i=k+1}^{m-2} \sum_{l=k+1}^{m-1} + \sum_{k=0}^{m-3} \sum_{l=k+1}^{m-2} \sum_{i=l+1}^{m-1} \right) 
\\ \times \langle \tau_i^2 \rangle \langle \tau_k \rangle \langle \tau_l \rangle (-r)^{2i+k+l} \nonumber \times \sum_{p=0}^{i} \sum_{q = 0}^{i} \sum_{w = 0}^{k} \sum_{s = 0}^{l} \langle \eta_p \eta_q \eta_r \eta_s \rangle \; (-r)^{-p-q-r-s} 
\\ \nonumber = 12 \;\langle \tau^2 \rangle \langle \tau \rangle^2 \langle \eta^2 \rangle^2 \left( \sum_{i=0}^{m-3} \sum_{k=i+1}^{m-2} \sum_{l=k+1}^{m-1} + \sum_{k=0}^{m-3} \sum_{i=k+1}^{m-2} \sum_{l=k+1}^{m-1} +  
\sum_{k=0}^{m-3} \sum_{l=k+1}^{m-2} \sum_{i=l+1}^{m-1} \right)
\\ \times \nonumber (-r)^{2i+k+l} \sum_{p=0}^{i} \sum_{q = 0}^{i} \sum_{ w = 0}^{k} \sum_{s = 0}^{l} \left[ \delta_{p,q} \delta_{w,s} + \delta_{p,w} \delta_{q,s} + \delta_{p,s} \delta_{q,w} \right]
\\ \hspace{-2.2cm} \nonumber  =  12 \;\langle \tau^2 \rangle \langle \tau \rangle^2 \langle \eta^2 \rangle^2 \left( \sum_{i=0}^{m-3} \sum_{k=i+1}^{m-2} \sum_{l=k+1}^{m-1} + \sum_{k=0}^{m-3} \sum_{i=k+1}^{m-2} \sum_{l=k+1}^{m-1} + \sum_{k=0}^{m-3} \sum_{l=k+1}^{m-2} \sum_{i=l+1}^{m-1} \right) (-r)^{2i+k+l}  \\ \hspace{-1.cm} \times \nonumber \left[ (i+1) (\textrm{min}(k, l) + 1) + 2 (\textrm{min}(i, k) + 1)\;(\textrm{min}(i, l) + 1) \right] \\
\simeq 
\left \{\begin{array}{ccc} 
\frac{36}{5} \left( \frac{\sigma^4 m^5}{\beta^4}\right) + \mathcal{O}(m^4) \qquad \textrm{for}\; r = -1 \\
-7 \left( \frac{\sigma m}{\beta}\right)^4 + \mathcal{O}(m^3) \qquad \textrm{for}\; r = 1
\end{array}
            \right. \label{eq38}.
\eea

\bea
\hspace{-2.5cm} \nonumber 
\langle x_m^4\rangle_{i \ne j \ne k \ne l} = 4! \sum_{i=0}^{m-4} \sum_{j=i+1}^{m-3} \sum_{k=j+1}^{m-2} \sum_{l=k+1}^{m-1} \langle \tau_i \tau_j \tau_k \tau_l \rangle 
\\ \times \nonumber \sum_{p=0}^{i} \sum_{q = 0}^{j} \sum_{ w = 0}^{k} \sum_{s = 0}^{l} \langle \eta_p \eta_q \eta_w \eta_s \rangle (-r)^{i+j+k+l-p-q-w-s} 
\\ \nonumber  = 4! \;\langle \tau\rangle^4 \langle \eta^2 \rangle^2 \sum_{i=0}^{m-4} \sum_{j=i+1}^{m-3} \sum_{k=j+1}^{m-2} \sum_{l=k+1}^{m-1} (-r)^{i+j+k+l} 
\\ \times \nonumber \sum_{p=0}^{i} \sum_{q = 0}^{j} \sum_{ w = 0}^{k} \sum_{s = 0}^{l} \left[ \delta_{p,q}  \delta_{w,s} + \delta_{p,w} \delta_{q,s} + \delta_{p,s} \delta_{q,w} \right] 
\\ \hspace{-2.5cm} \nonumber  = 4! \;\langle \tau\rangle^4 \langle \eta^2 \rangle^2 \sum_{i=0}^{m-4} \sum_{j=i+1}^{m-3} \sum_{k=j+1}^{m-2} \sum_{l=k+1}^{m-1} (-r)^{i+j+k+l} \left[ (1+i) (1+k) + 2 (1+i) (1+j) \right] \\ \hspace{-1.5cm}
\simeq 
\left \{\begin{array}{ccc} 
\frac{1}{3} \left( \frac{\sigma^4 m^6}{\beta^4}\right) + \mathcal{O}(m^5) \qquad \textrm{for}\; r = -1 \\
\frac{7}{4} \left( \frac{\sigma m}{\beta}\right)^4 + \mathcal{O}(m^3) \qquad \textrm{for}\; r = 1
\end{array}
            \right. \label{eq39}.
\eea
Now, adding Eqs.~\eref{eq35}, \eref{eq36}, \eref{eq37}, \eref{eq38}, and \eref{eq39} and keeping only leading order contributions, we  get
\bea
\langle x_m^4\rangle \simeq 
\left \{\begin{array}{ccc}
\frac{1}{3} \left( \frac{\sigma^4 m^6}{\beta^4}\right) + \mathcal{O}(m^5) & \textrm{for} \; r = -1 \\
\frac{7}{4} \left( \frac{\sigma m}{\beta}\right)^4 + \mathcal{O}(m^3) & \textrm{for} \; r = 1.
\end{array}
            \right. \label{x_m^4_app_rpm1}.
\eea
From the above calculation it is quite clear that the dominant contribution of $\langle x_m^4\rangle$ for $r = -1$ is coming from off-diagonal elements of the matrix $\Sigma_m$ which are different from each others, whereas for $r = 1$ dominant contributions are coming from the both diagonal and off-diagonal terms.

\section{Determination of higher order moments of position $x(t)$ for $r=1$ and $0<\alpha \leq 2$ in case III}
\label{HOM-r-1-case-III}

Moments of any order can be computed from the generating function $\tilde{P}_x(k,s)$ in Eq.~\eref{tildeP_k_s}. by taking derivatives with respect to $k$. The Laplace transform of the $2n^{\textrm{th}}$ order moment is given by
\bea
\hspace{-2.7cm}
\widetilde{\langle x^{2n}\rangle}(s) &= (-1)^n \frac{d^{2n}}{dk^{2n}} \tilde{P}_x(k,s) = \sum_{m=0}^\infty\sigma^{2n} \frac{1}{n!} \frac{(2n)!}{2^n} \left \langle \left ( \bar{\tau}_{m}^{T}\Sigma_{m}\bar{\tau}_{m} \right)^{n} e^{-s(\tau_0+\tau_1+...+\tau_{m-1}+\tau_m^*)} \right \rangle.
\label{tilde-x_m^2n(s)}
\eea
Using the explicit form of the matrix $\Sigma_m$ in Eq.~\eref{ele_m} one can write
\bea 
\hspace{-2cm}
\nonumber
\bar{\tau}_{m}^{T}\Sigma_{m}\bar{\tau}_{m}&=\left( \mathbb{I}(m \ge 1) \sum_{l=0}^{m-1} \tau^{2}_{l} \sum_{j=0}^{l} r^{2 j}  + {\tau^{*}_m}^{2} \sum_{l=0}^{m} r^{2 l}   
+ 2\;\mathbb{I}(m \ge 2) \sum_{l=0}^{m-1}  \tau_l\; \right. \\
& \left. \times \sum_{j=l+1}^{m-1} \tau_{j} \; (-r)^{j-l} \sum_{p=0}^{l} r^{2 p}  +  2 \; \mathbb{I}(m \ge 1) \; \tau^{*}_m \sum_{l=0}^{m-1}  \tau_l\; (-r)^{m-l} \sum_{p=0}^{l} r^{2 p} \right)
\eea
which for $r=1$ becomes
\bea
\hspace{-2cm}
\nonumber
\bar{\tau}_{m}^{T}\Sigma_{m}\bar{\tau}_{m}&=\left( \mathbb{I}(m \ge 1) \sum_{l=0}^{m-1} \tau^{2}_{l} (l+1)  + {\tau^{*}_m}^{2} (m+1)   
+ 2\;\mathbb{I}(m \ge 2) \sum_{l=0}^{m-1}  \tau_l \right. \\
& \left. \times \;\sum_{j=l+1}^{m-1} \tau_{j} \; (-1)^{j-l} (l+1) +  2 \; \mathbb{I}(m \ge 1) \; \tau^{*}_m \sum_{l=0}^{m-1}  \tau_l\; (-1)^{m-l} (l+1) \right).
\eea
Now we would like approximate this expression by using the fact that for $r=1$, as can be observed in fig.~\ref{pdf_pl_rpm1_a<1}($2^{\textrm{nd}}$ row), fig.~\ref{pdf_pl_r1_agt1}(b) and fig.~\ref{pdf_pl_r1_1020}, the position $x(t)$ for large $t$ gets dominant contribution from the jump of longest duration {\it i.e.} $x(t) \simeq \Delta_{\max}(t)$. Using this fact we identify the contribution from this jump only and disregard contribution from other steps or correlations with other steps. Hence retaining contributions from longest jumps only we get 
\bea
(\bar{\tau}_{m}^{T}\Sigma_{m}\bar{\tau}_{m})^n& \simeq \left( \mathbb{I}(m \ge 1) \sum_{l=0}^{m-1} \tau^{2n}_{l} (l+1)^n  + {\tau^{*}_m}^{2n} (m+1)^n \right ). 
\eea
The first term represents the contribution from the event in which the longest jump occurs at $l$th step [$l=0,1,...,(m-1)$] and the second term represents the event in which the longest jump occurs in the last incomplete step in a trajectory of duration $t$ having $m$ steps. 
Using this approximate expression of $(\bar{\tau}_{m}^{T}\Sigma_{m}\bar{\tau}_{m})^n$ in Eq.~\eref{tilde-x_m^2n(s)} and simplifying we get 
\bea
\hspace{-2cm} \nonumber
\widetilde{\langle x^{2n}\rangle}(s) \simeq \sum_{m=0}^\infty\sigma^{2n} \frac{1}{n!} \frac{(2n)!}{2^n} 
\\ \hspace{-2cm} \times \nonumber \left \langle  \left( \mathbb{I}(m \ge 1) \sum_{l=0}^{m-1} \tau^{2n}_{l} (l+1)^n  + {\tau^{*}_m}^{2n} (m+1)^n \right ) e^{-s(\tau_0+\tau_1+...+\tau_{m-1}+\tau_m^*)} \right \rangle 
\\ \hspace{-2cm} \nonumber
\simeq \sigma^{2n}\frac{(2n)!}{2^n n!} \sum_{m=0}^\infty \left( \mathbb{I}(m \ge 1) \tilde{\Psi}(s) \tilde{\psi}^{(2n)}(s) H_{m,-n} \tilde{\psi}(s)^{m-1} + (m+1)^n \Psi^{(2n)}(s)\tilde{\psi}(s)^{m}\right) 
\\ \hspace{-2cm} \nonumber
\simeq \sigma^{2n}\frac{(2n)!}{2^n n!} \left( \tilde{\Psi}(s) \tilde{\psi}^{(2n)}(s) \frac{Li_{-n}(\tilde{\psi}(s))}{(1-\tilde{\psi}(s))\tilde{\psi}(s)} + \Psi^{(2n)}(s)\frac{Li_{-n}(\tilde{\psi}(s))}{\tilde{\psi}(s)} \right) \\ \hspace{-2cm} 
\simeq \sigma^{2n}\frac{(2n)!}{2^n } \frac{Li_{-n}(\tilde{\psi}(s))(1-\tilde{\psi}(s))^{n+1}}{\tilde{\psi}(s) n!}\left( \frac{\tilde{\psi}^{(2n)}(s)+s~\Psi^{(2n)}(s)}{s(1-\tilde{\psi}(s))^{n+1}}\right),
\eea
where $H_{m,-n}$ is Harmonic number and $Li_{-n}(z)$ is PolyLog function. In going from the $3^{\textrm{rd}}$ line to the  $4^{\textrm{th}}$ line, we have used $\tilde{\Psi}(s) =\frac{1-\tilde{\psi}(s)}{s} $. Using the following $s \to 0$ limit 
$\frac{Li_{-n}(\tilde{\psi}(s))(1-\tilde{\psi}(s))^{n+1}}{\tilde{\psi}(s) n!} \to 1$, we get 
\bea
\widetilde{\langle x^{2n}\rangle}(s)    & \simeq \sigma^{2n}\frac{(2n)!}{2^n }\left( \frac{\tilde{\psi}^{(2n)}(s)+s~\Psi^{(2n)}(s)}{s(1-\tilde{\psi}(s))^{n+1}}\right). \label{tilde-x^2n(s)} 
\eea
We now use the small $s$ behaviour of $\tilde{\psi}(s)$ and $\tilde{\Psi}(s)$ in different ranges of $\alpha$ to compute the small $s$ behaviour of $\widetilde{\langle x^{2n}\rangle}(s) $.\\

\noindent
In this regime $\tilde{\psi}(s) \simeq 1- \Gamma(1-\alpha)s^\alpha + O(s)$ for small $s$ as can be seen from Eq.~\eref{psi_s->0}. Hence $\tilde{\Psi}(s) \simeq \Gamma(1-\alpha) s^{\alpha-1}$.  Using these approximation in 
Eq.~\eref{tilde-x^2n(s)} one gets $\widetilde{\langle x^{2n}\rangle}(s)  \propto s^{-(2+\alpha)n-1}$ for small $s$ which via Tauberian theorem provides us
\bea
\langle x^{2n}(t)\rangle \sim t^{2n(1+\alpha/2)},~~\textrm{for}~n=1,2,3,... \textrm{and}~~\alpha<1.
\eea

\noindent
In this regime of $\alpha$, $\tilde{\psi}(s) \simeq 1 - \frac{\alpha s}{\alpha - 1} - \Gamma(1-\alpha) s^{\alpha} + \mathcal{O}(s^2) $ implying $\widetilde{\langle x^{2n}\rangle}(s)  \propto s^{-(3n+2-\alpha)}$ which through Tauberian theorem provides 
\bea
\langle x^{2n}(t)\rangle \sim t^{(3n+1-\alpha)},~~\textrm{for}~n=1,2,3,... \textrm{and}~~1<\alpha<2.
\eea

\noindent
Similarly using small $s$ behaviour of $\tilde{\psi}(s)$ and $\tilde{\Psi}(s)$ in Eq.~\eref{tilde-x^2n(s)} for other different ranges of $\alpha$, one can obtain the moments $\langle x^{2n}(t)\rangle$ at large $t$. Below we summarise the results
\begin{align}
\langle x^{2n}(t)\rangle \sim 
\begin{cases}
t^{2n(1+\alpha/2)},~&\text{for}~\alpha<1 \\
 & \\
\frac{t^{3n}}{(\ln t)^{n+1}},~&\text{for}~\alpha=1 \\
 & \\
t^{(3n+1-\alpha)},~&\text{for}~1<\alpha<2  \\
 & \\
\frac{1}{4}t^2 \log(t) \;\delta_{n,1}  + \mathbb{I}(n \ge 2)\; t^{(3n+1-\alpha)},~&\text{for}~\alpha=2.
\end{cases}
\label{<x^2n>-vs-t-r_1}
\end{align}
Here $\delta_{n,m}$ is Kronecker delta function which yields $1$ if $n=m$, otherwise it is $0$. $\mathbb{I}(z)$ is the indicator function. 
Note that for $\alpha > 2$ contribution of total displacement arises from  all the `jumps' taken within a given time and hence  the approximation $x(t) \simeq \Delta_{\max}(t)$ no longer remain valid.

\vspace{1cm}
\bibliography{citation_dg} 
\bibliographystyle{ieeetr}

\end{document}